\documentclass[aps,a4paper]{quantumarticle}
\pdfoutput=1
\usepackage{mathrsfs}
\usepackage{swstyles}
\usepackage{physics}
\usepackage{swphysics}
\usepackage{tikz}
\usetikzlibrary{decorations.pathreplacing,calligraphy,math}

\usepackage[numbers]{natbib}
\newcommand{\onlinecite}[1]{~\hspace{-1ex}\nocite{#1}\citenum{#1}} 

\usepackage[dvipsnames]{xcolor} % nicer colors
\usepackage[normalem]{ulem} % for \sout
\definecolor{tolred}{RGB}{221,110,121}
\definecolor{tolgreen}{RGB}{67,134,62}
\definecolor{tolblue}{RGB}{80,118,166}
\definecolor{tolyellow}{RGB}{201,187,88}
\definecolor{tolsky}{RGB}{128,201,234}
\definecolor{tolpurple}{RGB}{157,60,117}
\definecolor{tolgrey}{RGB}{187,187,187}
\definecolor{tolgray}{RGB}{187,187,187}

\usepackage{hyperref}
\hypersetup{
    colorlinks = true,
    linkcolor = black,%
    citecolor = magenta,
    urlcolor = tolblue}%
\usepackage[capitalise]{cleveref} % for better referencing

\Crefname{figure}{Figure}{Figures}
\crefname{figure}{Fig.}{Figs.}
\Crefname{equation}{Equation}{Equations}
\crefname{equation}{Eq.}{Eqs.}
\Crefname{section}{Section}{Sections}
\crefname{section}{Sec.}{Secs.}
\Crefname{appendix}{Appendix}{Appendices}
\crefname{appendix}{Appendix}{Appendices}

\crefformat{equation}{Eq.~(#2\textcolor{MidnightBlue}{#1}#3)}
\Crefformat{equation}{Equation~(#2\textcolor{MidnightBlue}{#1}#3)}

\crefrangeformat{equation}{Eqs.~(#3\textcolor{MidnightBlue}{#1}#4) to~(#5\textcolor{MidnightBlue}{#2}#6)}
\Crefrangeformat{equation}{Equations~(#3\textcolor{MidnightBlue}{#1}#4) to~(#5\textcolor{MidnightBlue}{#2}#6)}

\crefmultiformat{equation}{Eqs.~(#2\textcolor{MidnightBlue}{#1}#3)}%
{ and~(#2\textcolor{MidnightBlue}{#1}#3)}%
{, (#2\textcolor{MidnightBlue}{#1}#3)}%
{, and~(#2\textcolor{MidnightBlue}{#1}#3)}

\Crefmultiformat{equation}{Equations~(#2\textcolor{MidnightBlue}{#1}#3)}%
{ and~(#2\textcolor{MidnightBlue}{#1}#3)}%
{, (#2\textcolor{MidnightBlue}{#1}#3)}%
{, and~(#2\textcolor{MidnightBlue}{#1}#3)}

\makeatletter
\renewcommand{\eqref}[1]{\textup{(\hyperref[#1]{\textcolor{MidnightBlue}{\ref*{#1}}})}}
\makeatother

\makeatletter
\newcommand{\subeqref}[2]{\textup{(\hyperref[#1]{\textcolor{MidnightBlue}{\ref*{#1}#2}})}}
\makeatother

\makeatletter
\newcommand{\nopareneqref}[1]{\textup{\hyperref[#1]{\textcolor{MidnightBlue}{\ref*{#1}}}}}
\makeatother

\usepackage{placeins}
\usepackage{subcaption}
\usepackage{booktabs}
\usepackage{amsmath}

\usepackage{amsthm}

\newtheorem{theorem}{Theorem}
\newtheorem{result}{Result}
\newtheorem{definition}{Definition}
\newtheorem{example}{Example}

\usepackage{cancel}

\tikzmath{\lat=1; \del=0.2;}

\newcommand{\upd}[1]{^\mathrm{#1}}
\newcommand{\ind}[1]{_\mathrm{#1}}

\newcommand{\sqdiamond}{\mathbin{\tikz[baseline=-0.7ex] \draw (0,0.9ex) -- (0.9ex,0) -- (0,-0.9ex) -- (-0.9ex,0) -- cycle;}}

\newcommand{\pic}[2][scale=1.2]{
    \vcenter{\hbox{\includegraphics[#1]{dg/#2.pdf}}}
}

\newsavebox{\myobjbox}
\newcommand{\punct}[3]{%
  \sbox{\myobjbox}{$\displaystyle #1$}% Put object in a math box to measure it
  \usebox{\myobjbox}% Print the object
  #2% Print the middle symbols
  \raisebox{\dimexpr -\dp\myobjbox +1ex\relax}{$#3$}% Drop the punct to the exact bottom of the box
}

\newcommand{\fpic}[1]{
    \pic[scale=0.5]{#1}
}

\makeatletter
    \newcommand{\showfont}{
            encoding: \f@encoding{}
            family: \f@family{}
            series: \f@series{}
            shape: \f@shape{}
            size: \f@size{}
    }
\makeatother

\begin{document}
\setcounter{tocdepth}{2}
% Todos

% \begin{itemize}
%     \item SW: Go over whole paper and find grammatical / structural errors, write little introductory paragraphs
%     \item capitalization: appendix
% \end{itemize}

% \clearpage
% \newpage

\title{Influence-solvability: a systematic theory of $(1+1)D$ solvability and its application to brickwork circuits}
\author{Friedrich Hübner}
\email{friedrich.huebner@phys.ens.fr}
\affiliation{Department of Mathematics, King's College London, Strand, London WC2R 2LS, United Kingdom}
\affiliation{Laboratoire de Physique de l’École Normale Superieure, CNRS, ENS \& Université PSL, Sorbonne Université, Université Paris Cité, 75005 Paris, France}
\author{Sun Woo P. Kim}
\thanks{\texttt{swk34} \texttt{[at]} \texttt{cantab} \texttt{[dot]} \texttt{ac} \texttt{[dot]} \texttt{uk}}
\affiliation{Department of Physics, King's College London, Strand, London WC2R 2LS, United Kingdom}
% \email{swk34@cantab.ac.uk}

\begin{abstract}
`Solvable' circuits, such as dual unitaries and its generalisations, have arisen as paradigmatic examples of tractable chaotic non-equilibrium dynamics, both in classical and quantum systems. However, while increasingly complicated sufficient conditions have been proposed, a systematic theory classifying and understanding general features of Solvable circuits is missing. In this work we develop such a theory by deriving the most general local necessary and sufficient conditions under which a $(1+1)D$ circuit is Solvable. Concretely, we define a circuit to be \emph{influence-solvable} if its influence matrix, representing the `bath' generated by its own evolution, is given by a uniform MPS with finite bond-dimension $\chi$. This property allows for efficient computation of subsystem dynamics and essentially contains all known examples of Solvable circuits. The set of necessary and sufficient local conditions follows from a new version of the fundamental theorem of MPS for open boundary conditions. Next, we apply our theory to brickwork circuits with $\chi=1$ influence-solvability and perform a systematic classification of classical brickwork circuits with local dimension up to $d=3$ and quantum brickwork circuits with $d=2$. Our search reveals new Solvable circuits that are not captured by known conditions.
%`Solvable' circuits, such as dual unitaries and its generalisations, have arisen as paradigmatic examples of tractable chaotic non-equilibrium dynamics, both in classical and quantum systems. However, while increasingly more complicated sufficient conditions have been proposed, a systematic theory classifying and understanding general features of solvable circuits is missing. We develop such a theory by introducing \emph{influence-solvable} circuits, defined as the class of $(1+1)D$ circuits whose influence matrix, which represents the `bath' generated by its own evolution, is given by a uniform MPS with finite bond-dimension $\chi$. This property allows for efficient computation of subsystem dynamics and essentially contains all known examples of solvable circuits. We derive a set of necessary and sufficient local conditions by using a version of the fundamental theorem of MPS for open boundary conditions. Next we apply our theory to brickwork circuits with $\chi=1$ influence-solvability and perform a systematic classification of classical brickwork circuits with local dimension up to $d=3$ and quantum brickwork circuits with $d=2$. Our search reveals new solvable circuits that are not captured by known solvability conditions.
\end{abstract}

\maketitle

\section{Introduction}
Construction of exact analytic solutions is the best-case scenario when studying systems in theoretical physics. They are of particular importance for many-body systems because, although we have access to vast numerical resources and methods for simulations today, it is still extremely difficult to obtain precise numerical simulations for large, strongly interacting systems, especially for quantum systems. Furthermore, there are no known frameworks to obtain exact analytic solutions for generic many-body systems, and examples of exact results are few and far between.

Therefore, when such results exist, they act as a lighthouse: (a) they act as an important check to compare asymptotic or more uncontrolled methods, such as perturbative approaches~\cite{altland2010condensed,kim2021real,coleman2015introduction}, Boltzmann equations, mean-field approximations~\cite{kardar2007statistical}, renormalisation group~\cite{kardar2007statistical,altland2010condensed}, replica tricks~\cite{mezard1988spin,PhysRevX.15.011015}, hydrodynamics~\cite{doyon2020lecture,kim2025circuits}, and more. (b) The framework surrounding the exact solution become a useful lens to view similar systems which may be no longer exactly solvable. Such singular examples of exactly solvable systems include all-to-all interacting models~\cite{sachdev1992gapless,mezard1988spin}, models defined on the Bethe lattice~\cite{baxter1985exactly,p2025planted,derrida1986solution}, and frustration-free projector Hamiltonians~\cite{kitaev2006anyons,levin2005string,affleck1988valence}.

Integrable~\cite{jscaux,caux2011remarks,arutyunov2020elements,Babelon_Bernard_Talon_2003,Korepin_Bogoliubov_Izergin_1993} and non-interacting (e.g. free fermions and bosons) models \cite{lieb1961two,kitaev2001unpaired,peschel2009reduced} are other families of many-body systems which admit exact microscopic solutions. However, due to the infinite conservation laws they possess, most aspects of their properties are non-generic; for example, they do not equilibrate to conventional thermal states~\cite{Vidmar_2016}.

This situation explains the excitement that was generated when recently, a flurry of ``Solvable'' many-body systems were found, with its start marked by the discovery of dual-unitary circuits~\cite{gopalakrishnan2019unitary,bertini2019exact,yx73-dk86}. Dual-unitary circuits are examples of circuit models, which are time-discretised versions of classical and quantum many-body dynamics composed of local components called `gates'. In the classical case, circuits are special types of cellular automata, which have been studied for a long time, and currently in renaissance, partially due to such Solvable examples. On the other hand, quantum circuits are natural ways to discretise time-continuous quantum systems for classical simulation (e.g. via time-evolving block-decimation, TEBD~\cite{PAECKEL2019167998,annurev:/content/journals/10.1146/annurev-conmatphys-040721-022705}), or digital quantum simulation on quantum computers~\cite{RevModPhys.86.153,Keenan2023}. In particular, Solvable circuits can be used to benchmark quantum computers against exact results~\cite{Fischer2026,zn6p-c8w2}.

Generically, circuit systems are as hard to solve as their time-continuous counterparts. Nevertheless, in these Solvable circuits, it is possible to derive exact microscopic dynamics of certain quantities for special initial states, e.g. the infinite temperature state for dual-unitaries. Construction of Solvable circuits has been based on simplifying the `evolution' in space instead of time, which in turn has been afforded by imposing certain local conditions in an ad-hoc fashion, followed by a search for gates that satisfy such conditions. After the introduction of dual-unitary circuits, generalisations such as $\mathrm{DU}(2)$~\cite{Yu2024hierarchical} and others have been found~\cite{PhysRevLett.126.160602,10.21468/SciPostPhys.11.6.106,10.21468/SciPostPhys.11.6.107,PhysRevLett.133.170402,PhysRevResearch.7.L012011,Kos2023circuitsofspacetime,PhysRevLett.126.100603,Claeys_2024,rampp2025solvablequantumcircuitsspacetime,wang2025temporalcomplexityhierarchiessolvable}.  Solvability appears to be independent of but not mutually exclusive to integrability, another notion of tractability in many-body systems. Due to these reasons, Solvable circuits have become standard systems to analytically study phenomena of non-integrable many-body systems~\cite{Bertini2021,Claeys2022emergentquantum,MontaLpez_2024,PRXQuantum.6.010324,PhysRevB.104.014301,PhysRevB.104.214303,PhysRevB.107.174311,PhysRevE.103.062133,PhysRevLett.121.264101,PhysRevLett.130.130402,PhysRevLett.131.180403,PhysRevLett.134.050405,PhysRevResearch.2.043403,PhysRevResearch.4.043212,PhysRevResearch.6.033271,PhysRevResearch.6.043077,PhysRevX.12.031016,kos2026vanishingcorrelationsbistochasticcontrolled,tan2026logarithmicgrowthoperatorentanglement,t6zs-3f8k,qpsz-pvxm,pathak2026mixedstateentanglementminimalmodel}. While dual-unitaries appear to be `generic' in one sense, as they display ergodic behaviour, they seem to be fine-tuned in another, with correlations being non-zero only on the light cone~\cite{bertini2019exact}. Other Solvable circuits do not share all the features of dual-unitaries: for example, $\mathrm{DU}(2)$ allow non-zero correlations away from the light cone, near the origin~\cite{Yu2024hierarchical}. 

What is missing, however, is a systematic theory of Solvable circuits. Given a circuit specifying a particular time-evolution, under which conditions can there exist Solvability, and under which others can Solvability be ruled out? What is the most general necessary and sufficient condition for Solvability? How can we classify all Solvable circuits, and to what extent are their dynamics generic or non-generic?

In this work, we build such a general theory for one-dimensional circuits, in both classical and quantum settings. We propose \emph{influence-solvability}, which we believe to be the most general definition of Solvability in $(1+1)D$ circuits. Influence-solvability is fulfilled if the influence matrix of the circuit (i.e. a cut of the circuit at a fixed position) has a particularly simple form for all $t$: a uniform matrix product state (MPS) of a finite bond dimension $\chi$, i.e. an MPS constructed by contractions of identical tensors $\bm{A}$ (see \Cref{fig:intro1}). We prove the most general necessary and sufficient \emph{local, finite-dimensional} conditions for influence-solvability, and this is our first main result. This is based on a novel open-boundary variant of the fundamental theorem of MPS~\cite{Fannes1992,annurev:/content/journals/10.1146/annurev-conmatphys-031016-025507,10.1063/1.5000784,RevModPhys.93.045003,floridollinas2025uniformmatrixproductstates}. While we illustrate our method on brickwork circuits, our approach and definitions are straightforwardly generalised to any one-dimensional circuit systems.

Our definition of influence-solvability contains virtually all known examples of Solvable circuits, including dual-unitaries, $\mathrm{DU}(2)$, $\mathrm{DU}(2)$-like circuits such as the Floquet East model~\cite{PhysRevLett.132.120402,PhysRevE.110.L022101,de2024exact}, circuits satisfying the zipper~\cite{PhysRevLett.133.170402} condition, and others like Rule 54~\cite{PhysRevLett.126.160602,10.21468/SciPostPhys.11.6.106,10.21468/SciPostPhys.11.6.107}.

If a circuit is influence-solvable, then the usually intractable `bath' generated by the bulk of the system of unbounded size can be replaced by an effective, finite one. Such dynamics can be solved for arbitrarily long times $t$ by diagonalisation of a finite-dimensional matrix, avoiding the usual exponential scaling with $t$. Our most general definition of influence-solvability allow for the exact computation of finite subsystem dynamics in many situations, illustrated in \Cref{fig:intro2}. Moreover, in some circuits, such as dual-unitaries, it is possible to obtain further results e.g. correlation functions.

Viewing the influence matrix as the bath generated by the circuit itself, $\chi=1$ influence-solvability corresponds to the case where the bath is exactly Markovian, which is the time-discrete equivalent of the effective Lindblad description in time-continuous systems. While the Lindbladian description can only be justified for certain scaling limits (a standard example is the Caldeira-Leggett model~\cite{CALDEIRA1983587}) for time-continuous systems, for these circuit systems, Markovianity is microscopically exact. Furthermore, $\chi>1$ can be viewed as finite memory that the bath retains, which means that the bath is an exact hidden Markov model~\cite{PhysRevLett.133.170402}.

The rest of the paper is structured as follows. In \cref{sec:overview} we give an overview over our results. In \Cref{sec:mot}, we motivate our definition of influence-solvability and review (brickwork) circuits. Next, we state a necessary and sufficient local condition for influence-solvability in \Cref{sec:FTMPS}. Then, in \Cref{sec:zerobond}, we focus on the case of exact Markovian baths ($\chi=1$) and use our theory to systematically classify all classical gates up to $d=3$ and quantum gates up to $d=2$. We further discuss other aspects of influence-solvability in \Cref{sec:prop}, including its relationship to conserved quantities, $\chi>1$, relationship between classical and quantum influence-solvability, and generalisation to other quantities such as entanglement entropy. We close with further discussion and outlook. 

\section{Overview of new contributions}\label{sec:overview}
\begin{figure}
    \centering
    \begin{align*}
    \cdots \pic{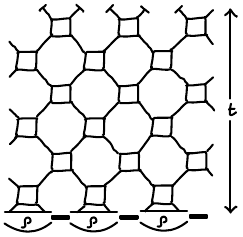} = \pic{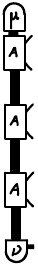}
    \end{align*}
    \caption{We say a circuit is influence-solvable if its influence-matrix is given by a uniform MPS of finite-bond dimension $\chi$ for all times $t$. Note that $\bm{A}$ is independent of $t$.}
    \label{fig:intro1}
\end{figure}

The main contribution of this work is to establish the most general local condition under which a brickwork circuit is Solvable. In the context of this paper we mean that a circuit is (left) Solvable if its (left) influence matrix has the form of a uniform MPS of finite bond dimension for all times $t$ as in \Cref{fig:intro1}. We call such circuits to be (left) \emph{influence-solvable} to avoid confusion with other systems that allow for exact or efficient solutions such as integrable~\cite{PhysRevLett.121.030606,PhysRevLett.130.260401,10.21468/SciPostPhys.16.3.078,10.21468/SciPostPhys.16.4.114,Pozsgay_2014,Pozsgay_2016,Piroli_2019} or non-interacting systems~\cite{10.1098/rspa.2008.0189,Eisler_2007,PhysRevLett.97.156403,PhysRevA.65.032325}. Similarly, one can define right influence-solvable circuits. This class of influence-solvable circuits contains, with very exceptions, all examples that have been recently identified under the umbrella term ``Solvable circuits'',  including dual-unitaries, $\mathrm{DU}(2)$, $\mathrm{DU}(2)$-like circuits such as the Floquet East model~\cite{PhysRevLett.132.120402,PhysRevE.110.L022101,de2024exact}, circuits satisfying the zipper~\cite{PhysRevLett.133.170402} condition, and others like Rule 54~\cite{PhysRevLett.126.160602,10.21468/SciPostPhys.11.6.106,10.21468/SciPostPhys.11.6.107}. The only exception are $\mathrm{DU}(n \geq 3)$ circuits~\cite{Yu2024hierarchical}. However, to best of our knowledge, these circuits do not allow computation of exact dynamics, and one can only show that the maximum speed of propagation is less than the light cone velocity. This is an exact analytical insight, but we do not consider it sufficient to label these gates Solvable.

\begin{figure}
    \centering
    \begin{align*}
 \pic{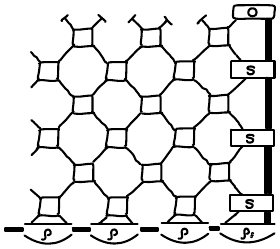} = \pic{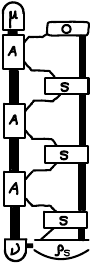} 
    \end{align*}
    \caption{Using influence-solvability one can simplify infinite circuits into an efficiently computable finite tensor network. Here, evolution of an observable $O$ of a joint system between $S$ coupled to an infinite circuit. The tensor $\bm{A}$ represents an exact finite memory bath generated by the circuit on the left.}
    \label{fig:intro2}
\end{figure}

When a circuit is influence-solvable one can indeed compute exact dynamics. For instance, when coupling the circuit to another system $S$ as in \Cref{fig:intro2}, one can simplify the tensor network to be one-dimensional with finite bond dimensions such that it can be exactly computed for arbitrarily long times $t$ by diagonalizing a finite-dimensional matrix. Physically, this corresponds to ``integrating out the bath'' represented by the circuit and replacing it by an effective finite dimensional (non-unitary) system. This can also be interpreted as the fact that bath can only store a finite amount of information, characterised by the bond dimension $\chi$ of the MPS. The setup of \Cref{fig:intro2} is one potential application of influence-solvability, but it can be applied to many more problems depicted in \Cref{fig:mot_infsolv_applications}, such as equilibration, autocorrelation functions, impurity problems and interfaces between different systems.

The condition that the influence matrix is a uniform MPS for all $t$ is difficult to check directly. In section \cref{sec:FTMPS} we prove a novel version of the fundamental theorem of MPS specifically for open boundary conditions (\cref{thm:local_condition_MPS}) and use it to derive a set of sufficient and necessary local conditions for a circuit to be influence-solvable, see \eqref{eq:FTMPS_FTMPS_local_conditions}. For a given bond dimension $\chi$, those conditions are finite dimensional equations. The importance of this is two-fold: first, given a guess for the tensor $\bm{A}$ in \Cref{fig:intro1}, one can easily verify that a circuit is influence-solvable and also find all compatible initial states (note that the notion of influence-solvability depends on the initial state of the circuit). Such complete classification of compatible initial states was so far only possible in dual unitary circuits~\cite{PhysRevB.101.094304}. The second advantage of our local conditions is that, for the first time, they can be used to show that a certain circuit cannot be influence-solvable. So far, all previous conditions were only sufficient and hence they cannot be used to rule out influence-solvability.

In \cref{sec:zerobond} we use these local conditions to systematically study the simplest case of bond dimension $\chi=1$. Physically, this means that the circuit can be integrated out to give an exact Markovian bath characterized by a quantum channel $\bm{A}$. We show (\cref{res:zerobond_space_inv}) that this is only possible if either the gate is not spatially-invertible or it satisfies a simple generalization of the dual-unitarity condition, both of which are easy to verify. We then exhaustively classify all $\chi=1$ influence-solvable classical ($d=2,3$) and quantum circuits ($d=2$) for local dimension $d$. We find that all classical $d=2$ and $\approx 32\%$ of the classical $d=3$ gates are influence-solvable for at least some initial state (\cref{res:zerobond_classical}). Among those are gates that are not covered by any previous construction of Solvable circuits. We also identify an influence-solvable circuit that is solvable for a whole family of initial states (\cref{ex:220318}) and another one which has only a single conserved quantity (\cref{ex:996}). The latter is particularly interesting since so far all examples of Solvable circuits either had zero or infinitely many conserved quantities~\cite{10.21468/SciPostPhys.8.4.068,HoldenDye2025fundamentalcharges}. In the case of quantum circuits we find that the only $\chi=1, d=2$ influence-solvable circuits are either dual-unitaries (although they might have further solvable initial states than the infinite temperature state) or certain controlled gates (\cref{res:zerobond_quantum}). We also provide an example of a $d=3$ gate that is not dual-unitary (\cref{ex:non-du-spatially-invertible}).

In \cref{sec:prop} we then establish and discuss general properties of influence-solvable circuits for any $\chi$. For instance, we show that the MPS tensor $\bm{A}$ can always be constructed from the circuit itself.
\\

\noindent \textit{Note added: Shortly after the first preprint of this manuscript was made public, an efficient algorithm was independently developed~\cite{yang2026solvingnonequilibriumdynamicsinfluence} that allows to check whether a circuit is influence-solvable and identifies the tensor $\bm{A}$. In their paper the authors then apply it to identify new solvable initial states for the rule 201 and rule 54 cellular automata, including one with $\chi=54$.}

\tableofcontents

\section{Motivation and definition of influence-solvability}\label{sec:mot}
The aim of this work is to introduce a general framework for the so-called ``Solvable'' circuit dynamics in the literature. There exist many examples of circuit which is referred to as Solvable, such as the dual unitary circuits~\cite{gopalakrishnan2019unitary,bertini2019exact}, its generalised hierarchy~\cite{Yu2024hierarchical} and many more~\cite{PhysRevLett.126.160602,10.21468/SciPostPhys.11.6.106,10.21468/SciPostPhys.11.6.107,PhysRevLett.133.170402,PhysRevResearch.7.L012011,Kos2023circuitsofspacetime,PhysRevLett.126.100603}.

Usually, it does not mean that all types of dynamics can be solved exactly, but rather that there exist certain time-evolved quantities, for instance the evolution of certain local observables or certain correlation functions, which can be. Crucially, the notion of solvable dynamics also depends on the initial state; for example, for dual-unitary circuits, the above quantities can only be computed if the initial state is the infinite temperature state. Solvable circuits are also usually distinguished from other traditional methods of analytically computable dynamics, such as integrable~\cite{PhysRevLett.121.030606,PhysRevLett.130.260401,10.21468/SciPostPhys.16.3.078,10.21468/SciPostPhys.16.4.114,Pozsgay_2014,Pozsgay_2016,Piroli_2019} or non-interacting systems~\cite{10.1098/rspa.2008.0189,Eisler_2007,PhysRevLett.97.156403,PhysRevA.65.032325}. We note that there are circuits that are both traditionally analytically computable and also Solvable, such as the brickwork Floquet XXZ model~\cite{10.21468/SciPostPhys.12.3.102} and Rule 54, which is both integrable and Solvable at certain parameters~\cite{Buca_2021,PhysRevLett.126.160602,10.21468/SciPostPhys.11.6.106,10.21468/SciPostPhys.11.6.107}.

So far, Solvable circuits have been identified by introducing some local conditions, such as dual unitarity, to individual components which allow for a reduction of the circuit into an analytically tractable object. After imposing such conditions, one tries to identify interesting circuits that fulfil them. What is lacking, however, is a general understanding of what are the minimal necessary conditions under which solvability is possible, or equivalently, understanding under which conditions solvability can be ruled out.

Any general study of ``Solvability'' requires a good mathematical and physical definition of what it means. In this work, we will define it as the condition that the influence matrix of the circuit can be written as a uniform MPS of finite bond dimension. We will call this specific type of Solvable circuits \emph{influence-solvable} circuits.

In the following we will motivate this particular choice by two reasons:
\begin{enumerate}
    \item As we will show, virtually all known examples of (one-dimensional) Solvable circuits fall into our definition of influence-solvability.
    \item If a circuit is influence-solvable, then one can use it to make concrete predictions of its dynamics, specifically on the evolution of local observables.
\end{enumerate}

We note here that our definition of influence-solvability is tailored to a particular quantity of interest. Therefore, while influence-solvable circuits allow us to analytically compute the evolution of certain local observables, they may not allow computation correlation functions in general. This is in contrast to dual unitary circuits. Thus, our definition of influence-solvable circuits is a superclass: all known Solvable circuits belong to it, but some classes within the influence-solvable superclass may allow the computation of more quantities.

\subsection{Review of quantum and classical circuit dynamics}
In this subsection, we review circuit dynamics. In general, circuits in the many-body physics context are time-discrete versions of many-body systems. It can describe both classical or quantum dynamics. Here, we restrict ourselves to one-dimensional two-site brickwork circuits, which are the most commonly studied types of circuits. Extension of our methods to other types of one-dimensional circuits, such as three-site brickwork circuits~\cite{10.21468/SciPostPhys.20.2.061}, hexagonal circuits~\cite{PhysRevResearch.7.L012011}, round-a-face~\cite{10.1063/5.0056970} or controlled circuits~\cite{Buca_2021}, should be straightforward.

\subsubsection{Classical brickwork circuits}
A classical brickwork circuit describes the dynamics of a $1D$ lattice $\mathbb{Z}$, where each lattice point $x\in\mathbb{Z}$ has local dimension $d$, i.e. on each lattice point we have a configuration labelled by  $a_x\in\Omega=\qty{0,1,\ldots,d-1}$. We will adopt the notation $\bm{a} := (a_x)_{x\in\mathbb{Z}}$ to denote the configuration on the whole lattice.

The circuit dynamics is defined via a gate, which is a map $g:\Omega^2\to\Omega^2$ acting on two neighboring lattice sites. It is graphically represented by
\begin{align}
    \pic{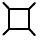}.\label{eq:mot_classic_gate}
\end{align}

The dynamics is now defined as follows. In the first step one applies the gate on sites $2x$ and $2x+1$, i.e. updating configurations $a_{2x}$ and $a_{2x+1}$ by $g_1(a_{2x},a_{2x+1})$ and $g_2(a_{2x},a_{2x+1})$ for all $x\in\mathbb{Z}$. In the second step, the above is applied on sites $2x+1$ and $2x+2$ for all $x$. In the third step, the first step is repeated, then the second step again and so on. Graphically, this can be represented as:
\begin{align}
    \cdots \punct{\pic{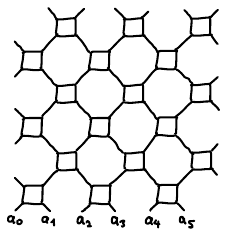}}{\cdots}{\,.} \label{eq:mot_classic_basic}
\end{align}

This provides the evolution for individual configurations. To study the evolution of distributions such configurations, we lift the map $g:\Omega^2\to\Omega^2$ to a linear map $\bm{G}:\mathbb{R}^d\otimes\mathbb{R}^d \to\mathbb{R}^d\otimes \mathbb{R}^d$ defined via
\begin{align}
    \bm{G} \keT{a}_\mathrm{l} \keT{a'}_\mathrm{r} = \keT{g_1(a,a')}_\mathrm{l} \keT{g_2(a,a')}_\mathrm{r},
\end{align}
where $\keT{a}$ is the $a$\textsuperscript{th} unit vector. Here, the tensor product is between the left ($\mathrm{l}$) and right ($\mathrm{r}$) spaces is implied, i.e. $\keT{a}_\mathrm{l} \keT{a'}_\mathrm{r} = \keT{a}_\mathrm{l} \otimes \keT{a'}_\mathrm{r}$.

We will also graphically represent $\bm{G}$ by \eqref{eq:mot_classic_gate}. Then we can represent a probability distribution over all configurations $p(\bm{a})$ as a vector 
\begin{align}
    \keT{\rho} &= \sum_{\bm{a}} p(\bm{a}) \qty[\cdots  \keT{a_{-1}} \keT{a_{0}} \keT{a_{1}} \cdots].  
\end{align}
At this point, the gate $\bm{G}$ can be generalised to probabilistic evolution, i.e. local Markov chains.

Similarly, we can express a general expectation value of some observable observable $O$ on the time-evolved distribution $\keT{\rho_t}$ as an inner product $\brAkeT{O}{\rho_t}$, where
\begin{align}
    \brA{O} &= \sum_{\bm{a}} O(\bm{a}) \qty[\cdots \brA{a_{-1}} \brA{a_0} \brA{a_1} \cdots],  
\end{align}
and $O(\bm{a})$ is the value of the observable for configuration $\bm{a}$. If $O$ is local, i.e. it only depends on a few neighbouring sites (say) $i:j := \{i, i+1, \dots, j\}$, then we can write
\begin{align}
    \brA{O} &= \cdots \brA{-}_{i-1} \qty[\sum_{\bm{a}_{i:j}} O(\bm{a}_{i:j}) \brA{a_i}  \cdots \brA{a_j} ] \brA{-}_{j+1} \cdots,
\end{align}
where we have introduced the $d$-dimensional `flat vector' $\brA{-}=(1,1,\ldots,1)$. 
% Its interpretation is that we discard all information on the site on which we apply it.
We denote it graphically by
\begin{align}
    \pic{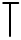}.\label{eq:mot_classicl_flat}
\end{align}
With these definitions we can now graphically represent the expectation value of $O$ after evolving $p$ for some time steps as
\begin{align}
    \cdots 
    \punct{\pic{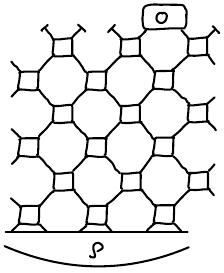}}{\cdots}{\;.} 
    % \raisebox{-6.2 em}{.}
    \label{eq:mot_classic_observable}
\end{align}
In tensor network diagrams, we will use a diagrammatic ket-like $\keT{\rho}$ notation to denote distributions that are properly normalised as $\Big(\cdots\brA{-}\brA{-}\cdots\Big){\rho} = 1$.

Then from the conservation of probability, we have the identity
\begin{align}
    \pic{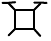} = \pic{flatbra} \; \pic{flatbra}.\label{eq:mot_classic_flatid_top}
\end{align}
Though our methods should hold for non-reversible or probabilistic gates, in this work, we will assume that gates are reversible in time, which implies the identity from the bottom,
\begin{align}
    \pic{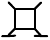} = \pic{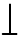} \; \pic{flatket}.\label{eq:mot_classic_flatid_bottom}
\end{align}
Note that (\ref{eq:mot_classic_flatid_top}, \ref{eq:mot_classic_flatid_bottom}) have different physical interpretations. \eqref{eq:mot_classic_flatid_top} denotes that the total probability is conserved, while \eqref{eq:mot_classic_flatid_bottom} denotes that the infinite temperate state, i.e. the distribution where all states are equally likely, is a stationary state of the dynamics. To emphasise this we will define the normalized flat vector or the single-site infinite temperature state as
\begin{align}
    \pic{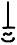} = \keT{\sim} := \frac{1}{d} \pic{flatket} = \keT{-} = \qty(\frac{1}{d}, \dots, \frac{1}{d})^\top.
\end{align}
and write \eqref{eq:mot_classic_flatid_bottom} as
\begin{align}
    \pic{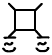} = \pic{tildeket} \quad \pic{tildeket}. \label{eq:mot_classic_flatid_bottom_normalized}
\end{align}

Classical gates that are reversible in time can be viewed as permutation matrices. Thus for each local dimension $d$ there are finitely many such gates which we index by $(d, \sigma)$, where $\sigma$ is the lexicographic index of the permutation~\footnote{
The label $\sigma$ that defines a gate is defined as follows. 
We first identify each two-site configuration $(a, a')$ in $\mathcal{D} \times \mathcal{D}$ with a number $A = da + a'$, such that $(0, 0) \leftrightarrow 0$, $(0, 1) \leftrightarrow 1$, until $(d-1, d-1) \leftrightarrow d^2-1$. Then the action of a gate corresponds to a permutation $(0, 1, \dts, d^2-1) \rightarrow A_{0:d^2-1} = (A_0, \dts, A_{d^2-1})$. The set of all permutations $\{A_{0:d^2-1}\}$ can be then indexed in lexicographic order: $\sigma=0$ corresponds to the trivial permutation $A_{0:d^2-1}=(0, 1, 2, \dts, d^2-3, d^2-2, d^2-1)$, $\sigma=1$ to $A_{0:d^2-1}=(0, 1, 2, \dts, d^2-3, d^2-1, d^2-2)$, and so on.}.
    
Note that the dynamics we described above are for an infinite system. 
% One can similarly define dynamics on a finite system.
One can similarly define dynamics for finite systems or semi-infinite systems. Let us consider the semi-infinite circuit infinite on the left, coupled to another system $\mathrm{S}$,
\begin{align}
    \cdots \punct{\pic{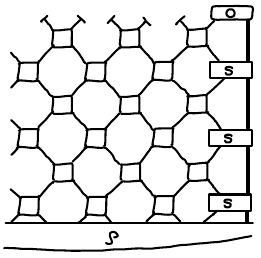}}{\;}{.} % \raisebox{-6.2 em}{.}
    \label{eq:mot_classic_circuit_system}
\end{align}
Here, we have used different thickness for the legs of the system to denote that it may have different dynamics and dimensions to that of the circuit extended to the left.

We may think of \eqref{eq:mot_classic_circuit_system} as joint system between a system $\mathrm{S}$ coupled to a semi-infinite bath given by the circuit specified by gate $g$. Here the system $\mathrm{S}$ can have any (finite or even infinite) number $d\ind{S}$ of internal states $\Omega\ind{S}$. Then the evolution on the system, $g\ind{S}:\Omega\times \Omega\ind{S} \to \Omega\times \Omega\ind{S}$ represented by $\pic{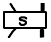}$,
can be an arbitrary map describing a single time step evolution of the system $\mathrm{S}$ taking into account the current value of the ``bath'' on its neighboring side. 

This setup will be the most basic setup to which influence-solvability can be applied. However, it can also be applied to other setups, discussed in \cref{fig:mot_infsolv_applications}.

\subsubsection{Quantum brickwork circuits}

Quantum brickwork circuits can be defined similarly to classical brickwork circuits. They are natural ways to discretise the continuous evolution of quantum system in time via trotterisation, commonly used to simulate such systems~\cite{PAECKEL2019167998}. For simulations on a classical computer, the standard TEBD algorithm computes the time-evolution as a quantum circuit with discretised timestep $\Delta t$. For a fixed $\Delta t$, one obtains a brickwork circuit, which one can be computed up to some required time $t = N \Delta{t}$. To obtain the result for the time-continuous system, one must verify convergence to the  $\Delta t\to 0$ limit, which is costly. On the other hand, if the dynamics is defined on a brickwork circuit to begin with, the verification of $\Delta t \rightarrow 0$ is not required, greatly reducing the numerical cost. Moreover, it is a natural way of simulating dynamics on quantum computers.  Since discrete-time dynamics often show the same features as time-continuous dynamics, in recent years, quantum circuits have emerged as a new paradigm for quantum many-body physics.

For quantum circuits, similar to a classical gate $g$ acting on two sites, one applies a two site unitary matrix $u :\mathbb{C}^d \otimes \mathbb{C}^d \to \mathbb{C}^d\otimes \mathbb{C}^d$ to them. This way one can represent the evolution of any wave function $\ket{\psi_t}$ of the quantum system as
\begin{align}
    \cdots \punct{\pic{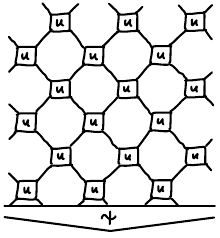}}{ \cdots \;}{.} % \raisebox{-6.2 em}{.}
\end{align}
However, in order to compute the expectation value of an observable $\hat O $ at time $t$, one has to compute $\bra{\psi(t)} \hat O \ket{\psi(t)}$. This can be written as \eqref{eq:unfoldedevolvedop}:
\begin{align} 
    \cdots \pic{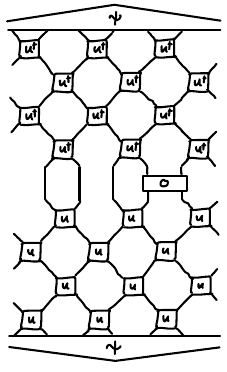} \cdots \qquad \label{eq:unfoldedevolvedop}
\end{align}
\begin{align} 
    \qquad \longrightarrow \quad \cdots \punct{\pic{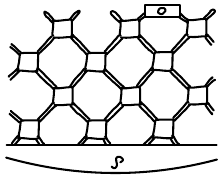}}{ \cdots}{.} \label{eq:foldedevolvedop}
\end{align}

To bring it back into the form that is similar to the classical evolution \eqref{eq:mot_classic_observable} it is common to ``fold'' the upper part onto the lower part as \eqref{eq:foldedevolvedop}. This way one obtains an evolution for the density matrix $\hat{\rho} = \ketbra{\psi}$ in terms of the folded gate
\begin{align}
    \bm{G} = \pic{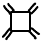} = \hat{u} \otimes \hat{u}^* = \pic{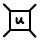} \otimes \pic{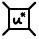}\,.
\end{align}
At this point there is no reason to restrict to pure states: in general $\hat \rho$ can be any mixed state.

The quantum analogue of the ``flat vector'' \eqref{eq:mot_classicl_flat} is given by the identity matrix,
\begin{align}
    \pic{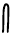} = \pic{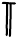} = \brA{\mathbb{1}_d} = \brA{-},
\end{align}
which links the forward and backward evolution copies together. The maximally mixed state on a single site, $\mathbb{1}_d / d$, will be denoted as
\begin{align} \label{eq:tilde-state}
    \pic{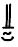} \, = \keT{\sim} = \, \frac{1}{d} \, \pic{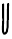} = \frac{1}{d} \, \keT{-}.
\end{align}
Note that any equalities with flat vectors $\keT{-}$ on both the LHS and the RHS can be replaced with that of $\keT{\sim}$.

To unify notation, we will denote both classical vectors and folded quantum vectors by a rounded ket $\keT{\cdot}$. In the quantum case, we will use a hat ($\; \hat{} \;$) to denote an operator acting on the non-folded Hilbert space. For example, for the density matrix operator $\hat{\rho} = \ketbra{\psi}$ the folded ket is $\keT{\rho} = \ket{\psi} \otimes \ket{\psi^*}$. We will use boldface to denote both linear maps acting on probability vectors in the classical case and linear maps acting on operators (which are vectorised in the folded space) in the quantum case.

Although we have composed the evolution on density matrices from unitary evolution, $\pic{channelgate}$ could be more generally a quantum channel acting on two sites. For any channel, the completely-positive trace-preserving (CPTP) property gives
\begin{align}
    \pic{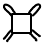} = \pic{identitybra} \quad \, \pic{identitybra} \,.
\end{align}

Though our methods also generalise to channels, we will only consider unitary gates, which has the further property that
\begin{align}
    \pic{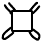} = \pic{identityket} \quad \, \pic{identityket} 
    \quad \Longleftrightarrow \quad \pic{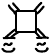} = \pic{foldedtildeket} \quad \pic{foldedtildeket}.
\end{align}

This way, we can see that classical and quantum circuits can be brought into very similar forms and one can foresee that they could be studied by similar methods. Therefore, unless required, we will derive results using the classical circuit notation, which will directly apply also to quantum circuits in the folded picture. In cases where it is required, or only holds for the quantum case, we will explicitly use the folded diagrammatic notation.

\subsection{Influence-solvability and finite-memory circuits}
In general, obtaining analytical results for circuit systems without further imposed structures is as hard as their time-continuous counterparts, in both classical and quantum settings. Apart from non-interacting and integrable systems, time-continuous dynamics is largely analytically intractable. Therefore, we also expect the same for circuits. However, for some circuits, such as dual unitaries, it is possible to exactly compute some aspects of their microscopic dynamics. This is a special feature of circuit dynamics, which does not have a time-continuous counterpart.

Let us restrict to initial states given by a semi-infinite uniform MPS state with bond dimension $\chi_\rho$ (see Appendix \ref{app:mps_space}). In the case of \eqref{eq:mot_classic_circuit_system}, this is given by
\begin{equation}
\begin{aligned}
    & \cdots \pic{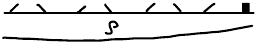} \\
    & \quad \; = \; \cdots \pic{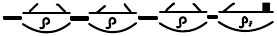} \; .
\end{aligned}
\end{equation}
We will further assume that the tensor $\bm{\rho}$ generates a clustering\footnote{Clustering means that connected correlation functions decay to zero at large distances.} translation invariant MPS state~\cite{Fannes1992}; see Appendix \ref{app:mps_space} for a review of their properties. Examples of such states are product states $\chi_\rho = 1$ (e.g. global infinite temperature state), states prepared by some finite depth circuit starting from a product state, or Gibbs states of finite-ranged Hamiltonians (e.g. that of the 1D transverse field Ising model).

To define influence-solvability we consider cutting the infinite circuit at some point in space, obtaining the following object,
\begin{align}
    \cdots \punct{\pic{influence-matrix-left-t}}{}{,}\label{eq:mot_infsolv_influencematrix}
\end{align}
which is a vector in the spatial direction. This vector is commonly called the ``influence matrix'' of a circuit~\cite{PhysRevX.11.021040,li2026lowrankstructurereduced,carignano2025overcomingentanglementbarriersampled,Carignano2025loschmidtecho,PhysRevResearch.6.033021,cerezoroquebrun2026mesoscopicregimestemporalentanglement} (because in a quantum system it is actually a matrix due to the doubled Hilbert space). For simplicity we will adopt the the name ``influence matrix'' even though it should be referred to as an ``influence vector'' in our formalism. The influence-matrix can be viewed as the time-discrete analog of the influence-functional in the path integral formalism~\cite{CALDEIRA1983587}.

If this influence matrix is known explicitly and has a simple form, then we can efficiently compute circuits such as \eqref{eq:mot_classic_circuit_system}, because this can be written as an inner product between the two vectors,
% \begin{equation}
\begin{align}
    & \cdots \!\!\!\!\!\!\!\! \pic{mpsbathonS} \\
    & = \qty(\cdots \!\!\!\!\!\!\!\! \pic{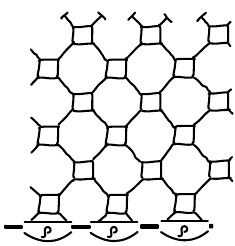}) \bm{\cdot} \punct{\qty(\pic{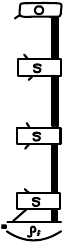})}{}{.}
    \nonumber
\end{align}
% \end{equation}

Note that for any finite $t$, the influence matrix is not dependent on the entire semi-infinite system on the left: one can use the identity  \eqref{eq:mot_classic_flatid_top} to reduce it to
\begin{equation}
\begin{aligned}
    & \pic{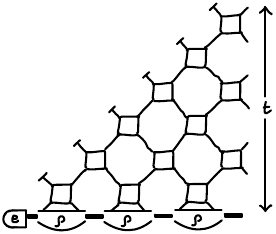} \\
    & \quad  = \punct{\pic{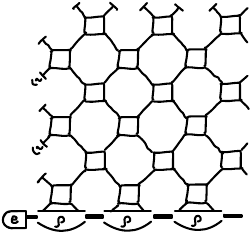}}{}{,}\label{eq:mot_infsolv_triangle}
\end{aligned}
\end{equation}
see Appendix \ref{app:mps_space} for the definition of $\pic{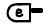}$. On the RHS, we have used \eqref{eq:mot_classic_flatid_bottom_normalized} to write the influence matrix as a uniform MPS.
% \begin{align}
%     \includegraphics[width=0.5\linewidth]{figs/fred/Screenshot 2026-03-04 182702}\label{eq:mot_infsolv_triangle}
% \end{align}
Because of this, \eqref{eq:mot_infsolv_triangle} is also the exact influence-matrix for a finite system of length $\ell_\mathrm{left}$ for times $t<\ell_{\mathrm{left}}$.

For small $t$ one can compute the influence matrix this way for any system. Alternatively one can obtain it as an eigenvector of the spatial transfer matrix, which we will define and discuss in \cref{sec:FTMPS_transfer_matrix}. However, as time increases the dimension of the influence matrix generically increases exponentially~\cite{SONNER2021168677}. Hence, this is in general not a generally efficient way to calculate long time dynamics.

However, there are cases in which the complexity of the influence matrix does not explode with time. Loosely speaking, this is the case for Solvable circuits. For concreteness, in this work we study the case where the influence matrix is a uniform finite-bond dimension MPS of the form \eqref{eq:mot_infsolv_def}:
\begin{definition}[Influence solvability]\label{def:mot_infsolv_def}
    We call a circuit (left) influence solvable if and only if for all times $t\geq 0$ its influence matrix is given by a uniform MPS of constant bond dimension $\chi$,
    \begin{align}
    \cdots \pic{influence-matrix-left-t} = \punct{\pic{solvabilitydefRHS}}{\;}{.}
    \label{eq:mot_infsolv_def}
    \end{align}
Here the MPS is constructed from the tensor $\bm{A}$, and boundary tensors $\bm{\mu}$ and $\bm{\nu}$ which are independent of time $t$. The thicker vertical line denotes the bond of bond-dimension $\chi$.

\end{definition}

Why do we call the circuit solvable in this case? The reason is that assuming \eqref{eq:mot_infsolv_def} greatly simplifies time-evolved quantities into a finite bond-dimensional object. For instance, \eqref{eq:mot_classic_circuit_system} simplifies to
\begin{align}
    \pic{mpsbathonS} = \punct{\pic{solvablebathonS}}{\;}{.} \label{eq:mot_infsolv_circuit_system_simp}
\end{align}
% \begin{align}
%     \includegraphics[width=0.5\linewidth]{figs/fred/Screenshot 2026-03-23 140058}
% \end{align}

The interpretation of result is as follows: the dynamics inside the system S, originally described by the full circuit \eqref{eq:mot_classic_circuit_system} can alternatively described by effective dynamics of a finite system given by the effective time evolution operator $\bm{G}\ind{eff}$,
\begin{align}
    \pic{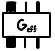} \; = \; \punct{\pic{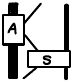}}{\;}{.}\label{eq:mot_infsolv_Geff}
\end{align}

Such dynamics can be computed efficiently \emph{for arbitrarily long times $t$} by diagonalizing the effective time evolution operator $\bm{G}_\mathrm{eff} = \sum_i\lambda_i\keTbrA{\psi_i}{\psi_i}$. The time-evolution of the expectation value is given by
\begin{align}
    \expval{O(t)} &= \sum_i \lambda_i^t \brAkeT{O}{\psi_i}\brAkeT{\psi_i}{\rho\ind{S}}\label{eq:mot_is_G_Ot}.
\end{align}

Note that in general there is no reason why $\bm{G}\ind{eff}$ should be diagonalisable. However, even it is not diagonalisable, one can still write it in a Jordan normal form and obtain a formula similar to \eqref{eq:mot_is_G_Ot}.

One way to think about $\bm{G}\ind{eff}$ is that, going from \eqref{eq:mot_classic_circuit_system} to \eqref{eq:mot_infsolv_circuit_system_simp}, we have ``integrated out the bath'', i.e. replaced the infinite circuit by effective simple dynamics. The tensor $\bm{A}$ describes all information of the bath generated by circuit that is relevant to the system. In particular, the bond dimension of $\bm{A}$ represents the finite memory time of the bath. In the simplest case when the bond dimension is $\chi=1$ the bath has no memory and the effective dynamics is exactly Markovian. This can be seen as a time-discrete version of integrating out a bath and obtaining an effective Lindblad description in a time-continuous system\footnote{This can be for instance done in a certain limit of a bath composed from harmonic oscillators, see Caldeira-Leggett model~\cite{CALDEIRA1983587}.}. Note that there is no reason for the bath to be unitary or even invertible, and in general, it will induce non-unitary dynamics.

We would like to state three remarks exemplifying the power of having an influence-solvable circuit.

The first is that if we have influence-solvability we can efficiently compute not just the evolution of observables, but also any autocorrelation function of the system $\mathrm{S}$,
\begin{multline}
    \expval*{O^{(n)}_\mathrm{S}(t_n)O^{(n-1)}_\mathrm{S} (t_{n-1})\cdots O^{(1)}\ind{S}(t_1)}\\
    = \brA{-}\bm{O}^{(n)}\ind{S} \bm{G}\ind{eff}^{t_n-t_{n-1}}\bm{O}^{(n-1)}\ind{S}\cdots \bm{O}^{(1)}\ind{S}\bm{G}\ind{eff}^{t_1}\keT{\rho}.
\end{multline}
Here $\bm{O}^{(n)}\ind{S}$ is a matrix that implements the observable $O^{(n)}\ind{S}$. Diagramatically, this would be inserting a few tensors $\{\bm{O}^{(n)}\ind{S}\}_n$ into the RHS of \eqref{eq:mot_infsolv_circuit_system_simp}.

\begin{figure*}[t]
\centering
\newcommand{\motappscale}{0.7}
\newcommand{\motappsizethree}{5cm}
\newcommand{\motappsizetwo}{8cm}
\begin{subfigure}{\motappsizethree}
    $\pic[scale=\motappscale]{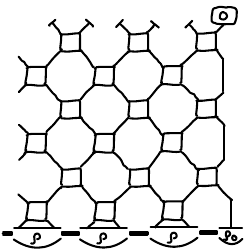} = \pic[scale=\motappscale]{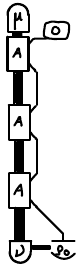}$
    \caption{Observable evolution in semi-infinite systems with open boundary conditions}
\end{subfigure}
\begin{subfigure}{\motappsizethree}
    $\pic[scale=\motappscale]{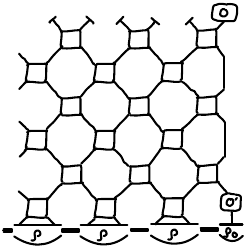} = \pic[scale=\motappscale]{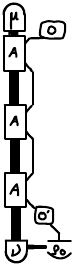}$
    \caption{Autocorrelation function in semi-infinite systems with open boundary conditions}
\end{subfigure}
\begin{subfigure}{\motappsizethree}
     $\pic[scale=\motappscale]{mpsbathonS} = \pic[scale=\motappscale]{solvablebathonS}$
    \caption{Evolution in a system $\mathrm{S}$ coupled to a semi-infinite circuit}
\end{subfigure}
\begin{subfigure}{\motappsizetwo}
    $\pic[scale=\motappscale]{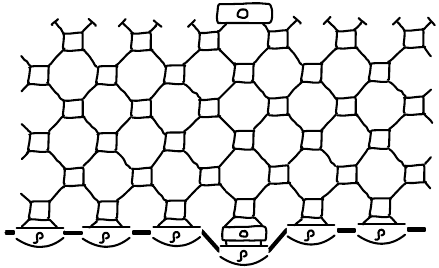} = \pic[scale=\motappscale]{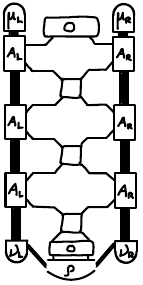}$
    \caption{Evolution in infinite systems, assuming left and right influence-solvability}
\end{subfigure}
\begin{subfigure}{\motappsizetwo}
    $\pic[scale=\motappscale]{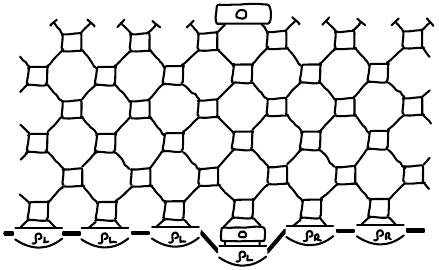} = \pic[scale=\motappscale]{matappriemannRHS}$%\includegraphics[width=\linewidth]{figs/fred/Screenshot 2026-03-23 140701}
    \caption{Evolution after bipartitioning protocols (domain wall states), assuming left and right influence-solvability}
\end{subfigure}
\begin{subfigure}{\motappsizetwo}
    $\pic[scale=\motappscale]{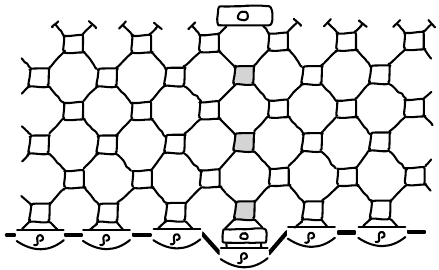} = \pic[scale=\motappscale]{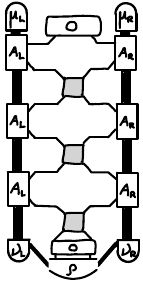}$%\includegraphics[width=\linewidth]{figs/fred/Screenshot 2026-03-23 140720}
    \caption{Evolution at an impurity (shaded box), assuming left and right influence-solvability}
\end{subfigure}
\begin{subfigure}{\motappsizetwo}
    $\pic[scale=\motappscale]{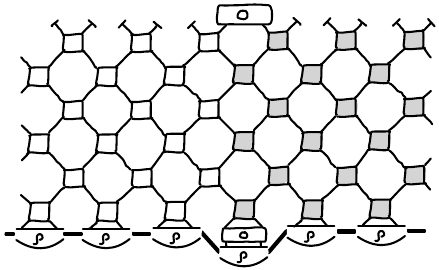} = \pic[scale=\motappscale]{matappimpurityRHS}$
    \caption{Evolution at an interface between two different infinite systems, assuming left and right influence-solvability}
\end{subfigure}
\caption{Examples of scenarios applicable to influence-solvability. Any time-evolved observable $\ev*{O_x(t)}$ or auto-correlator of the form $\ev*{O^{(n)}_x(t_n) \cdots O^{(1)}_x(t_1)}$ are tractable in the above scenarios. Other applications include computation of entanglement entropies (see \cref{sec:prop_entanglement}), computation of correlation functions (see \cref{sec:prop_corr}) and potentially computing diffusion constants (see \cref{sec:prop_CQ}).}
\label{fig:mot_infsolv_applications}
\end{figure*}

The second is that influence-solvability is not restricted to the case of a finite system $\mathrm{S}$ coupled to a bath given by the circuit. We give an overview over further possible applications of influence-solvability in \cref{fig:mot_infsolv_applications}.

The third is that, while in this work we restrict to reversible gates, influence-solvability and all the constructions in this work can straightforwardly be extended to circuits constructed from non-reversible gates, i.e. gates that are local Markov matrices in the classical case (which we will refer to as `classical channels') or CPTP channels in the quantum case.

\subsection{Explicit example: Dual unitary circuits}
In the following sections, we will develop a systematic theory to study influence-solvability. Here, we show an example of a circuit with additional local properties that is sufficient for influence-solvability. An example of this are circuits that in addition to \eqref{eq:mot_classic_flatid_top} and \eqref{eq:mot_classic_flatid_bottom} satisfy

\begin{align}
    \pic{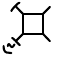} = \pic{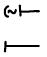} \, .\label{eq:mot_DU_flatid_left}
    %\pic{twoleftflatongate} = \pic{twoleftflat} \, .\label{eq:mot_DU_flatid_left}
    % \includegraphics[width=2cm]{figs/fred/Screenshot 2026-03-22 175749}.
\end{align}

In the quantum case this condition is the famous dual-unitarity condition~\cite{gopalakrishnan2019unitary,bertini2019exact}\footnote{In the quantum unitary case, the left condition implies the right condition and vice versa. If both hold in the classical case, they are dual-symplectic circuits~\cite{10.21468/SciPostPhys.16.2.049}.}
\begin{align}
    \pic{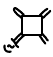} = \pic{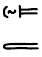}\,\,, \quad \quad \pic{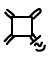} = \pic{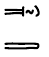}\,\,,\label{eq:mot_DU_flatid_left_quantum}
\end{align}
which were the first examples of solvable circuits. This condition is usually written as
\begin{align}
    \pic{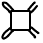} = \pic{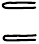}\,\,, \quad \quad \pic{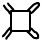} = \pic{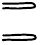}\,\,.
\end{align}
We will use \eqref{eq:mot_DU_flatid_left_quantum} to highlight that $\keT{\sim}$ is a normalized state transported upwards in the space-time picture (i.e. forward in time), while $\brA{-}$ is an identity matrix transported downwards in the space-time picture (i.e. backwards in time). In both the classical and quantum case the influence matrix for the infinite temperature initial state can easily be derived as follows,
\begin{align}
    \cdots \pic{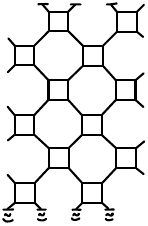} = \pic{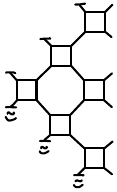} = \punct{\pic{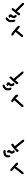}}{}{,}\label{eq:mpt_DU_IM}
\end{align}
% \begin{align}
%     \includegraphics[width=\linewidth]{figs/fred/Screenshot 2026-03-22 180757}\label{eq:mpt_DU_IM}
% \end{align}
from which it is clear that circuits satisfying the condition \eqref{eq:mot_DU_flatid_left} are influence-solvable for the infinite temperature state with a factorized MPS given by $\pic{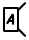} = \pic{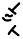}$.

Importantly, the above derivation only works for the initial state $\keT{\sim}$. For generic initial states the circuit is not influence-solvable, although further solvable initial states may exist: those can have the same influence matrix (for example, for the quantum dual unitary case, such cases always exist and were fully classified in Ref.~\onlinecite{PhysRevB.101.094304}) or a completely different influence matrix.

\section{Necessary and sufficient local conditions for influence-solvability}\label{sec:FTMPS}
\eqref{eq:mot_infsolv_def} is a global condition that becomes more and more complicated to verify as $t\to \infty$. Like for dual unitaries and its extensions it would be good to have a local condition that is easier to check that imply influence-solvability. Such a condition is what we derive in this section, see \cref{sec:FTMPS_FTMPS}. We prove that the local conditions we derive are equivalent to \cref{def:mot_infsolv_def}, i.e. they are both sufficient and necessary conditions. 

\subsection{Eigenvector of the spatial transfer matrix}\label{sec:FTMPS_transfer_matrix}
It is a well known fact that the influence matrix can be obtained as an eigenvector of the so-called transfer matrix $\bm{M}_t$, see e.g. Ref. \onlinecite{10.21468/SciPostPhys.11.6.106}. Here, the transfer matrix $\bm{M}_t$ is defined as a vertical column of the circuit,
\begin{align}
    \pic{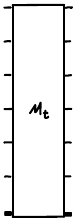} = \punct{\pic{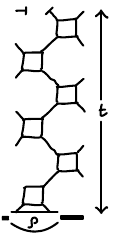}}{\,}{,}
\end{align}
which can be viewed as an ``evolution matrix'' of the circuit in space. Note that since we are studying a semi-infinite system, we can view the influence matrix \eqref{eq:mot_infsolv_influencematrix} as a very large power of the transfer matrix $\lim_{L\to\infty} \bm{M}_t^{L}$. Note that by taking any matrix to a very large power we obtain a matrix dominated by the projector on the eigenspace corresponding to the largest eigenvalue,
\begin{align}
    \lim_{L\to\infty} \bm{M}_t^{L} \sim \qty(\max_i\abs{\lambda_i})^L \vec{v}\ind{R}\vec{v}\ind{L}^\top .
\end{align}
Here, $\vec{v}\ind{R}$ and $\vec{v}\ind{L}$ are the right and left eigenvalues corresponding to the eigenvalue of largest magnitude, assuming there is only one.
Thus, the influence matrix is the eigenvector of the transfer matrix of this eigenvalue, i.e. $\vec{v}\ind{L}^\top$.

Fortunately, it follows from the cluster assumption of the initial state that the transfer matrix $\bm{M}_t$ has exactly one eigenvalue $1$ which also has the largest magnitude. The argument goes as follows: consider $\tr\bm{M}_t^L$, which graphically can be represented as a system of size $L$ with periodic boundary conditions,
\begin{equation}
\begin{aligned}
    \tr[M^L_t] & = \pic{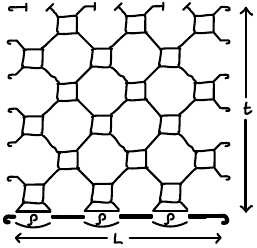} \\
    & = \pic{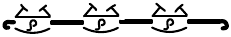} \to 1.
    \label{eq:FTMPS_eigvec_periodic}
\end{aligned}
\end{equation}
As shown in \eqref{eq:FTMPS_eigvec_periodic} this circuit can be simplified to time $t=0$:
\begin{align}
    \tr[\bm{M}_t^L] = \tr[\bar{\bm{\rho}}^L] \to 1,
\end{align}
where $\pic{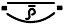} = \pic{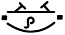}$.
The RHS must approach $1$ as $L\to \infty$ due to clustering and normalisation, see Appendix \ref{app:mps_space}. This implies that there is a single eigenvalue $1$. In general, as discussed in Appendix \ref{app:mps_space}, it is not necessarily true that $\tr \bar{\bm{\rho}}^L=1$ for finite $L$. However, it holds for certain states, such as factorised states (i.e. $\chi_\rho=1$). In this case $\bm{M}_t$ is not diagonalisable. Instead, in addition to the single eigenvalue $1$ it is composed of Jordan blocks with eigenvalue $0$~\cite{10.21468/SciPostPhys.11.6.107}.

The fact that $\bm{M}_t$ has a single largest eigenvalue $1$ is convenient, since we only need to find this one eigenvector corresponding to eigenvalue $1$ to obtain the influence-matrix.

Due to our ansatz \eqref{eq:mot_infsolv_def} for the influence matrix we have:
\begin{theorem}
(Left) influence-solvability (\cref{def:mot_infsolv_def}) is equivalent to the (left) eigenvector of the spatial transfer matrix $\bm{M}_t$ corresponding to eigenvalue $1$ being given by the RHS of \eqref{eq:mot_infsolv_def} for all $t\geq 0$, i.e.
\begin{align}
    \pic{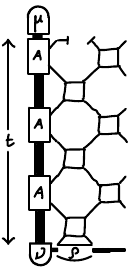} = \punct{\pic{solvabilitydefRHS}}{}{.}
    \label{eq:FTMPS_eigvec}
\end{align} 
\end{theorem}

\subsection{Application of the fundamental theorem of MPS}\label{sec:FTMPS_FTMPS}

At this point we make the following observation: the RHS of \eqref{eq:FTMPS_eigvec} is a uniform MPS by our ansatz, but the LHS is also a uniform MPS composed from the tensor $\bm{B}$ given by
\begin{align} \label{eq:Btensor}
\pic{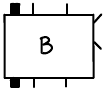} := \punct{\pic{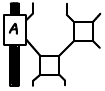}}{\;}{.}
\end{align}

So, the condition \eqref{eq:FTMPS_eigvec} is the same as the condition for two uniform MPS with open boundary conditions to be equivalent for any length $t\geq 0$.

In the physics literature, such statements are commonly called fundamental theorem of MPS~\cite{Fannes1992,annurev:/content/journals/10.1146/annurev-conmatphys-031016-025507,10.1063/1.5000784,RevModPhys.93.045003,floridollinas2025uniformmatrixproductstates}. The standard version of fundamental theorem of MPS considers uniform MPS with periodic boundary conditions. For open boundary conditions the situation is slightly different. Before stating the general theorem let us briefly discuss canonical examples of equivalent MPS, i.e.\ $\brA{\phi\ind{A}^{\alpha_0}}\bm{A}^{\alpha_1}\ldots \bm{A}^{\alpha_T}\keT{\psi\ind{A}^{\alpha_{T+1}}} =   \brA{\phi\ind{B}^{\alpha_0}}\bm{B}^{\alpha_1}\ldots \bm{B}^{\alpha_T}\keT{\psi\ind{B}^{\alpha_{T+1}}}$ for all $T\in\mathbb{N}$:
\begin{itemize}
    \item Gauge transformation: If the bond dimension of both MPS are equal then one possible option for equivalence is the existence of an invertible map $\bm{V}$ s.t.
    \begin{equation}
        \begin{aligned}
        \bm{A}_{\alpha} = \bm{V}^{-1}\bm{B}_{\alpha}\bm{V}, & \quad \keT{\psi\ind{A}^{\alpha}} = \bm{V}^{-1}\keT{\psi\ind{B}^\alpha}, \\ \quad \brA{\phi\ind{A}^\alpha}  &= \brA{\phi\ind{B}^\alpha}\bm{V}. \label{eq:FTMPS_FTMPS_gauge}
        \end{aligned}
    \end{equation}
    This simply corresponds to a change of basis (a ``gauge transformation'').
    \item Auxillary dimensions: if the two bond dimensions are not equal it can happen that some dimensions are never explored. A simply example of this is 
    \begin{equation}
    \label{eq:auxillary-dimensions}
    \begin{aligned} 
			\vb{B}^{\alpha} = \begin{pmatrix}\vb{A}^\alpha\\
			& 0\end{pmatrix}, \quad & \keT{\psi\ind{B}^\alpha} = \begin{pmatrix}\keT{\psi\ind{A}^\alpha}\\0\end{pmatrix}, \\
            \brA{\phi\ind{B}^\alpha} &= \begin{pmatrix}\brA{\phi\ind{A}^\alpha} & 0\end{pmatrix}.
		\end{aligned}
    \end{equation}
        \item Orthogonal dimensions: consider the trivial MPS with bond dimension $\chi\ind{A}=1$ \begin{align}
			\vb{A}^\alpha &= 1, & \keT{\psi\ind{A}^\alpha} &= 1, & \brA{\phi\ind{A}^\alpha} &= 1. \end{align} 
            This is equivalent to the following MPS,
            \begin{align}
			\vb{B}^{\alpha} &= \begin{pmatrix}1&0\\1& 1\end{pmatrix}, & \keT{\psi\ind{B}^\alpha} &= \begin{pmatrix}1\\1\end{pmatrix}, & \brA{\phi\ind{B}^\alpha} &= \begin{pmatrix}1 & 0\end{pmatrix}.
		\end{align}
        Note that when we apply any number of $\vb{B}^\alpha$ on $\keT{\psi\ind{B}^\alpha}$ we explore the whole bond dimension. However, when eventually contracting with $\brA{\phi\ind{B}^\alpha}$ we realize that the second dimension is obsolete as it can never affect the value of the MPS.
\end{itemize}

These basic examples highlight the reasons why two MPS can be equivalent. In general, the equivalence of two MPS will be a combination of these effects. The most general statement is as follows:
\begin{theorem}[Fundamental theorem of MPS for open boundary conditions]\label{thm:local_condition_MPS}
    Two MPS
        \begin{align}
    \brA{\phi\ind{A}^{\alpha_0}}\bm{A}^{\alpha_1}\ldots \bm{A}^{\alpha_T}\keT{\psi\ind{A}^{\alpha_{T+1}}} &=   \brA{\phi\ind{B}^{\alpha_0}}\bm{B}^{\alpha_1}\ldots \bm{B}^{\alpha_T}\keT{\psi\ind{B}^{\alpha_{T+1}}}
	\end{align}
    are equal for all $T \in \mathbb{N}$ if and only if there exists a map $\bm{W}$ and two projectors $\bm{L}$ and $\bm{R}$ satisfying

\begin{subequations}
    \begin{align}
        \bm{L}\bm{B}^\sigma\bm{W} &= \bm{W}\bm{A}^\sigma\bm{R},\label{eq:main_MPS_bulk}\\
        \bm{L}\bm{B}^\sigma &= \bm{L}\bm{B}^\sigma\bm{L},\label{eq:main_MPS_bulk_PL}\\
        \brA{\phi\ind{B}^\alpha}\bm{L} &= \brA{\phi\ind{B}^\alpha},\label{eq:main_MPS_leftP}\\
        \brA{\psi\ind{B}^\alpha}\bm{W} &= \brA{\phi\ind{A}^\alpha}\bm{R},\label{eq:main_MPS_leftW}\\
        \bm{W}\keT{\psi^\alpha\ind{A}} &= \bm{L}\keT{\psi^\alpha\ind{B}},\label{eq:main_MPS_rightW}\\
        \bm{A}^\sigma\bm{R} &= \bm{R}\bm{A}^\sigma\bm{R},\label{eq:main_MPS_bulk_PR}\\
        \bm{R}\keT{\psi\ind{A}^\alpha} &= \keT{\psi\ind{A}^\alpha}\label{eq:main_MPS_rightP}.
    \end{align}
    
\end{subequations}
    \end{theorem}  
    \cref{thm:local_condition_MPS} is proved in Appendix \ref{sec:fundamental-theorem}. Let us briefly explain the intuition behind these conditions. Without the projectors $\bm{L}$ and $\bm{R}$, \eqref{eq:main_MPS_bulk} is simply \eqref{eq:FTMPS_FTMPS_gauge} generalized to non-invertible $\bm{V}$. Hence, $\bm{W}$ maps the bond space of $\bm{A}^\sigma$ to that of $\bm{B}^\sigma$. This map cannot be invertible unless the two bond dimensions are equal. Let $\mathscr{L}$ be defined as the invariant subspace of $\bm{B}^\sigma$ generated from $\brA{\phi^\alpha\ind{B}}$ by successive application of $\bm{B}^\sigma$. It can be obtained via the iteration
    \begin{align} \label{eq:def-of-Ln}
        \mathscr{L}_0 & = \mathrm{span}_\alpha \qty{\brA{\phi\ind{B}^\alpha}}, \, \nonumber\\
        \mathscr{L}_{n+1} & = \mathrm{span}_\sigma \qty{\brA{\phi\ind{B}} \bm{B}^\sigma : \brA{\phi\ind{B}}\in \mathscr{L}_n},
    \end{align}
    which must terminate for some $N \leq \chi\ind{B}$, such that $\mathscr{L}_{n \geq N} = \mathscr{L}_{N} =: \mathscr{L}$. $\bm{L}$ is then a projector that projects any bra onto $\mathscr{L}$. Then (\nopareneqref{eq:main_MPS_bulk_PL},~\nopareneqref{eq:main_MPS_leftP}) holds by definition. Similarly, $\mathscr{R}$ is defined as the invariant subspace of $\bm{A}^\sigma$ generated from $\keT{\psi^\alpha\ind{A}}$. It is associated with $\bm{R}$ and (\nopareneqref{eq:main_MPS_bulk_PR},~\nopareneqref{eq:main_MPS_rightP}) hold by definition. The role of these projectors is to discard the auxiliary bond dimensions as in the second example \eqref{eq:auxillary-dimensions}.

    We will now apply \cref{thm:local_condition_MPS} onto the circuit setting we are studying. Since we are free to choose the bond dimension of $\bm{A}$, we can always choose the MPS to have minimal bond dimension and hence the projector $\bm{R} = \mathbb{1}$ can be set to be the identity matrix.

    In the graphical notation we can write these conditions as
    \begin{theorem}[Solvability for brickwork circuits for general $\chi$]\label{thm:thm_local_condition_circuit}
        A circuit is (left) influence-solvable if and only if there exists tensors $\bm{A}$, $\bm{\mu}$, $\bm{\nu}$, $\bm{W}$ and a projector $\bm{L}$ such that
\begin{subequations} \label{eq:FTMPS_FTMPS_local_conditions}
\begin{equation}
    \pic{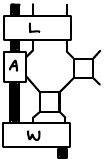} = \pic{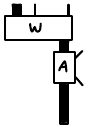},\label{eq:circuit_MPS_bulk}
\end{equation}
\begin{equation}
    \pic{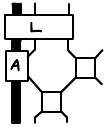} = \pic{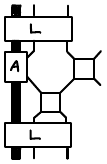},\label{eq:circuit_MPS_bulk_PL}
\end{equation}
\begin{minipage}{0.54\linewidth}
    \begin{equation}
        \pic{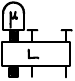} = \pic{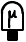} \pic{flatbra} \pic{flatbra},\label{eq:circuit_MPS_leftP}
    \end{equation}
\end{minipage} % \hfill pushes the two minipages away from each other
\begin{minipage}{0.45\linewidth}
    \begin{equation}
        \pic{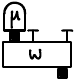} = \pic{betatensor}\label{eq:circuit_MPS_leftW},
    \end{equation}
\end{minipage}
% \begin{equation}
%     \pic{solv_c_LHS} = \pic{betatensor} \pic{flatbra} \pic{flatbra}
%     % \pic{solv_c_RHS}
% \end{equation}
% \begin{equation}
%     \pic{solv_d_LHS} = \pic{solv_d_RHS}
% \end{equation}
\begin{equation}
    \pic{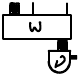} = \pic{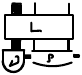}\label{eq:circuit_MPS_rightP}.
\end{equation}
\end{subequations}
    %     \begin{align}\label{eq:FTMPS_FTMPS_local_conditions}
        % \includegraphics[width=\linewidth]{figs/fred/Screenshot 2026-03-22 231753}
    % \end{align}
    \end{theorem}

    Note that if \eqref{eq:FTMPS_FTMPS_local_conditions} are assumed then it is easy to see that influence-solvability follows:
\begin{equation}
\begin{aligned}
    & \pic{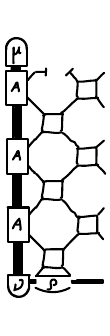} \overset{\eqref{eq:circuit_MPS_leftP}}{=} \pic{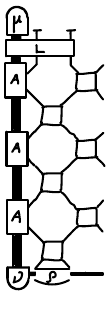} \overset{\eqref{eq:circuit_MPS_bulk_PL}}{=} \pic{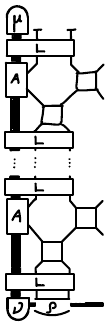} \\ & \quad \quad  \overset{\eqref{eq:circuit_MPS_rightP}}{=} \pic{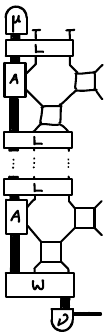} \overset{\eqref{eq:circuit_MPS_bulk}}{=}  \pic{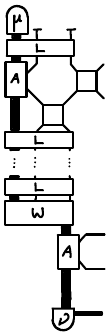} \overset{\eqref{eq:circuit_MPS_leftW}}{=}  \punct{\pic{solvabilitydefRHS}}{\,}{.}
\end{aligned}
\end{equation}

Showing that \eqref{eq:FTMPS_FTMPS_local_conditions} is also a necessary condition for \eqref{eq:FTMPS_eigvec} is more complicated and the essence of \cref{thm:local_condition_MPS}.

These local conditions \eqref{eq:FTMPS_FTMPS_local_conditions} have two major advantages over the global condition \eqref{eq:FTMPS_eigvec}.
\begin{enumerate}
    \item The conditions \eqref{eq:FTMPS_FTMPS_local_conditions} are finite dimensional matrix equations. Using these equations it is easy to establish whether a given $(\bm{A}, \bm{\mu}, \bm{\nu})$ will work or not. 
    \item The conditions \eqref{eq:FTMPS_FTMPS_local_conditions} separate into bulk (\nopareneqref{eq:circuit_MPS_bulk}-\nopareneqref{eq:circuit_MPS_bulk_PL}) and boundary conditions (\nopareneqref{eq:circuit_MPS_leftP}-\nopareneqref{eq:circuit_MPS_rightP}). This will be important: because of this one can first try to find non-trivial solutions to the bulk conditions (\nopareneqref{eq:circuit_MPS_bulk}-\nopareneqref{eq:circuit_MPS_bulk_PL}) and then find \textbf{all} initial conditions $\bm{\rho}$ that are compatible with it by solving the boundary conditions (\nopareneqref{eq:circuit_MPS_leftP}-\nopareneqref{eq:circuit_MPS_rightP}). This way we can turn the logic around: instead of searching for a suitable $(\bm{A}, \bm{\mu}, \bm{\nu})$ for a given gate and $\bm{\rho}$, we can try to find all solvable $\bm{\rho}$ given the gate and a guess for $(\bm{A}, \bm{\mu}, \bm{\nu})$.
\end{enumerate}

Let's suppose that we would like to check if a guess for $(\bm{A}, \bm{\mu}, \bm{\nu})$ is compatible with influence-solvability. Though \eqref{eq:FTMPS_FTMPS_local_conditions} are local conditions, they involve unknown $\bm{L}$ and $\bm{W}$. Let us show how in principle, we can restate \eqref{eq:FTMPS_FTMPS_local_conditions} into a form that does not explicitly depend on them.

The projector $\bm{L}$ can be removed as follows. Recall that we can define the spaces $\mathscr{L}_n$ by \eqref{eq:def-of-Ln}, i.e.
\begin{align}
    \mathscr{L}_0 = \mathrm{span} \qty{\pic{betatensor} \pic{flatbra} \pic{flatbra}}, \; \mathscr{L}_1 = \mathop{\mathrm{span}}_{ab} \qty{\pic{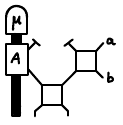}},
\end{align}
and so on. For brevity, we will use the following graphical shorthand,
\begin{align} \label{eq:L1-shorthand}
    \mathscr{L}_1 = \mathrm{span} \punct{\qty{\pic{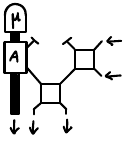}}}{\;}{.}
\end{align}
Then, $\mathscr{L}_2$ is the $\bm{B}$ tensor \eqref{eq:Btensor} applied below \eqref{eq:L1-shorthand} and so on.

After $N \leq \chi d^2$ steps, the iteration must stop. Therefore \eqref{eq:circuit_MPS_bulk} is equivalent to
\begin{align} \label{eq:t=2N+1-solvability}
    \pic{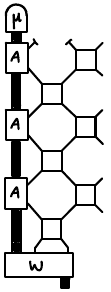} = \punct{\pic{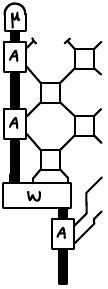}}{\;}{.}
\end{align}
Hence, if \eqref{eq:t=2N+1-solvability} is true for $t\leq 2N+2$, then \eqref{eq:t=2N+1-solvability} must be true for any $t > 2N+2$ and hence imply influence-solvability.

$\bm{W}$ can be removed as follows. Define the space $\mathscr{R}\upd{B}$ recursively from the bottom, i.e.
\begin{equation} \label{eq:R_B-space}
\begin{aligned} 
    \mathscr{R}\upd{B}_0 &= \mathrm{span}\qty{\pic{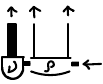}}\\ \mathscr{R}\upd{B}_1 &= \mathrm{span}\punct{\qty{\pic{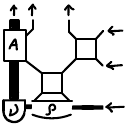}}}{\;}{,}
    % \mathscr{R}\upd{B}_2 &= \ldots\nonumber
\end{aligned}
\end{equation}
and so on. Note that this is distinct from $\mathscr{R}$. This iteration must also stop after some $N\ind{R}\leq \chi d^2$ steps. Then, if we can show influence-solvability \eqref{eq:FTMPS_eigvec} for time $t= 2(N+N\ind{R})$,
% \begin{align}
%     \includegraphics[width=\linewidth]{figs/fred/Screenshot 2026-03-23 002139},
% \end{align}
then it must follow for all times $t\in\mathbb{N}$. This is a consequence of the proof of \Cref{thm:local_condition_MPS}: at time $t= 2(N+N\ind{R})$ the bond dimension of $\bm{B}^\sigma$ is fully explored, both from above and below, implying that the tensor $\bm{A}$ is entirely determined from the data at time $t= 2(N+N\ind{R})$. Since $N,N\ind{R} \leq \chi\ind{B} =\chi d^2$, establishing influence-solvability for time $t=4\chi d^2$ implies influence-solvability for all times. Though finite, the dimensions of these equations can unpractically large at $d^{4\chi d^2}$. We believe it will always be significantly more efficient to check the local conditions \eqref{eq:FTMPS_FTMPS_local_conditions}, as applied in \cref{sec:zerobond} for $\chi=1$.

\subsection{Relation to known types of solvability}\label{sec:FTMPS_rel}
In this section, we connect our notions of influence solvability with existing notions of solvable circuits.

\subsubsection{Dual unitary-like condition}\label{sec:FTMPS_rel_DU_like}
We have already established the influence-solvability of circuits satisfying the classical/quantum version of the dual unitarity condition \eqref{eq:mot_DU_flatid_left} for the infinite temperature state. We can also easily check that indeed \eqref{eq:mpt_DU_IM} is the eigenvector of the transfer matrix:
\begin{align}
    \pic{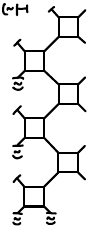} = \pic{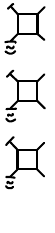} = \punct{\pic{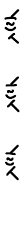}}{\;}{.}
\end{align}
Alternatively, we can also establish this using the local conditions \eqref{eq:FTMPS_FTMPS_local_conditions}. First, we need to construct $\mathscr{L}$, which turns out to be equal to $\mathscr{L}_1$. Indeed $\mathscr{L}_1$ is given by
\begin{align}
    \mathscr{L}_1 = \mathrm{span}\qty{\pic{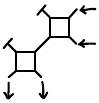}}
\end{align}
and one can easily check that
\begin{align}
    & \mathscr{L}_2 = \mathrm{span}\qty{\pic{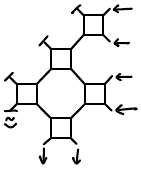}} \label{eq:FTMPS_rel_DU_L2} \\ 
    & = \mathrm{span}\qty{\pic{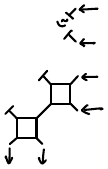}} = \mathrm{span}\qty{\pic{page27/Solvabilitypage274}} = \mathscr{L}_1, \nonumber 
\end{align}
and hence $\mathscr{L}=\mathscr{L}_1$. With this one can easily check condition \eqref{eq:circuit_MPS_bulk}
\begin{align}
    \pic{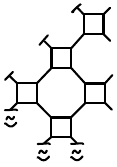} =\pic{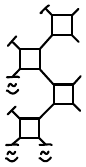} = \punct{\pic{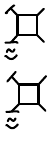}}{\;}{.} \label{eq:FTMPS_rel_DU_W}
\end{align}

\subsubsection{Second level dual unitary-like condition $\mathrm{DU}(2)$}
Another well-studied condition is the $\mathrm{DU}(2)$ condition~\cite{Yu2024hierarchical}. There are two different conditions for the second level of dual-unitarity,

\begin{subequations}\label{eq:FTMPS_rel_DU2}
\noindent
\begin{minipage}{0.45\linewidth}
  \begin{equation}\label{eq:FTMPS_rel_DU2_a}
    \pic{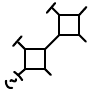} = \pic{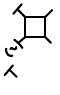}
  \end{equation}
\end{minipage}%
\hfill $\text{or}$ \hfill%
\begin{minipage}{0.45\linewidth}
  \begin{equation}\label{eq:FTMPS_rel_DU2_b}
    \pic{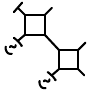} = \pic{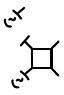}.
  \end{equation}
\end{minipage}
\end{subequations}
% \begin{subequations}
    
% \end{subequations}
% \begin{align}\label{eq:FTMPS_rel_DU2}
%     \pic{page27/Solvabilitypage2710} &= \pic{page27/Solvabilitypage2711} &&\mathrm{or} & & \pic{page27/Solvabilitypage2712} &= \pic{page27/Solvabilitypage2713}.
% \end{align}
%\includegraphics[width=\linewidth]{figs/fred/Screenshot 2026-03-21 123133}
Any one of the two is sufficient for influence-solvability with no bond dimension and tensor
$\pic{0bdA} = \pic{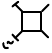}$
One can easily check that those are indeed the eigenvectors of the transfer matrix, however in a slightly different way
\begin{align}
    \pic{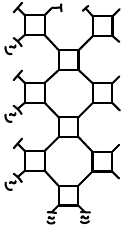} &= \pic{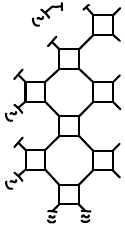} \overset{\eqref{eq:FTMPS_rel_DU2_a}}{=} \pic{page27/Solvabilitypage272}
\end{align}
for the first condition, and
\begin{align}
    \pic{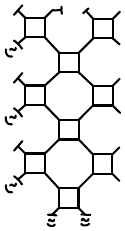} &\overset{\eqref{eq:FTMPS_rel_DU2_b}}{=} \pic{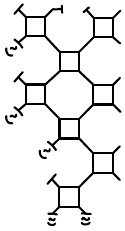} \overset{\eqref{eq:FTMPS_rel_DU2_b}}{=} \pic{page27/Solvabilitypage272}
    \label{eq:FTMPS_rel_DU2_bottom_simplify}
\end{align}
for the second.

This is also reflected in two different ways to establish influence-solvability using the local conditions \eqref{eq:FTMPS_FTMPS_local_conditions}. In the first case \eqref{eq:FTMPS_rel_DU2_a} one has $\mathscr{L}=\mathscr{L}_1$ similar to \eqref{eq:FTMPS_rel_DU_L2} and also \eqref{eq:circuit_MPS_bulk} can be shown analogously to \eqref{eq:FTMPS_rel_DU_W}. For \eqref{eq:FTMPS_rel_DU2_b} observe that, because $\mathscr{L}$ is defined recursively, \eqref{eq:circuit_MPS_bulk} is equivalent to
\begin{align}
    \pic{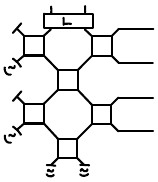} &=\pic{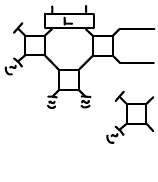},
\end{align}
which can easily be shown as
\begin{equation}
\begin{aligned}
    \pic{page27/Solvabilitypage2718} &= \pic{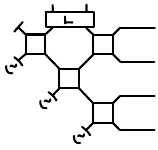}\\
    &= \pic{page27/Solvabilitypage2719},
\end{aligned}
\end{equation}
independently of the details of the space $\mathscr{L}$. Note that equivalently one can study case \eqref{eq:FTMPS_rel_DU2_b} as a time-reflection of case \eqref{eq:FTMPS_rel_DU2_a}, meaning that one would build the space $\mathscr{L}$ from below instead of from above~\footnote{This is equivalent to the construction of $\mathscr{R}\ind{B}$ of \eqref{eq:R_B-space} in \cref{sec:FTMPS_FTMPS}.}.

\subsubsection{Examples with bond dimension $\chi>1$}\label{sec:FTMPS_higher_bond_dim}
So far we only considered cases where the influence-matrix factorizes, i.e. the bond dimension is $\chi=1$. They are also the most studied. However, there also exists examples with bond dimension $\chi>1$.

Ref. \onlinecite{PhysRevLett.133.170402} studies circuits where $\pic{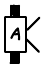} = \pic{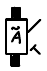}$ which satisfy
\begin{align}
    \pic{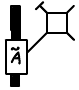} = \pic{page29/page2914}.
\end{align}
It is easy to see that this indeed satisfies \eqref{eq:FTMPS_eigvec} in the bulk. Hence, together with appropriate boundary conditions these circuits are influence-solvable. Note that as a straightforward generalization we can also consider the ``zipper condition''~\cite{annurev:/content/journals/10.1146/annurev-conmatphys-031016-025507,PhysRevLett.128.220401} 
\begin{align}
    \pic{page29/page2913} = \pic{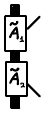},
\end{align}
together with
\begin{align}
    \pic{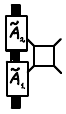} = \pic{page29/page2916}.
\end{align}

In the literature there is also a more general construction: in Refs.~\onlinecite{PhysRevLett.126.160602,10.21468/SciPostPhys.11.6.106,10.21468/SciPostPhys.11.6.107}, the influence-matrix of the quantum version of the Rule 54 cellular automata was derived for a whole family of initial states. Note that Rule 54 is not a brickwork circuit but has a slightly different circuit architecture, hence the result cannot be directly expressed in our notation. However, in spirit, their construction splits
\begin{align}
    \pic{page29/page2913} = \pic{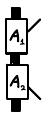}
\end{align}
and says that there exists an ``intertwiner'' $\bm{C}$ satisfying
\begin{align}\label{eq:FTMPS_katja}
    \pic{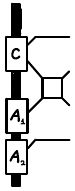} = \punct{\pic{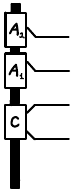}}{\;}{.}
\end{align}
together with some suitable boundary conditions, for instance
\begin{align}
    \pic{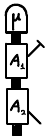} &= \pic{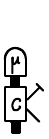} & \pic{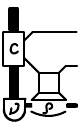} &= \pic{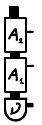} \nonumber\\
    \pic{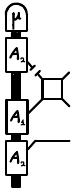} &= \pic{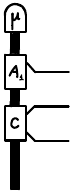} & \pic{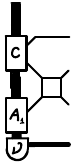} &= \punct{\pic{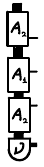}}{\;}{.}
\end{align}
More sophisticated boundary conditions are also possible. 

Note that the intertwiner $\bm{C}$ has a very similar effect to the $\bm{W}$ in \eqref{eq:circuit_MPS_bulk}. However, the main difference is that $\bm{W}$ simplifies two layers of the spatial evolution of the circuit at once, while $\bm{C}$ simplifies the two layers one by one, at the expense of $\bm{C}$ having physical legs as opposed to $\bm{W}$.

It would be interesting to see whether the conditions \eqref{eq:FTMPS_FTMPS_local_conditions} for two layers of the spatial evolution can always be recasted into conditions similar to \eqref{eq:FTMPS_katja} applying only to one spatial layer. This is plausible because apart from the boundary the two layers are identical up to a shift by one side. Such reasoning has been used to reduce the problem of finding conserved quantities (which is similar to finding an eigenvector of the temporal evolution) from a two layer problem to a single layer one~\cite{kim2025circuits,sharipov2026ergodicbehaviorsreversible3state,prosen2007chaos}.

Before completing this section, we would like to emphasise that all the previous notions of Solvability do not study it accounting for $\mathscr{L}$. Our analysis involving $\mathscr{L}$ allows us to uncover new types of local solvability conditions, which we present in \cref{sec:zerobond_newloccond} even for the simplest case of $\chi=1$. We expect our approach to be fruitful in general.

\section{$\chi=1$ influence solvability as exact Markovian baths}\label{sec:zerobond}

In this section, we study the simplest case where the the tensor $\bm{A}$ in MPS \eqref{eq:mot_infsolv_def} has bond dimension $\chi=1$. First, we present a necessary and sufficient condition for spatially invertible gates for general $d$, applicable to both quantum and classical settings (\cref{sec:zerobond_spatial_inv_gate}). This condition is reminiscent but distinct from the dual-unitarity condition, as we show by an example of a non dual-unitary gate that fulfills this condition (\cref{ex:non-du-spatially-invertible}). Next, we categorise all classical gates up to local dimension $d=3$ (\cref{sec:classic-classification}), and all quantum gates of local dimension $d=2$ (\cref{sec:quantum-qubits-classification}). In particular, we find new local conditions for $\chi=1$ influence-solvability that has a different form to dual-unitary-like and $\mathrm{DU}(2)$-like conditions (\nopareneqref{eq:zerobond_novel_cond1}) and (\nopareneqref{eq:zerobond_novel_cond2}). Many of these new local conditions allow for calculation of correlation functions near the left light cone \eqref{eq:light-cone} that is generically non-vanishing. In \cref{sec:zerobd-further} we discuss two further examples of tunable exact Markovian baths and hydrodynamic bipartitioning protocol.

For $\chi=1$ influence-solvability, the influence matrix exactly factorises, i.e. 
\begin{align}
    \cdots \pic{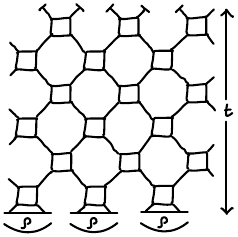} = \punct{\pic{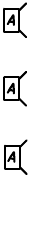}}{}{.}
    \label{eq:zerobond_infsolv_def}
\end{align}
Physically, this means that the bath generated by the circuit is an exact Markovian bath. At each time-step the interaction with the bath is equivalent to applying the classical or quantum channel $\bm{A}$ to the system. In quantum systems this is the time-discrete equivalent of a Lindblad equation.

For $\chi=1$ we can use the fact that influence-solvability must also hold for the particular time $t=2$ to conclude
\begin{align}
    \pic{0bdA} = \pic{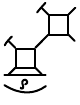} = \raisebox{-0.3em}{$\pic{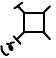}$},
\end{align}
where $\keT{\gamma}$ is defined as
\begin{align}\label{eq:zerobond_gamma_def}
    \pic{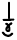} := \pic{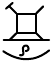}.
\end{align}
Hence, instead of requiring the search over $d^2$ unknown degrees of freedom, $\bm{A}$ is fully determined by $\keT{\gamma}$ which requires the search over $d$ degrees of freedom. This is a significant reduction of complexity which will allow us to classify all influence-solvable gates for small $d$, see \Cref{sec:classic-classification,sec:quantum-qubits-classification}.

In general, it is clear from \eqref{eq:zerobond_gamma_def} that $\keT{\gamma}$ is a proper state, i.e. it is normalized and non-negative\footnote{For classical circuits $\keT{\gamma}$ is a vector with non-negative entries summing to $1$ and for quantum circuits $\keT{\gamma}$ represents a positive semi-definite matrix with trace $1$.}.

Applying \cref{thm:thm_local_condition_circuit} we thus obtain the following local conditions for influence-solvability:
\begin{align}\label{eq:zerobond_cond_b}
    \pic{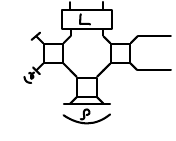} = \pic{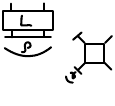},
\end{align}
where $\mathscr{L}$ is constructed by demanding that the flat state is in $\mathscr{L}$, i.e.
\begin{align}
    \pic{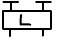} = \pic{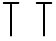},
\end{align}
and then recursively applying $\bm{B}$, i.e.
\begin{align}
    \pic{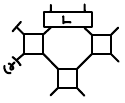} = \pic{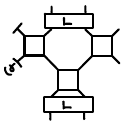}.\label{eq:zerobond_classic_L2_a}
\end{align}
Note that in \eqref{eq:zerobond_infsolv_def} we have restricted to a factorized initial state. This is restriction is possible because if we find a solution to (\nopareneqref{eq:circuit_MPS_bulk}-\nopareneqref{eq:circuit_MPS_leftW}) we can always use $\keT{\rho} = \keT{W}$ as a solvable initial state with bond dimension $\chi_\rho=1$. Further solutions for any bond dimension can then be found by solving \eqref{eq:circuit_MPS_rightP}. In particular, if there is no solvable factorized initial state, then there is no solvable initial state for any bond dimension $\chi_\rho\geq 1$.

Observe that for the classical or quantum version of dual unitary-like and $\mathrm{DU}(2)$-like conditions discussed in \cref{sec:FTMPS_rel}, $\keT{\gamma}$ is given by the infinite temperature state
\begin{align}
\pic{gammaupright} = \pic{tildeket} \quad \text{ or } \quad \pic{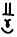} = \pic{foldedtildeket}.
\end{align}
Using our formalism we can immediately generalise these sufficient conditions for general $\keT{\gamma}$:
\begin{result}\label{res:zerobond_DU1_DU2}
    For a general $\keT{\gamma}$ defined by \eqref{eq:zerobond_gamma_def}, the circuit is influence-solvable for some initial state $\keT{\rho}$ if it satisfies one of the following generalisations of the $\mathrm{DU}(2)$ condition
    \begin{subequations}
    \label{eq:zerobond_DU2_gen}
        \begin{align}
            \pic{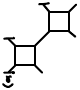} & = \pic{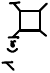} \\
            & \text{or} \nonumber \\
            \pic{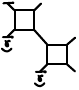} & = \pic{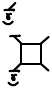},
        \end{align}
    \end{subequations}
    \
    %     \includegraphics[width=\linewidth]{figs/fred/Screenshot 2026-03-20 090434}\label{eq:zerobond_DU2_gen}
    % \end{align}
    or the following generalization of the dual unitarity condition
    \begin{align}
        \pic{gamma-lower-left-gate} = \pic{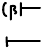} \quad \text{ and } \quad \pic{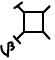} = \pic{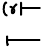}\,.\label{eq:zerobond_DU1_gen}
    \end{align}
    for some $\keT{\beta}$. 
\end{result}
Furthermore, we will show in \cref{sec:zerobond_spatial_inv_gate} that \eqref{eq:zerobond_DU1_gen} is the only possible option if the gate is invertible in space. Note that, unless $\keT{\beta}=\keT{\gamma}$, the dual unitarity-like condition, \eqref{eq:zerobond_DU1_gen}, is contained in the $\mathrm{DU}(2)$-like condition, \eqref{eq:zerobond_DU2_gen}. One can easily to show that these conditions are sufficient for influence-solvability by replacing the infinite-temperature state $\keT{\sim}$ with $\keT{\gamma}$ in \cref{sec:FTMPS_rel_DU_like}. For the right condition of \eqref{eq:zerobond_DU2_gen} one must choose an appropriate initial state satisfying for instance (see Appendix \ref{app:DU2_bottom_initstate})
\begin{align}
    \pic{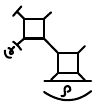} = \pic{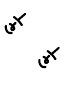}.
\end{align}

We will encounter examples for all of these conditions throughout this section. In fact for the cases we study the absolute majority of gates which are influence-solvable satisfy one of the conditions in \cref{res:zerobond_DU1_DU2}. However, there are also some gates which do not satisfy any of these conditions, which are discussed in \cref{sec:zerobond_newloccond}.

For  \subeqref{eq:zerobond_DU2_gen}{b}, \eqref{eq:zerobond_DU1_gen} (but not \subeqref{eq:zerobond_DU2_gen}{a}) and the new conditions \eqref{eq:zerobond_novel_cond1} and \eqref{eq:zerobond_novel_cond2}, any observables of the form
\begin{equation} \label{eq:light-cone}
    \pic{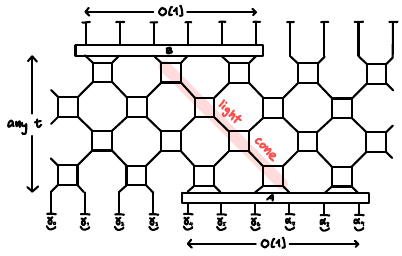}
\end{equation}
can be computed exactly by diagonalisation of an $O(1)$-sized matrix. Here, $\{\keT{\gamma_i}\}_i$ are any solvable states that we discuss, while $\{\keT{\alpha_i}\}_i$ can be any state at all, and local operators $A$ and $B$ may have support up to $O(1)$ distance from the light cone. Such observables include (a) expectation values of observables starting from a uniform non infinite-temperature state ($A=1$ and $\keT{\alpha_i}$ to some $\keT{\gamma}$), (b) expectation values of observables starting from a bi-partitioning protocol ($A=1$ and any $\keT{\alpha_i}$), (c) and correlation functions on the above settings. Specialising to single-site observables, correlation functions and observables do not generically vanish at finite distance away from the light cone, unlike dual-unitary and $\mathrm{DU}(2)$ gates.

\subsection{Spatially invertible gates}\label{sec:zerobond_spatial_inv_gate}
    In this subsection, we will show the following general result: given a gate that, when viewed as a matrix in space, is invertible, the bath generated by the circuit cannot be Markovian unless the gate satisfies a simple generalisation of the dual-unitarity condition. Diagrammatically, the `spatial invertibility' corresponds to the existence of $\pic{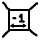}$ such that 
    \begin{align}
        \pic{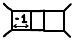} = \pic{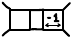} = \pic{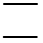}.
    \end{align}

    % A simple example of such a gate is SWAP. Intuitively, we expect that space-invertible gates to be rarely influence-solvable, because they can transport information deep into the bath. 
    For such gates we show (see Appendix \ref{app:spat-inv})
    \begin{result}\label{res:zerobond_space_inv}
        Consider a gate with local dimension $d\geq 2$ that is spatially invertible and $\chi=1$ influence-solvable. Then we have
        \begin{align}
            \mathscr{L} = \mathscr{L}_1,
        \end{align}
        and the $\chi=1$ influence-solvability condition is equivalent (i.e. necessary and sufficient) to the existence of $\keT{\gamma}$ and $\keT{\beta}$ such that
        \begin{align} 
            \pic{gamma-lower-left-gate} = \pic{beta-upper-left-identity} \,, \quad \quad \pic{beta-lower-left-gate} = \pic{gamma-upper-left-identity}\,,
            \label{eq:gamma-beta-equation}
        \end{align}
        Here, $\keT{\beta}$ can be equal to $\keT{\gamma}$. 
    \end{result}
    
    Although for generality, we have adopted notation agnostic of classical and quantum gates, it is rare for deterministic and reversible classical gates to be spatially invertible. Such gates are permutation matrices and viewing a matrix in space is equivalent to a reshuffling of the entries. Therefore it is easy to obtain a full row of zeros upon reshuffling. In fact, the only spatially invertible deterministic and reversible classical gates are dual-permutation gates. On the other hand, unless fine-tuned, a two-qudit gate (which is a $d^2 \times d^2$ unitary matrix) is invertible in space. Non-invertibility of the gate viewed from left to right requires that the determinant is equal to zero, which is a single constraint and is generically not satisfied.
     
    For quantum unitary gates, one can use subadditivity and invariance under unitaries of entanglement entropy to show that (Appendix \ref{apdx:proof-of-gamma-and-beta-from-below})
    \begin{result}
        Consider unitary gates that are both spatially invertible and $\chi=1$ influence-solvable. Then we have
        \begin{align} \label{eq:gamma-and-beta-from-below}
        \pic{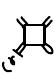} = \pic{identityket} \quad \pic{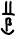} \,, \quad         \pic{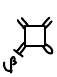} = \pic{identityket} \quad \pic{gammafoldedupright} \,.
        \end{align}
    \end{result}
    In \cref{sec:quantum-qubits-classification}, we will show that all quantum $d=2$ unitary gates that are spatially invertible are dual-unitary. Here, we provide an example of a unitary gate that is spatially invertible but not dual-unitary for $d \geq 3$.

    \begin{example} \label{ex:non-du-spatially-invertible}
        For $d \geq 3$, there exist a spatially invertible and $\chi=1$ solvable unitary gate that is \textbf{not} dual unitary. The gate is defined as follows. Choose some computational basis, and let $\hat{\gamma} = \hat{\beta} = \ketbra{0}{0}$. Then consider a unitary gate that acts as a SWAP in the subspace $\mathrm{span}\{\ket{0, i}, \ket{0, i} \forall i \in 0\colon(d-1) \}$, and a generic unitary elsewhere. For example, for $d=3$, we have
        \begin{align} \label{eq:non-du-spatially-invertible}
            u = 
            \left(
            \scalebox{0.85}{
            \begin{tabular}{c|ccccccccc}
               &$00$&$01$&$10$&$02$&$20$&$11$&$12$&$21$&$22$ \\
            \hline
            $00$ & $1$&  &  &  &  &  &  &  &   \\
            $01$ &  &  & $1$&  &  &  &  &  &   \\
            $10$ &  & $1$&  &  &  &  &  &  &   \\
            $02$ &  &  &  &  & $1$&  &  &  &   \\
            $20$ &  &  &  & $1$&  &  &  &  &   \\
            $11$ &  &  &  &  &  & $*$& $*$& $*$& $*$ \\
            $12$ &  &  &  &  &  & $*$& $*$& $*$& $*$ \\
            $21$ &  &  &  &  &  & $*$& $*$& $*$& $*$ \\
            $22$ &  &  &  &  &  & $*$& $*$& $*$& $*$ \\
            \end{tabular}
            } % Don't forget to close the math mode and scalebox
            \right).
        \end{align}
        In general the matrix is a block diagonal matrix of a SWAP-like part of size $(d+1) \times (d + 1)$ and an arbitrary part of size $(d^2 - (d+1)) \times (d^2 - (d+1))$.  
        
        The starred entries in \eqref{eq:non-du-spatially-invertible} can be an arbitrary unitary matrix, as long as it is not singular in space (left to right spatially) nor dual-unitary itself.
    \end{example}

\subsection{Classification of $\chi=1$ solvable classical deterministic reversible gates up to $d=3$}\label{sec:classic-classification}
    Since there is a finite number $(d^2)!$ of classical deterministic and reversible gates with local dimension $d$, we perform an exhaustive search and identify all influence-solvable gates. From this search we conclude that 
    \begin{result}\label{res:zerobond_classical}
        For classical deterministic and reversible brickwork circuits, all $(2^2)! = 24$ of $d=2$ gates are left (and right) influence-solvable for at least one initial state. For $d=3$, $115080$ out of $(3^2)! = 362880$ are influence-solvable (and $36084$ are simultaneously left and right influence-solvable).
    \end{result}
    
    We now describe how we performed this systematic search: The idea is that if we have a solution to equation \eqref{eq:FTMPS_eigvec} for all times $t$, then we also have a solution to equation \eqref{eq:FTMPS_eigvec} for any particular $t$. For example, for $t=6$, this is equivalent to \eqref{eq:zerobond_cond_b}, but with $\mathscr{L}$ restricted to $\mathscr{L}_2$:
    \begin{align} \label{diag:L2}
    \mathscr{L}_{2} &=\mathrm{span}\punct{\qty{\pic{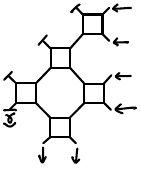}}}{\;}{.}
    \end{align}
    
    Explicitly, this means that we require that
    \begin{align} \label{eq:zerobond_classic_L2_b}
    \pic{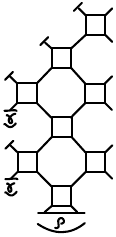} = \punct{\pic{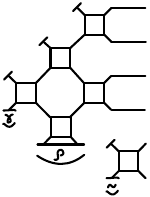}}{}{.}
    \end{align}
    
    For a given $\keT{\gamma}$ we can easily find all solutions $\rho$ to this numerically as the kernel of a rectangular matrix. This matrix is given by subtracting the RHS from the LHS of \eqref{eq:zerobond_classic_L2_b} and plus the additional condition that $\keT{\rho}$ has to be normalized to $1$). However, for generic $\keT{\gamma}$ there will be no solution. Thus we want to identify the special points where this kernel is not empty. To do this numerically we compute the smallest singular value (which we call residual) of the matrix for all values of $\keT{\gamma}$. Then we can identify points that meet this necessary condition for influence-solvability as zeros of this function.
    
\begin{figure}
        \centering
        \includegraphics{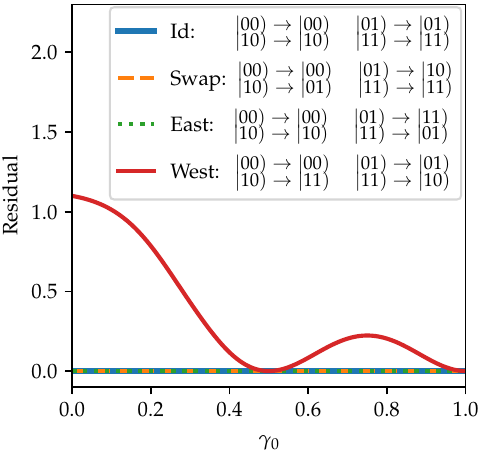}
        \caption{Residual as function of $\gamma_0=1-\gamma_1$ for some classical $d=2$ gates. The gate is influence-solvable for $\gamma^\top=(\gamma_0\;\gamma_1)$ at zeros of this function. The identity, SWAP and east gate (or CNOT gate) are solvable for all $\gamma$, but the west gate is only solvable for $\gamma^\top=(1\;0)$ and $\gamma^\top=(1/2\;1/2)$.}
        \label{fig:markov_search_classic_L2_d2}
    \end{figure}

\begin{figure}[t]
        \centering
        \begin{subfigure}{\linewidth}
            \includegraphics[width=\linewidth]{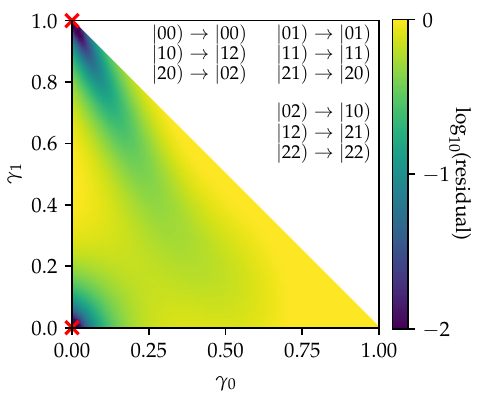}
            \caption{Classical gate $(d,\sigma)=(3,996)$}
        \end{subfigure}
        \begin{subfigure}{\linewidth}
            \includegraphics[width=\linewidth]{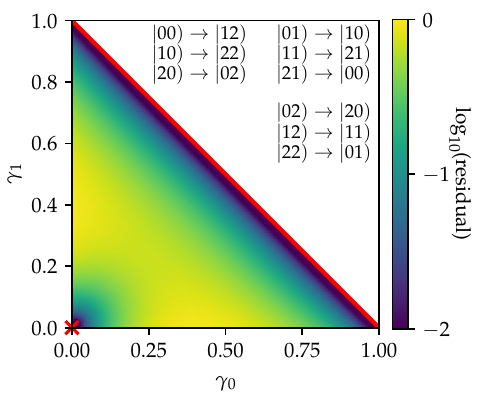}
            \caption{Classical gate $(d,\sigma)=(3,220318)$}
        \end{subfigure}
        \caption{Residual as function of $\gamma_0$ and $\gamma_1$ ($\gamma_2$ is fixed by $\gamma_2=1-\gamma_0-\gamma_1$ for two classical $d=3$ gates that we discuss in the text. The red crosses and lines indicate values of $\keT{\gamma}$ for which the gate is influence-solvable. }\label{fig:markov_search_classic_L2_d3_996}
    \end{figure}

    We give a plot of this for some $d=2$ gates in \cref{fig:markov_search_classic_L2_d2} and in \cref{fig:markov_search_classic_L2_d3_996}. Note that in these plots, if the value of the residual is not $0$, then we can rigorously exclude the possibility of influence-solvability. If the residual vanishes then we know that we only have a solution to \eqref{eq:zerobond_classic_L2_b}, but not necessarily a solution to \eqref{eq:zerobond_cond_b}. To check \eqref{eq:zerobond_cond_b} we then check larger $\mathscr{L}_n$ until convergence to $\mathscr{L}$.
    
    Let us briefly discuss the case $d=2$. For $d=2$ gates are either solvable for any $\keT{\gamma}$ or if not always solvable for the infinite temperature state $\keT{\gamma}=(1/2, 1/2)^\top$ and potentially also for $\keT{\gamma}=(1, 0)^\top$ or $\keT{\gamma}=(0, 1)^\top$. It turns out that they all satisfy at least one of the conditions in \cref{res:zerobond_DU1_DU2}, hence are of known types of solvability. Let us discuss the examples depicted in \cref{fig:markov_search_classic_L2_d2}, which exemplify the most basic behavior of influence-solvability
    \begin{itemize}
        \item The identity gate is always (trivially) solvable for any $\keT{\gamma}$ with $\pic{0bdA} = \pic{gamma-lower-left-gate} = \pic{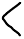}$.
        \item The SWAP gate is also solvable for any $\keT{\gamma}$ with $\pic{0bdA} = \pic{gamma-lower-left-gate} = \pic{gamma-upper-left-identity}$. This describes a bath which practically ``destroys'' all information sent from the system into it, i.e. the information is unrecoverable for the system and replaces it by a state $\keT{\gamma}$, which is transported to the system from the left.
        \item The Floquet East gate (or CNOT gate with control on the right) is also solvable for any $\keT{\gamma}$ with $\pic{0bdA} = \pic{gamma-lower-left-gate} = \pic{legTRBR}$. This is influence-solvable because while information from the system can affect the bath, the bath cannot affect the system.
        \item The Floquet West gate (or CNOT gate with control on the left) is only influence-solvable for $\keT{\gamma}=(1, 0)^\top$ with $\pic{0bdA} = \pic{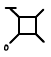} = \pic{legTRBR}$ and $\keT{\gamma}=(1/2, 1/2)^\top$ with $\pic{0bdA} = \pic{flatandtildeleftongate} = \pic{tildeflatleft}$. The former is influence-solvable because the control is always turned off, thereby not influencing the system. The latter is solvable because it describes a bath that flips or does not flip the state with equal probability, which means that there is no way to predict what the output bit on the system side is, and therefore we can replace it with an infinite temperature state.
    \end{itemize}
    We give a list of all solvable $\keT{\gamma}$ for all $d=2$ gates in Appendix \ref{app:classic_d2}.
    
    For local dimension $d=3$ there is a significantly larger amount of possibilities. We briefly discuss two examples, which we will further investigate later in sections \ref{sec:zerobond_partitioning} and \ref{sec:zerobond_markov}:

    \begin{example}\label{ex:996}
        The classical deterministic reversible gate $(d, \sigma) = (3, 996)$ with rules given in \cref{fig:markov_search_classic_L2_d3_996} (a), has only one local conserved quantity: the number of $1$'s on the even sites minus the number of $0$'s on the odd sites. It is influence-solvable from the left with $\keT{\gamma} = \keT{1} = (0, 1, 0)^\top$ and $\keT{2} = (0, 0, 1)^\top$. Furthermore, it is also influence-solvable from the right with $\keT{\gamma} = \keT{0} = (1, 0, 0)^\top$.
    \end{example}

    We chose to present this gate because not only is it influence-solvable, but it also has exactly one conserved quantity with non-trivial hydrodynamics (see Ref.~\onlinecite{kim2025circuits} for an investigation into its hydrodynamics). This demonstrates that solvability is also possible for chaotic circuits with conservation laws. To best of our knowledge, all the examples of solvable circuits either have non or infinitely many conservation laws. In particular, it is known that if dual unitary circuits have conservation laws they can be decomposed into solitons (or gliders): those travel with exact velocity $\pm 1$ independent of the state of the system~\cite{10.21468/SciPostPhys.8.4.068,HoldenDye2025fundamentalcharges}. Furthermore, if there exists one such soliton one can construct arbitrarily many conserved quantities from it. In fact, we are even able to compute certain partitioning protocols for this gate, see section \ref{sec:zerobond_partitioning}.

    The gate $(d, \sigma) = (3, 996)$ is only solvable for isolated points in $\keT{\gamma}$. On the other hand, there are also gates for which an whole line is solvable:
    \begin{example}\label{ex:220318}
        $(d,\sigma)=(3,220318)$ has a full line $\keT{\gamma}=(\gamma_0,1-\gamma_0,0)^\top$ of solvable states, see \cref{fig:markov_search_classic_L2_d3_996} b). 
        
        We will discuss later in section \ref{sec:zerobond_markov} that the corresponding $\bm{A}$ is a Markov matrix with one eigenvalue being controlled by the free parameter $\gamma_0$. Hence, in this model the bath generated by the circuit can be controlled by choosing different initial states.
    \end{example}
    
\begin{figure}[!h]
        \centering
        \includegraphics[width=\linewidth]{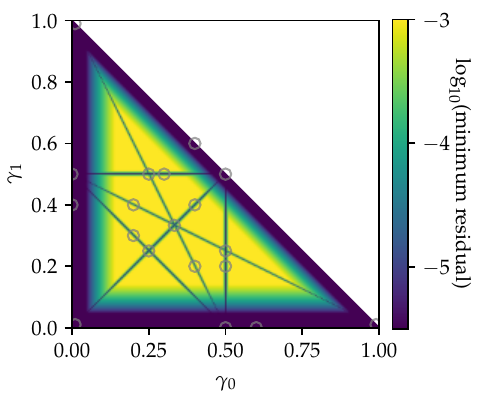}
        \caption{Residual as function of $\gamma_0$ and $\gamma_1$ ($\gamma_2$ is fixed by $\gamma_2=1-\gamma_0-\gamma_1$ minimized over all $d=3$ gates (excluding those where the residual vanishes for all $\keT{\gamma}$). Only the lower left corner corresponds to physical $\keT{\gamma}$. If this function is not zero for a specific $\keT{\gamma}$, it means that there is no $d=3$ gate that is solvable for this $\keT{\gamma}$ (apart from those that are always solvable). The gray circles mark the values of $\keT{\gamma}$ that were used to check solvability.}
        \label{fig:markov_search_classic_L2_d3_min}
    \end{figure}

    One interesting observation is that, unless the gate is solvable for all $\keT{\gamma}$, the possible values $\keT{\gamma}$ can only lie in a very restricted set. In \cref{fig:markov_search_classic_L2_d3_min}, we give a plot of the residual minimized over all gates (excluding those that are zero for all $\keT{\gamma}$). The way to interpret this plot is as follows. The regions where the minimal residual does not vanish rules out $\chi=1$ influence-solvability for any $d=3$ classical gate, excluding those that are trivially solvable for all states, such as the identity and SWAP. 
    From this figure we can clearly see that the only possible values for $\keT{\gamma}$ lie in the following set
    \begin{equation}
    \begin{aligned}
        \keT{\gamma} \in \Bigg\{(\gamma_0,1-\gamma_0,0)^\top, (\gamma_0,1/2-\gamma_0,1/2)^\top & , \\
        \quad \quad \quad \quad (\gamma_0,\gamma_0,1-2\gamma_0)^\top : \gamma_0 \in [0, 1] & \Bigg\},
    \end{aligned}
    \end{equation}
    and permutations thereof.
    
    In order to check which of these $\keT{\gamma}$ give rise to solvability for a particular gate we proceed as follows: First, we pick a particular value of $\keT{\gamma}$ (marked as gray circles in \cref{fig:markov_search_classic_L2_d3_min}). For each of these values, we find the space $\mathscr{L}$ by iteratively constructing the spaces $\mathscr{L}_n$ as explained in \cref{sec:FTMPS_FTMPS}. Then, we find all solutions $\keT{\rho}$ (if any) to the linear equation \eqref{eq:zerobond_cond_b}. This way, for each gate we can check at which $\keT{\gamma}$ it is $\chi=1$ influence-solvable.

    We see that while it is rare that a gate is influence-solvable for a fixed initial state, it is common to find \emph{at least one} initial influence-solvable state for a given gate. This demonstrates the usefulness of generalising solvability to arbitrary states. Note that here we only discussed the simplest case of influence-solvability with bond dimension $\chi=1$. It seems reasonable to expect that there may be even more options for influence-solvability if one considers larger bond dimensions.

\subsection{Classification $\chi=1$ solvable quantum unitary gates for $d=2$} \label{sec:quantum-qubits-classification}
    Unlike classical gates, where there are only finitely many gates for a given local dimension $d$, there exist an infinite number of quantum gates. Furthermore, one has to work in the folded space. These two considerations make analysis significantly harder, and even for local dimension $d=2$, it is not realistic to perform a systematic search numerically as we did for classical gates. Therefore, we analyze their influence-solvability by hand.

    \begin{result} \label{res:zerobond_quantum}
        All $d=2$ unitary gates that are quantum $\chi=1$ influence-solvable are either dual-unitary or they are controlled gates (up to a gauge choice). In the former, certain parameters allow for more solvable states than the infinite temperature state in the form
        \begin{align}
            \pic{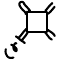} &= \pic{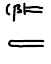} & \pic{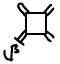} &= \pic{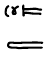},
        \end{align}
        where $\keT{\beta}$ is not necessarily equal to $\keT{\gamma}$.
        
        In the latter, they satisfy at least one of the two conditions reminiscent of the $\mathrm{DU}(2)$ condition \eqref{eq:zerobond_DU2_gen}:
        \begin{align}
        \pic{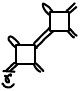} = \pic{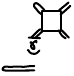} \, , \;\; \text{ or } \;\; \pic{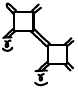} = \pic{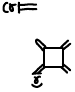} \, ,
            \label{eq:zerobond_DU2_gen_quantum}
        \end{align}
        which are the quantum versions of \eqref{eq:zerobond_DU2_gen}.
    \end{result}

    Based on these conditions one can solve the circuit analogously to the $\mathrm{DU}(2)$ case.
    
    Therefore, a generic unitary gate can only be influence-solvable if it satisfies \eqref{eq:zerobond_DU2_gen_quantum}. Since this is a fairly strong condition, we do not expect the generic unitary gate to be influence-solvable.
    
    For our analysis, we will use the following fact that any two-qubit gate can be decomposed as \cite{bertini2019exact,vatan2004optimal,kraus2001optimal}
    \begin{align} 
        \pic{unitary} = \pic{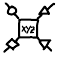}.\label{eq:zerobond_quantum_decomp_XYZ}
    \end{align}
    Here each $\pic{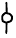}, \pic{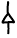}, \pic{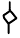}, \pic{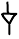}$ are independent single-site unitaries, and
    \begin{align}
    	\pic{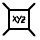} = 
        \hat{u}^\mathrm{XYZ}(x,y,z) & = \exp(-ix \hat{\sigma}_\mathrm{x}^{\otimes 2} -i y \hat{\sigma}_\mathrm{y}^{\otimes 2}  -i z \hat{\sigma}_\mathrm{z}^{\otimes 2} ).\label{eq:zerobond_quantum_XYZ}
    \end{align}
    \paragraph*{Space-invertible gates.} Here, we detail \Cref{res:zerobond_quantum} for spatially invertible gates. Combining \Cref{res:zerobond_space_inv} with the parametrisation \eqref{eq:zerobond_quantum_decomp_XYZ}, we prove in \cref{sec:d=2-space-invertible} the following.
    
    All $d=2$ spatially-invertible and $\chi=1$ (left) influence-solvable unitary gates are dual unitary. Within these gates, there are also gates that have other solvable states than the maximally mixed state, $\keT{\gamma} \neq \keT{\sim}$.
  s
    Parametrising a spatially invertible gate $\hat{u}$ as \eqref{eq:zerobond_quantum_XYZ}, they are only influence-solvable for these states if
    \begin{align} \label{eq:xyz-in-x-y-z}
        \pic{XYZ} \in \left\{ \pic{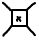}, \pic{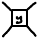}, \pic{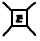} \right\}
    \end{align}
    and $\pic{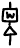} = \left(\pic{rhombus}\right)^\dag$, where $\hat w$ is defined below. For the case
    \begin{align}
        \pic{z} = \hat{u}^\mathrm{XYZ}(\pi/4,\pi/4,z),
    \end{align}
    $\hat w$ is either of the form $\hat w_+ = e^{i \phi \hat Z}$ and $\hat w_- = e^{i \frac{\pi}{2}(\hat X \cos \phi + \hat Y \cos \phi)}$. the solvable states are any of the form
    $\pic{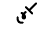} = \pic{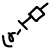}$ where 
    \begin{align}
        \hat \gamma' = \frac{1}{2} \hat I + p \hat \sigma_\mathrm{z},
    \end{align}
    where $p$ is interpreted as probability to be in the $\ketbra{0}{0}$ state.
    For the case $\hat w_+$, $\hat \beta = \hat \gamma$. For the case $\hat w_-$, $\hat \beta \neq \hat \gamma$ and
    \begin{align}
        \hat \beta' = \frac{1}{2} \hat I - p \hat \sigma_\mathrm{z},
    \end{align}
    where $\hat \beta'$ is similarly defined as $\hat \gamma'$. If $\pic{XYZ} = \pic{x}$ or $\pic{y}$, which are similarly defined as $\pic{z}$, $\hat{\gamma}'$ is diagonal in the $X$ and $Y$ basis, and $\hat w_\pm$ commutes/anticommutes with $\hat \sigma_\mathrm{x}$ and $\hat \sigma_\mathrm{y}$, respectively. 
   
    \paragraph*{Space-non-invertible gates.}
    As discussed in Appendix \ref{app:quantum_d2_space_notinv} the only non-invertible gates in space are controlled gates of the form
    \begin{align}
    	\pic{unitary} &= \pic{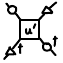} \; ,\label{eq:zerobond_quantum_gauge}
    \end{align}
    where
    \begin{equation} \label{eq:zerobond_quantum_space_invertible_draw}
    \begin{aligned} 
        \begin{aligned}
            \hat{u}'  = \ketbra{0} \otimes \hat{\Gamma}_0 + \ketbra{1} \otimes \hat{\Gamma}_1
        \end{aligned}
         \\
         \begin{aligned}
        = \pic{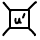}  = \pic{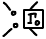} + \pic{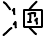} \; .
         \end{aligned}
    \end{aligned}
    \end{equation}
    In the notation of \eqref{eq:zerobond_quantum_decomp_XYZ} they correspond to Ising gates $\hat{u}\upd{XYZ}(0,0,z)$. Note that the single site unitaries $\bm{\triangle}$ and $\bigcirc$ are just a Gauge transformation: they cancel locally when building a brickwork circuit from \eqref{eq:zerobond_quantum_space_invertible_draw}. Therefore, we can set $\bm{\triangle}=\bigcirc=\mathbb{1}$ in which case the gates act as follows: if the left bit (the control bit) is $0$, then we apply the unitary matrix $\hat{\Gamma}_0$ to the right side and if the left bit is $1$ then we apply the unitary matrix $\hat{\Gamma}_1$.
    
    As we show in Appendix \ref{app:quantum_d2_space_notinv} the only possible options for $\hat{\gamma}$ are the following:
    \begin{itemize}
        \item Any $\hat{\gamma}$ in the trivially solvable case $\hat{\Gamma}_0 = \hat{\Gamma}_1$. In this case the circuit is influence-solvable for any initial state.
        \item The pure state $\bm{\triangle}^\dagger\hat{\gamma}\bm{\triangle}=\ketbra{0}{0}$ in case $\hat{\Gamma}_0$ is diagonal or equivalently $\bm{\triangle}^\dagger\hat{\gamma}\bm{\triangle}=\ketbra{1}{1}$ in case $\hat{\Gamma}_1$ is diagonal. In this case the quantum channel $\bm{A}$ is unitary and is given by $\hat{\Gamma}_0$ or $\hat{\Gamma}_1$ respectively.
        \item The infinite temperature state $\hat{\gamma} = \tfrac{1}{2}\hat{\mathbb{1}}$ (or equivalently any $\hat{\gamma}$ such that $\tr (\bm{\triangle}^\dagger\hat{\gamma}\bm{\triangle}\hat{\sigma}\ind{z}) = 0$) in case $\hat{\Gamma}_1 = \hat{\sigma}_{\vec{n}} \hat{\Gamma}_0$, where $\hat{\sigma}_{\vec n} = n_1\hat{\sigma}\ind{x} +n_2\hat{\sigma}\ind{y}+n_3\hat{\sigma}\ind{z}$ and $\sum_i n_i^2 = 1$. The only solvable gates are those where $\bra{0}\hat{\sigma}_{\vec{n}}\ket{0} = 0$ or $\bra{0}\hat{\Gamma}_0^\dagger\hat{\sigma}_{\vec{n}}\hat{\Gamma}_0\ket{0} = 0$. Here the quantum channel $\bm{A}$ is not unitary (and in fact not invertible): it is given by a unitary channel $\hat{\Gamma}_0$ followed by a projector on $\mathbb{1}$ and $\hat{\sigma}_n$. 
    \end{itemize}
    
   Note that these are the only influence-solvable two-qubit unitary gates which are not invertible in space. All of them satisfy at least one of the conditions of $\mathrm{DU}(2)$ gates, which is sufficient to be $\chi=1$ influence-solvable.

\subsection{New local conditions for $\chi=1$ influence-solvability}\label{sec:zerobond_newloccond}
From the previous two sections, we found that all classical and quantum $d=2$ gates satisfy one of the relations of \cref{res:zerobond_DU1_DU2}, which are generalizations of the dual unitary and $\mathrm{DU}(2)$ conditions. These conditions so far were not systematically analyzed apart from the standard case of $\keT{\gamma} = \keT{\sim}$. In particular, we have observed that only very special values of $\keT{\gamma}$ are possible. Nonetheless, these circuits can be studied similarly to dual unitary and $\mathrm{DU}(2)$ circuits. This leaves the question of whether there are any new forms of local solvability. 

We answer this question in the affirmative: we find that for classical $d=3$ gates, out of the 115080 influence-solvable gates 106728 gates satisfy at least one of the conditions \cref{res:zerobond_DU1_DU2} for at least one $\keT{\gamma}$. However, the remaining $8352$ gates are \emph{not} influence-solvable in a known way, to best of our knowledge. 

All of those possess solutions of \eqref{eq:zerobond_cond_b} with $\mathscr{L} = \mathbb{C}^{3^2}$, i.e. the entire space. Based on this we conclude that they must satisfy the following local condition,
\begin{align}
    \pic{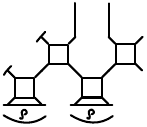} \; = \; \punct{\pic{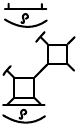}}{}{.}
    \label{eq:zerobond_novel_cond1}
\end{align}

As an example,
\begin{example}
    The $(d,\sigma)=(3,725)$ given by the rules
    \begin{center}
        \begin{tabular}{ccc}
    $\keT{{00}}\to\keT{{00}}$&$\keT{{01}}\to\keT{{01}}$&$\keT{{02}}\to\keT{{10}}$\\
    $\keT{{10}}\to\keT{{02}}$&$\keT{{11}}\to\keT{{11}}$&$\keT{{12}}\to\keT{{12}}$\\
    $\keT{{20}}\to\keT{{22}}$&$\keT{{21}}\to\keT{{21}}$&$\keT{{22}}\to\keT{{20}}$\end{tabular}
    \end{center}
    is $\chi=1$ influence-solvable for the initial state 
    \begin{align}
        \pic{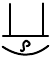} = \pic{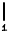} \quad \pic{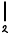}
    \end{align}
    and satisfies \eqref{eq:zerobond_novel_cond1}.
\end{example}

However, among the 106728 gates that satisfy \cref{res:zerobond_DU1_DU2} for some $\keT{\gamma^{(1)}}$, there are also 5742 gates for which there exists another $\keT{\gamma^{(2)}}$ and the corresponding $\keT{\rho^{(2)}}$ such that none of the conditions in \cref{res:zerobond_DU1_DU2} and not necessarily the new condition \eqref{eq:zerobond_novel_cond1} are satisfied, yet it is $\chi=1$ influence-solvable, obeying the local condition
\begin{align}
    \pic{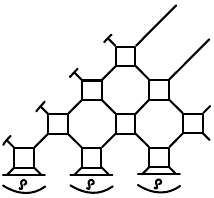} = \punct{\pic{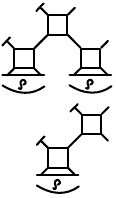}}{}.\label{eq:zerobond_novel_cond2}
\end{align}

For instance, 
\begin{example}
    The gate $(d,\sigma)=(3,44)$ given by
    \begin{center}
        \begin{tabular}{ccc}
    $\keT{{00}}\to\keT{{00}}$&$\keT{{01}}\to\keT{{01}}$&$\keT{{02}}\to\keT{{02}}$\\
    $\keT{{10}}\to\keT{{10}}$&$\keT{{11}}\to\keT{{12}}$&$\keT{{12}}\to\keT{{22}}$\\
    $\keT{{20}}\to\keT{{20}}$&$\keT{{21}}\to\keT{{11}}$&$\keT{{22}}\to\keT{{21}}$\end{tabular}
    \end{center}
    is solvable with respect to the initial state 
    \begin{align}
        \pic{rho} = \pic{1ketupright} \quad \pic{1ketupright},
    \end{align}
    obeying \eqref{eq:zerobond_novel_cond2}.
\end{example}
It turns out that \eqref{eq:zerobond_novel_cond2} is indeed the most general condition for the classical $d=3$ gates:
\begin{result}
    All influence-solvable classical $d=3$ gates for all of their solvable $\keT{\gamma}$ satisfy \eqref{eq:zerobond_novel_cond2} for at least one $\keT{\rho}$.
\end{result}

Note that for any gate satisfying \eqref{eq:zerobond_novel_cond1} or \eqref{eq:zerobond_novel_cond2} one can compute correlation functions along the left light cone as in \eqref{eq:light-cone}.

\subsection{Further examples of $\chi=1$ influence-solvable gates} \label{sec:zerobd-further}
In this subsection we discuss two gates may be of interest in detail.
\subsubsection{Tunable exact Markovian baths}\label{sec:zerobond_markov}
\begin{figure*}[t]
    \centering
    \includegraphics{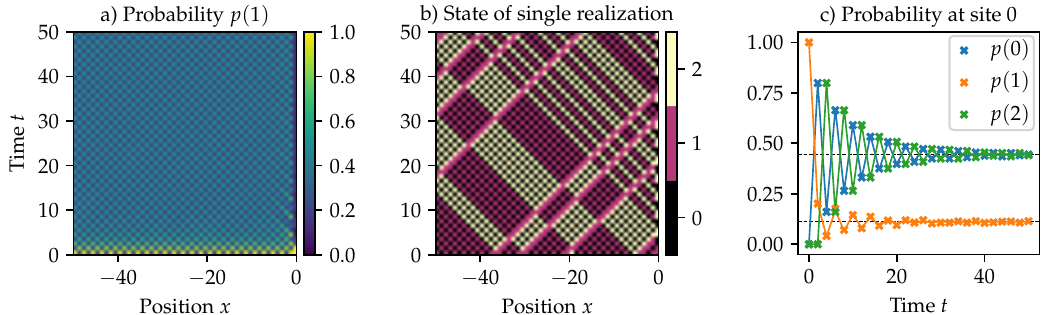}
    \caption{Simulation of half-system setup \cref{fig:mot_infsolv_applications} a) for $(d,\sigma)=(3,220318)$ starting from state $\keT{\rho} =\tfrac{23}{30} \keT{01}+\tfrac{5}{30}\keT{11}+\tfrac{1}{30}\keT{02}+\tfrac{1}{30}\keT{12}$: a) probability of observing $1$ as function of space and time b) the evolution of a single configuration c) the probabilities at the boundary site $x=0$ (crosses) together with the exact result from influence-solvability (line).}
    \label{fig:zerobond_example_220318}
\end{figure*}

The first example is the classical gate $(d,\sigma)=(3,220318)$ whose rules are given in \cref{fig:markov_search_classic_L2_d3_996} (b). This rule has the special feature that its bath is tuneable, i.e. different initial states inside the bath give rise to different influence-matrices. Note that this is technically true also for the SWAP gate: for any $\keT{\gamma}$, SWAP gate transports a train of $\keT{\gamma}$ towards the system and transports any information coming from the system into bath infinitely far away from the system. Therefore, for SWAP the channel $\bm{A}$ given by $\pic{0bdA} =\pic{gamma-upper-left-identity}$ effectively kills information sent from the system into the bath and replaces it by $\keT{\gamma}$. In particular, $\bm{A}$ is a projection, i.e. it has only eigenvalues $0$ and $1$. 

The gate $(d,\sigma)=(3,220318)$ is an example where the bath interacts with the system (not just erasing information), but still in an initial-state dependent manner. In particular, the gate is solvable for a family of $\keT{\gamma} = (\gamma_0,1-\gamma_0,0)^\top$ with a non-trivial $\keT{\gamma}$-dependent $\bm{A}$. Note that this gate satisfies the right condition of \eqref{eq:zerobond_DU2_gen}.

For $(d,\sigma)=(3,220318)$ the $\bm{A}$ is given by the matrix
\begin{align}
    \vb{A} = \begin{pmatrix} 
				0 & \gamma_0 & \gamma_0\\
				0 & 1-\gamma_0 & 1-\gamma_0\\
				1 & 0 & 0
			\end{pmatrix}
\end{align}
with eigenvalues $\lambda = 0,1,-\gamma_0$. Here $0\leq \gamma_0\leq 1$ depends on the initial state. In \cref{fig:zerobond_example_220318} we demonstrate this against numerical simulations for the semi-infinite circuit setup \cref{fig:mot_infsolv_applications} (a). We can analytically predict the evolution at the boundary using influence-solvability. As $t\to \infty$, the state at this boundary site $x=0$ is given by $\keT{\rho_{x=0,\infty}} = \tfrac{1}{1+1/\gamma_0}(1,1/\gamma_0-1,1)^\top$. In \cref{fig:zerobond_example_220318} (b) we also give the deterministic evolution of a single sample. From this we can see that the gate behaves similarly to SWAP: it transports information with velocity $1$ either towards the system or away from the system. In each time-step the gate closest to the boundary applies a deterministic map on the boundary state that depends on which state is currently present in the bath next to it. Since this state is random, the map is also random, making the dynamics of the boundary state non-deterministic.

\subsubsection{Exact microscopic solution to hydrodynamic partitioning protocol}\label{sec:zerobond_partitioning}
\begin{figure*}[t]
    \centering
    \includegraphics{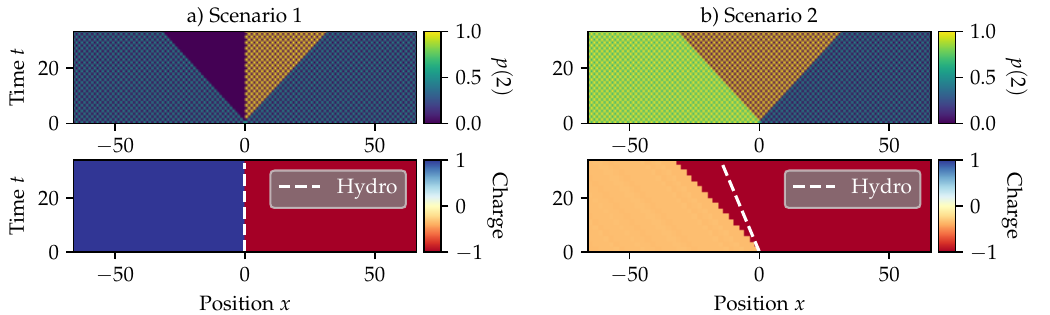}
    \caption{Simulation of the two different partitioning setups \cref{fig:mot_infsolv_applications} e) for $(d,\sigma)=(3,220318)$ discussed in the text: (a) left and right states Gibbs states (b) left and right states are not Gibbs states. The top plots show the probability of observing a $\keT{2}$ at a specific space-time point. The bottom plots show the corresponding charge density. Since the gate satisfies the generalization of $\mathrm{DU}(2)$ circuits all interfaces can either spread with velocity $-1,0$ or $1$. In the second plot this is not compatible with the expected hydrodynamic velocity (white dashed line), implying that hydrodynamics fails to describe this setup.}
    \label{fig:zerobond_example_996}
\end{figure*}
The second example is a further discussion of gate $(d,\sigma)=(3,996)$, given by the rules in \cref{fig:markov_search_classic_L2_d3_996} (a). As mentioned previously, this gate has a single conserved charge (and only one)~\cite{kim2025circuits}: the number of $1$ on even sites minus the number of $0$ on odd sites, i.e.
\begin{align}
    Q[\bm{a}] &= \sum_x \delta_{a_{2x}=1}-\delta_{a_{2x+1}=0}.
\end{align}
Furthermore, it has a non-trivial Euler hydrodynamics
\begin{align}
    \partial_t q + \partial_x j(q) = 0\label{equ:zerobond_hydroequation}
\end{align}
with current~\cite{kim2025circuits}:
\begin{align}
    j(q) &= \tfrac{1}{3}\qty(2-\sqrt{9q^2+16}).
\end{align}
This model is also influence solvable from the left with $\keT{\gamma\ind{L}}=\keT{1}$ or $\keT{\gamma\ind{L}}=\keT{2}$ and from the right with $\keT{\gamma\ind{R}}=\keT{0}$. In all cases the gate satisfies the right condition of \eqref{eq:zerobond_DU2_gen}, i.e.
\begin{align}
    \pic{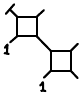} &=   \pic{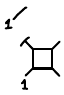} & \pic{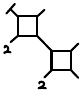} &=   \pic{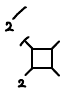}\\
    \pic{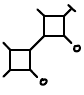} &=   \punct{\pic{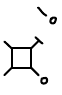}}{\;}{.}
\end{align}

We study partitioning protocols, i.e. setup \cref{fig:mot_infsolv_applications} (e) where the initial states on the left and the right differ. We consider two scenarios:
\begin{itemize}
    \item Scenario 1: The initial states are $\keT{\rho\ind{L}^{(1)}}=(\keT{11}+\keT{12})/2$ and $\keT{\rho\ind{R}^{(1)}}=(\keT{00}+\keT{20})/2$, corresponding to $\keT{\gamma\ind{L}^{(1)}}=\keT{1}$ and $\keT{\gamma\ind{R}^{(1)}}=\keT{0}$. These states are exact Gibbs states $\sim e^{-\beta Q}$ with $\beta=\pm \infty$ respectively.
    \item Scenario 2: The initial states are $\keT{\rho\ind{L}^{(2)}}=p\ind{L} \keT{22}+(1-p\ind{L})\keT{12})$ and $\keT{\rho\ind{R}^{(2)}}=p\ind{R}\keT{00}+(1-p\ind{R})\keT{20})$, corresponding to $\keT{\gamma\ind{L}^{(2)}}=\keT{2}$ and $\keT{\gamma\ind{R}^{(2)}}=\keT{0}$. Here $0\leq p\ind{L},p\ind{R}\leq 1$. These states are \underline{not} Gibbs states.
\end{itemize}
The charge density in these states are
\begin{align}
    q\ind{L}^{(1)} &= 1 & q\ind{R}^{(1)} &= -1 & q\ind{L}^{(2)} &= p\ind{L}-1 & q\ind{R}^{(2)} &= -1.
\end{align}

We give numerical simulations of these two setups in \cref{fig:zerobond_example_996}. Influence-solvability allows us to compute the exact effective time-evolution operator $\bm{G}\ind{eff}$ for the boundary site $x=0$. They are given by
\begin{align}
    \bm{G}\ind{eff}^{(1)} &= \pic{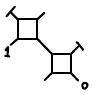} = \pic{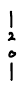}+\pic{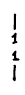}+\pic{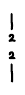} = \begin{pmatrix}
        0 & 0 & 0\\
        0 & 1 & 0\\
        1 & 0 & 1
    \end{pmatrix},\\
    \bm{G}\ind{eff}^{(2)} &= \pic{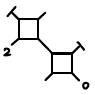} = \pic{page28/Solvabilitypage2818}+\pic{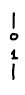}+\pic{page28/Solvabilitypage2820} = \begin{pmatrix}
        0 & 1 & 0\\
        0 & 0 & 0\\
        1 & 0 & 1
    \end{pmatrix}.
\end{align}
for scenario 1 and 2, respectively. Note that $\bm{G}\ind{eff}$ in both cases only has Jordan blocks with eigenvalues $0$ or $1$. Hence, instead of an exponential decay, the dynamics at site $0$ becomes stationary after $t=2$ and $t=4$ time steps. However, because this gate satisfies the generalization of $\mathrm{DU}(2)$ we are also able to compute the state of the system at any $x,t$. This can be done similar to the computation of correlation functions in $\mathrm{DU}(2)$ circuits~\cite{bertini2019exact}. Similar to $\mathrm{DU}(2)$ we find that the state is piece-wise constant and can only jump on the rays $v=-1,0$ and $1$. This is in agreement with the numerical simulations in \cref{fig:zerobond_example_996}.

Since the states on the left and the right have different charge, the same problem can also be studied from a hydrodynamic perspective, i.e. solving \eqref{equ:zerobond_hydroequation} with initial condition $q(t=0,x) = q\ind{L}\mathbb{1}_{x<0} + q\ind{R}\mathbb{1}_{x>0}$. The solution is a that the jump moves moves with velocity~\cite{kim2025circuits,10.1093/oso/9780198507000.001.0001} $v\ind{hydro} = (j(q\ind{R})-j(q\ind{L}))/(q\ind{R}-q\ind{L})$. This velocity is indicated in \cref{fig:zerobond_example_996} as a white dashed line. In Scenario 1, where the states are Gibbs states, we have $v\ind{hydro}=0$ in agreement with the microscopic solution. However, in Scenario 2, while $-\tfrac{1}{3} \leq v\ind{hydro} \leq 0$, the microscopic solution predicts the the interface moves with $v\ind{micro}=-1$. Thus hydrodynamics fails to describe this scenario. The reason for this is that the initial states are neither Gibbs states nor equilibrate to Gibbs states. This is easiest to see in the case where $p\ind{L} = 1$, i.e. $\keT{\rho\ind{L}^{(2)}}=\keT{22}$. Due to the rules of this gate, see \cref{fig:markov_search_classic_L2_d3_996}, this state is a locally stationary state of the dynamics.

\section{Further aspects of influence-solvability}\label{sec:prop}
In this section, we discuss further aspects of influence-solvability not restricted to unit bond dimension $\chi=1$.

\subsection{Influence-solvability and  conservation laws} \label{sec:prop_CQ}
After narrowing down solutions by checking necessary conditions for influence-solvability, let us now discuss under which cases we can rule out solvability.

For this note that if a circuit is influence-solvable then we can integrate out the bath and obtain an effective finite dimensional evolution matrix $\bm{G}\ind{eff}$. 

For simplicity let us assume that $\bm{G}\ind{eff}$ is diagonalizable (otherwise one has to slightly adapt the following argument using its Jordan normal form): Diagonalizing $\bm{G}\ind{eff}$ it has a finite number of eigenvalues $\lambda_i=e^{-\omega_i}$ and therefore the evolution of any expectation value is given by \eqref{eq:mot_is_G_Ot}
\begin{align}
    \expval{O(t)} &= \sum_i \lambda_i^t\brAkeT{O}{\psi_i}\brAkeT{\psi_i}{\rho\ind{S}} = \sum_i c_i e^{-\omega_i t}
\end{align}
where the $c_i$ depend on the observable $O$ and the initial state. Therefore, generically, we see that the evolution is a sum of a finite number of exponential decays. In particular, in a typical setup we image that the evolution exponentially decays into the subspace of $\abs{\lambda_i}=1$ (there is always one $\lambda_i=1$ due to conservation of total probability). 

However, there are many examples of systems where we know that the decay is not exponential. A standard example for that are systems with diffusive transport (similar argument applies for super- and subdiffusive transport). Here connected auto-correlation functions of the density of the conserved quantity $q$ decay as a power law,
\begin{align}
    \expval{q(t,x=0)q(t=0,x=0)}\upd{c} \sim 1/\sqrt{t}.
\end{align}
Hence, the circuit
\begin{align}
    \punct{\cdots\pic{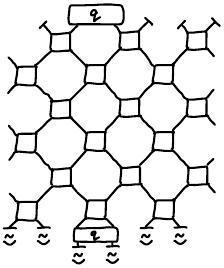}\cdots}{\;}{,}
\end{align}
which represents this auto-correlation function in the infinite temperature state, cannot be influence-solvable from both the left and the right.

While this cannot be described with a finite bond dimension MPS, it could be possible to describe this with a MPS of infinite bond dimension. Hence, it would be useful to develop a suitable mathematical theory extending the results of this paper to infinite bond dimension. While at first it may seem that an MPS composed of infinite dimensional matrices is not efficiently computable in practice, it may still be possible to find good finite dimensional approximations to it. Alternatively, some infinite dimensional MPS have the property that at any finite length they reduce to a low bond dimension (inhomogeneous) MPS (see for instance the ones used in ~\cite{PhysRevLett.107.137201} to describe steady states of Lindblad dynamics). In this case one can efficiently compute finite time dynamics, however it may hard to obtain analytic predictions on asymptotic behaviour. 

However, note that the above argument implies the absence of solvability for certain initial states, but not necessarily for all initial states. In particular, let us consider again a diffusive system\footnote{The following idea was pointed out to us by Luca Capizzi.}: while the two-point correlation function
\begin{align}
    \expval{q(t,0)q(0,x)}\upd{c} &\approx \frac{C}{\sqrt{4\pi Dt}}e^{-x^2/(4Dt)} 
\end{align}
decays algebraically in time at $x=0$, its Fourier transform
\begin{align}
    \sum_x e^{ikx} \expval{q(t,x)q(0,0)}\upd{c} &\sim e^{-Dk^2t}
\end{align}
decays exponentially in $t$. Furthermore, one can easily express the Fourier transform $\keT{\tilde{q}_k} = \sum_x e^{ikx} \keT{q_x}$, where $\keT{q_x}$ is the conserved density at site $x$, as a translation invariant MPS with bond dimension $2$,
\begin{align}
    \pic{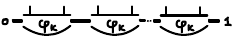},
\end{align}
where
\begin{align}
    \pic{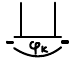} = e^{ik} \pic{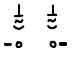} + \pic{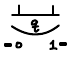} + \pic{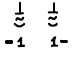}.
\end{align}

The idea behind this MPS is that one will start with a state $0$ in the bond from the left and end with a state $1$ in the bond on the right. Since the MPS bond can only switch once from $0$ to $1$ one obtains the superposition $\keT{\tilde{q}_k} = \sum_x e^{ikx} \keT{q_x}$. Therefore, we know that the following circuit
\begin{align}
    \cdots\pic{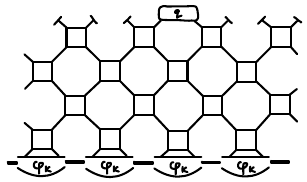}\cdots
\end{align}
will decay exponentially with rate $Dk^2$. This means that in principle it could be influence-solvable. If it was indeed possible to establish influence-solvable for some value of $k$ then we could read off the exact diffusion constant from its decay rate $D = -\tfrac{1}{k^2}\log\max_i{\abs{\lambda_i}}$ of the effective evolution. Given the great interest in developing numerical schemes to accurately compute the diffusion constant~\cite{PhysRevB.105.075131,10.21468/SciPostPhys.20.2.061,PhysRevX.9.041017,PhysRevB.110.104413,pinna2025approximationtheorygreensfunctions}, we believe that it will be an exciting line of research to understand whether and under what conditions it is possible to extract the diffusion constant using influence-solvability.

Beyond computing diffusion constants it would also be interesting to understand the general relation between the presence of influence-solvability and conserved quantities. While for $d=2$ dual unitary circuits it is known that they can only possess either no conservation laws or infinitely many conservation laws that evolve trivially~\cite{10.21468/SciPostPhys.8.4.068,HoldenDye2025fundamentalcharges} (for instance the SWAP gate), we do not see any relation between the presence of conservation laws and the presence of solvability. For instance, consider the model $(d,\sigma)=(3,996)$ of \Cref{ex:996}. This model has a single conserved quantity with a non-trivial hydroynamics, but still it is influence-solvable for some initial state. Those initial states even have different amount of conserved quantities allowing us to compute exact hydrodynamic partitioning protocols as we did in \cref{sec:zerobond_partitioning}. 

Nonetheless, we believe that there should be some general consequences restricting influence-solvability that can be drawn from hydrodynamic arguments in circuits with conservation laws, such as the argument in the beginning of the section. It would be interesting to explore these consequences further.

\subsection{Computation of correlation functions}\label{sec:prop_corr}
We have seen that the most general notion of influence-solvability allows us to compute the evolution of observables or autocorrelation functions. However, in many of the previously known examples of solvable circuits such as dual-unitary or $\mathrm{DU}(2)$ circuits, one is additionally also able to compute observables at other points, for instance correlation functions at different space-time points.

Earlier, we showed that for many of the $\chi=1$ influence-solvable gates discussed in the paper that are `bottom-left' solvable (and for `bottom-right' solvable), correlation functions near the left light cone \eqref{eq:light-cone} are tractable. However, away from this light cone, for $\chi=1$, i.e. Markovian baths, it is also possible to compute certain observables at different space points. These are observables that live in the space $\mathscr{L}$. This can be seen as follows: if $\brA{O} = \brA{O}\bm{L}$, then similar to \eqref{eq:FTMPS_eigvec} we find that
\begin{align}
    \pic{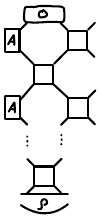} = \punct{\pic{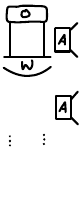}}{\;}{.}
\end{align}

Note that $\brAkeT{O}{w} = \brA{O}\bm{L}\keT{\rho} = \brAkeT{O}{\rho}$ is simply the expectation value of $O$ in the initial state.

From this it becomes clear that we can simplify a circuit of the form
\begin{align}
    \pic{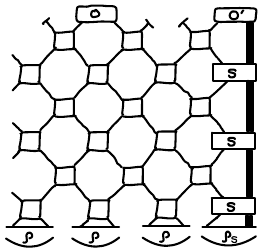} = \punct{\pic{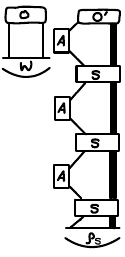}}{}{.}
\end{align}
Hence, we find that the expectation values of all observables $\brA{O}\in\mathscr{L}$ that are in the bulk of the circuit factorize from the expectation value in the system of interest.

Beyond this it would be interesting to systematically understand under which conditions influence-solvability allows for the computation of correlation functions and spatially dependent objects.

As a proposal, if one is interested in the correlation function along a specific space-time ray, one could define a ``tilted'' transfer matrix as
\begin{align}
    \cdots \punct{\pic{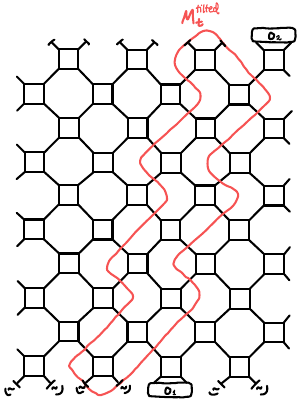}}{\cdots}{.}
\end{align}
Then one could make a similar MPS ansatz for the eigenvector of the tilted transfer matrix and study it using similar techniques to the one we used for the spatial transfer matrix.

\subsection{Computation of entanglement entropy in quantum circuits}\label{sec:prop_entanglement}
The evolution of the entanglement entropy after a quantum quench is an important problem in quantum many-body physics, since it relates to simulatibility~\cite{vidal2003efficient,PhysRevLett.129.140503}. We would like to briefly show that one can analytically compute this evolution in case the quantum gate is both left and right influence-solvable. Note that the following approach has already been applied to study entanglement dynamics of known types of solvable circuits~\cite{PhysRevX.9.021033,PhysRevB.101.094304,PhysRevLett.132.250402,10.21468/SciPostPhys.8.4.067,Foligno2025entanglementof,10.21468/SciPostPhys.11.6.107,PhysRevLett.133.170402,PhysRevResearch.7.L012011}. 

Specifically, we are interested in a situation where the initial state is given by a translation invariant pure MPS state, in which case the time-evolved pure state is given by:
\begin{align}
    \punct{\pic{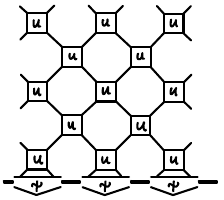}}{}{.}
\end{align}
We assume that the corresponding folded circuit is both left and right influence-solvable, i.e.
\begin{equation}
\begin{aligned}
    \cdots\hspace{-0.3cm}\pic{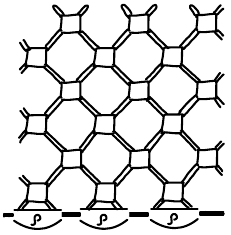} &= \punct{\pic{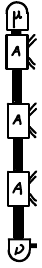}}{}{,} \\
    \pic{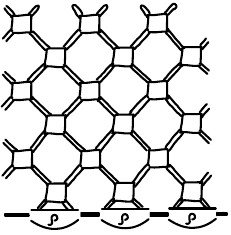}\hspace{-0.3cm}\cdots &= \punct{\pic{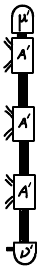}}{\;}{.}
\end{aligned}
\end{equation}

We want to study how the entanglement between the right $\Omega=(0,\infty)$ and the left half $\Omega\upd{c}=(-\infty,0)$ of the system evolves in time. A common measure is to compute the entanglement entropy $S_\Omega(t) = -\tr \hat\rho_\Omega(t) \log \hat\rho_\Omega(t)$ of the reduced density matrix $\hat\rho_\Omega$. Since this is not easy to compute directly, one instead uses the replica trick, formally obtaining $S_\Omega(t)$ as
\begin{align}
    \lim_{n\to 1} S^{(n)}_\Omega(t) &= S_\Omega(t)
\end{align}
from the Renyi entropies
\begin{align}
    S^{(n)}_\Omega(t) &= \frac{1}{1-n}\log \tr\rho_\Omega(t)^n.\label{eq:prop_ent_renyi}
\end{align}

For integer $n$ one can represent $S^{(n)}_\Omega(t)$ as a $2n$ times folded circuit as follows:
\begin{align}
    S^{(n)}_\Omega(t) = \punct{\pic{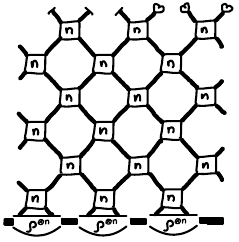}}{}{.}
\end{align}
Here, the two states on top
\begin{align}
    \pic{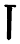}&=\pic{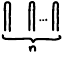} & \pic{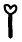}&=\pic{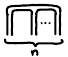}
\end{align}
represent different contractions in replica space. This is a ``partitioning protocol/domain wall'' that can naturally be studied using influence-solvability. In particular, since the initial state is pure, the left and right influence-matrix of the $2n$ times folded circuit are just products of the influence matrix of the $2$ times folded circuit. By assumption they are explicitly given by \eqref{eq:mot_infsolv_def}. Hence, we have
\begin{align}
    S^{(n)}_\Omega(t) = \punct{\pic{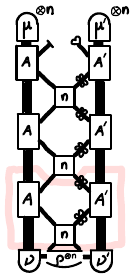}}{}{,}\label{equ:prop_entropy_solv}
\end{align}
where
\begin{align}
    \pic{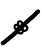} &= \pic{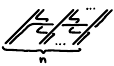}
\end{align}
is a permutation in replica space. For each $n$ this is an effective finite dimensional system with a finite dimensional time-evolution matrix indicated by the red shape. By diagonalizing it one can easily compute the $S^{(n)}_\Omega(t)$ exactly for all times $t$. This has been already studied for dual unitary circuits and $\mathrm{DU}(2)$ circuits~\cite{PhysRevB.101.094304,PhysRevLett.132.250402}. However, under this condition \eqref{eq:prop_ent_renyi} factorises in time and thus
\begin{align}
    S^{(n)}_\Omega(t) = \alpha t,
\end{align}
for some gate-dependent constant $\alpha$, which can be shown to be independent of $n$. In this case the limit $n\to 1$ can easily be taken and obtain $S^{(n)}_\Omega(t) = \alpha t$ as well. Note that the same result also holds for all the $\chi=1$ influence-solvable $d=2$ gates since they satisfy generalizations of the dual unitary and $\mathrm{DU}(2)$ conditions, see \cref{res:zerobond_quantum}.

However, if one can find a quantum gate which is influence-solvable but not (a generalization of) $\mathrm{DU}(2)$, then it is very likely that the dynamics of $S^{(n)}_\Omega(t)$ is more complicated, though still computable. In this case as $n$ increases the effective evolution matrix becomes larger and thus can show more complicated behaviour, implying that the limit $n\to 1$ is non-trivial. This has been carried out in the quantum Rule 54 circuit~\cite{10.21468/SciPostPhys.11.6.107}, where it was shown that the entanglement entropy follows the quasi-particle picture~\cite{Calabrese_2005} (which is an effective picture of entanglement dynamics for integrable systems).

Therefore, influence-solvable quantum circuits allow not only for the computation of the time-evolution of observables, but also the time-evolution of entanglement. It would be interesting to identify increasingly more complex quantum influence-solvable gates beyond the $\chi=1$ case we study in this work. For larger $\chi$ the tensor network \eqref{equ:prop_entropy_solv} becomes more complicated and thus should also the corresponding entanglement dynamics.

Similarly, one can also study the entanglement dynamics of partitioning protocols, i.e. when both halfs of the system are initialized in different states, or when two different systems are joined together.

\subsection{Construction of the MPS tensor $\bm{A}$ for $\chi > 1$}\label{sec:prop_gen_construction}
In the case of factorized influence matrix, i.e. bond dimension $\chi=1$ we have seen that the tensor $\bm{A}$ can be directly expressed in terms of the gate and the initial state.

In this section we will show that the tensor $\bm{A}$ of a given bond dimension $\chi$ can also be explicitly constructed using the circuit and its initial state. As it will become clear in the following, we do not expect the construction to be useful in practice. Nonetheless, it is of  conceptual interest. The construction is as follows:

As we have established in section \ref{sec:FTMPS_FTMPS}, to show influence-solvability for all times $t$, it is sufficient to establish that the MPS composed of $\bm{A}$ is an eigenvector of the transfer matrix for time $t = 4\chi\ind{B} = 4\chi d^2$. This transfer matrix is then of size $d^{t}\chi_\rho\times d^{t}\chi\rho$. We also know that it has only a single eigenvalue $1$ and otherwise Jordan blocks with eigenvalue $0$. Hence, if we take the transfer matrix to a very large power $N$, then $\bm{M}_t^{N}$ will project onto the influence matrix. This power is at most the dimension of the matrix $N\leq d^{4\chi d^2}\chi_\rho$, however $N=8\chi d^2 + \chi_\rho$ is sufficient due to the representation \eqref{eq:mot_infsolv_triangle}. Thus we conclude that we can write the transfer matrix as
\begin{align}
    \!\! \pic{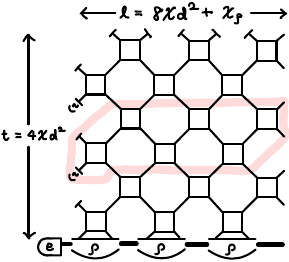} \! = \!\! \pic{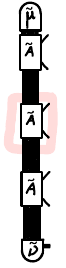} \!\!\!\! = \!\! \punct{\pic{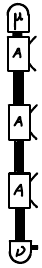}}{\;}{.}
\end{align}
Each row of this circuit can be viewed as an MPS $\tilde{\bm{A}}$ (red circle) with a very large (but finite) bond dimension. This MPS can be compressed into an equivalent MPS which smallest possible bond dimension $\bm{A}$. Note that this procedure only works if the compressed MPS $\bm{A}$ actually has bond dimension smaller or equal to the one assumed initially. Otherwise, it means that this circuit is not influence-solvable with bond dimension $\chi$.

In practice, given the gate and the initial state, one can simply compute the influence matrix explicitly for a small time $t$. Then ``cutting'' the influence matrix at some time-slice $0<t<t'$ one can perform a Schmidt decomposition of the influence matrix. The number of non-zero Schmidt values gives the minimally possible bond dimension across the ``cut''. As time $t$ increases this number will generically grow (very fast), in which case influence-solvability can be ruled out. However, if the number of non-zero Schmidt values remains finite across every cut, then the circuit is influence-solvable. 

\subsection{Relation between classical and quantum influence-solvability}
Since the notion of classical and quantum influence-solvability are very similar, one may ask whether there is any connection. In particular, note that since the classical gates are permutation gates they are also unitary. Hence, one can construct quantum circuits out of classical gates.

In fact, there exists classical influence-solvable gates that are also quantum influence-solvable when viewed as a unitary gate. A (trivial) example of this is the identity gate which is solvable for any $\keT{\gamma}$ with
\begin{align}
\pic{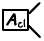} \, = \, \pic{gamma-lower-left-gate} \, = \, \pic{legTRBR}, \quad \quad \pic{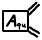} \, = \, \pic{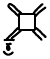} \, = \, \pic{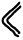}.
\end{align}
Another example is the east gate (or CNOT with control bit on the right, see \cref{fig:markov_search_classic_L2_d2}), which is also solvable for any $\keT{\gamma}$ in the classical case and any diagonal $\hat{\gamma}$ in the quantum case
\begin{align}
\pic{Acl} \, = \, \pic{legTRBR}, \quad \quad \pic{Aqu} \, = \,  \pic{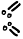} + \pic{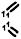}.
\end{align}
In fact they both satisfy the left condition of \eqref{eq:zerobond_DU2_gen}.

However, note that while $\bm{A}\ind{cl}$ is the identity matrix for both the classical identity and classical east gate, their quantum $\bm{A}\ind{qu}$ are different. This hints that there is no standard dictionary to convert classical influence-solvability into quantum influence-solvability.  

To prove this, we can give an explicit example of a gate that is classical influence-solvable, but not quantum influence-solvable:
\begin{example}
    The gate $(d,\sigma)=(3,70204)$ with rules
    \begin{center}
        \begin{tabular}{p{2cm}p{2cm}p{2cm}}$\keT{00}\to\keT{01}$&$\keT{01}\to\keT{20}$&$\keT{02}\to\keT{22}$\\$\keT{10}\to\keT{11}$&$\keT{11}\to\keT{00}$&$\keT{12}\to\keT{02}$\\$\keT{20}\to\keT{21}$&$\keT{21}\to\keT{10}$&$\keT{22}\to\keT{12}$
    \end{tabular}
    \end{center}
    is classical $\chi=1$ influence-solvable for $\keT{\gamma}=\keT{0}$, but not quantum $\chi=1$ influence-solvable for $\hat{\gamma}=\ketbra{0}{0}$.
\end{example}

To understand this, observe that this gate of the form
\begin{align}
    \pic{gate} = \sum_{a=0}^2 \pic{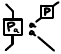},
\end{align}
where $P_a$ and $P$ are some permutations of $0,\ldots,d-1$. Note that it satisfies $\pic{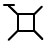} = \pic{flatbra}\pic{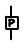}$, from which it easily follows that the classical gate satisfies the left condition of \eqref{eq:zerobond_DU2_gen} for any $\keT{\gamma}$. However when viewed as a unitary gate we instead have
\begin{align}
    \pic{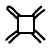} = \sum_{a,b=0}^3 \pic{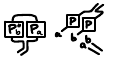}.
    %\pic{gate} = \sum_{a=0}^{2}  \\ \includegraphics[width=0.5\linewidth]{figs/fred/Screenshot 2026-03-26 234729}.
\end{align}
If $a=b$, then we have indeed $$\pic{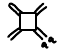} = \sum_{a,b=0}^3 \pic{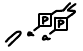},$$
but if $a\neq b$ then we obtain something else, which prevents the gate from satisfying the quantum version of \eqref{eq:zerobond_DU2_gen}. From this it is easy to understand where the problem is: if we could restrict to diagonal states, i.e. states that are equal in the forward and backward replicas, then classical influence-solvability would indeed imply quantum influence-solvability. However, since the states in both replicas can be different, classical influence-solvability does not imply quantum influence-solvability.

There is an alternative way of converting a classical gate into a quantum one: before and and after the application of each gate, project onto the diagonal states in both replicas. This will not give a circuit of unitary gates, but instead of CPTP gates. This is equivalent to treating them as quantum channels that are completely dephasing in the computational basis states. In this case classical influence-solvability directly implies quantum influence-solvability of the CPTP circuit.

\FloatBarrier

\section{Discussion and outlook}
In this work, we introduced a very general notion of Solvable circuits, which we call \emph{influence-solvability}. This notion includes virtually all known examples of Solvable circuits, such as dual-unitaries, $\mathrm{DU}(2)$ and $\mathrm{DU}(2)$-like gates such as Floquet East, and more complicated cases such as Rule 54, etc. In fact, we believe it to be the most general notion of Solvability for $(1+1)D$ circuits where one can exactly compute time-evolved quantities. Although we have focused on observables in brickwork circuits, we believe that our approach can be applied for general quantities, such as the Renyi entropies as we have shown \cref{sec:prop_entanglement}, OTOCs~\cite{PhysRevX.15.011015}, loschmidt echos, and even calculating partition functions of general $2D$ stat-mech models.

Based on this definition we found the most general local condition for influence-solvability. The local conditions are separated into ``bulk'' and ``boundary'' conditions. This enables us to first find gates that could be Solvable and then find all compatible initial states. 

We demonstrated our approach by studying the simplest case of bond dimension $\chi=1$, corresponding to an exact Markovian bath. We showed that if the gate is invertible in space, the only option for influence-solvability is to satisfy a condition reminiscent of the dual-unitarity condition. We also performed a in-depth investigation into classical circuits with local dimension $d=2$ (all of which are $\chi=1$ influence-solvable) and $d=3$ ($\approx 32\%$ are influence-solvable) and well as $d=2$ quantum circuits, where only dual unitaries and controlled gates are $\chi=1$ influence-solvable. 

We believe that the intuitive mechanism of influence-solvability is the following: influence-solvability allows only for finite amount of information (determined by the bond dimension $\chi$) that entered the bath from the system to ever influence the system again. In the $\chi=1$ case we have identified two ways. The first way is if the information that enters the bath from the system is transported away deep inside the bath and replaced by uncorrelated information which is transported from the initial state far inside the bath towards the system. This is the mechanism in which dual unitary circuits are influence-solvable, for the case of infinite-temperature initial states. The SWAP gate is an example where this mechanism holds for any initial state. The second way is if information cannot enter the bath. If the gate is not invertible in space then it prevents certain information from ever entering the bath. For example, if there is a control gate from the right to the left and the system is on the left, the system cannot affect the bath. This means that significantly less memory is required to describe the bath.

The obvious next step would be to understand under what conditions do there exists solutions to the local conditions \eqref{eq:FTMPS_FTMPS_local_conditions} (and in which cases they do not), in particular for $\chi > 1$. We know that it is possible, since there already exist examples for $\chi > 1$ influence-solvability as discussed in \cref{sec:FTMPS_higher_bond_dim}. Furthermore, shortly after the first preprint of this work was published on arXiv, an explicit solution to \eqref{eq:FTMPS_FTMPS_local_conditions} for $\chi>1$ in the Rule 201 cellular automata was made public~\cite{yang2026solvingnonequilibriumdynamicsinfluence}. Given the sparsity of exact microscopic results in both classical and quantum many-body systems, this is an exciting and timely direction. Though we have suggested a finite-dimensional method to determine if a circuit is influence-solvable, we expect it to be not practical, so a more efficient method would be required. It would also be interesting to gain a better intuitive pictures of different cases in which finite memory baths can occur.

In the case $\chi=1$, we have observed in \Cref{sec:zerobond} in both the quantum and the classical setups that only very restricted values of $\keT{\gamma}$ lead to influence-solvability. For now this is an observation and we do not have an explanation for this. We think that understanding the reason for this restriction and extending it to higher local dimension $d$ this might lead a systematic way to classifying all influence-solvable circuits, at least in the case $\chi=1$.

Another direction we find promising is to study gates that are invertible in space. This is a condition that is easy to check and we have already seen for the $\chi=1$ case that it dramatically restricts the possible options for influence-solvability. The advantage of gates that are invertible in space is that we can explicitly compute $\mathscr{L}$ since it is equal to
\begin{align}
    \mathscr{L} &= \mathrm{span}\qty{\pic{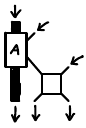}}
\end{align}
for any bond dimension. This can be derived in a similar fashion as for $\chi=1$, see \cref{sec:zerobond_spatial_inv_gate}. With that \eqref{eq:circuit_MPS_bulk} is given by \begin{align}
    \pic{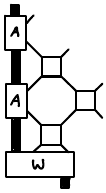} &= \punct{\pic{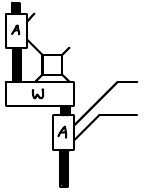}}{\;}{.}
\end{align}

The third direction is to study gates that have low rank as matrix in space. The simplest case of a rank $1$ matrix in space is always trivially influence-solvable, e.g. the identity gate. Similarly, gates with spatial-rank $2$ or $3$ should have a high chance of being influence solvable.

In analogy to integrability-breaking, it would be interesting to see whether there is a notion of ``almost influence-solvable'' circuits, i.e. circuits that can be well approximated using influence-solvable ones~\cite{PhysRevX.11.011022,riddell2024structural,PhysRevLett.130.130402,yadalam2026processtensorapproachcountingstatistics}. In this context it would also be important to understand whether it is always possible to represent the influence-matrix by a MPS of infinite bond dimension: of course, it is always possible to represent the influence matrix by a MPS with increasing bond dimension in time, which formally gives rise to an infinite bond dimension MPS~\footnote{Such infinite bond dimension MPS have been used to find steady-states of dynamics~\cite{PhysRevLett.107.137201}}. However, this is not useful unless the tensor $\bm{A}$ has some additional structure, such as describing bounded or compact operators on some Banach or Hilbert space. Such structures might then allow to find good finite-bond dimension approximations.

The problem of equivalence of uniform MPSs also appear in many other contexts such as finding exact steady-states of Lindblad dynamics~\cite{10.1007/978-3-319-66839-0_11,zhang2026exactnonequilibriumsteadystates}, steady states of stochastic systems such as the celebrated steady state of SSEP~\cite{PhysRevLett.87.150601}, ground states of Hamiltonians such as the AKLT model~\cite{affleck1988valence,Wei2022} or scar states~\cite{gehrmann2026exactquantummanybodyscars}. It would be interesting to see if our local conditions for influence-solvability can be used in the spatial direction to give new solutions to steady states.

Another natural generalisation would be to see whether the notion of influence-solvability, as it is possible for dual-unitarity, can be extended to higher dimensions~\cite{Suzuki2022computationalpower,10.21468/SciPostPhys.18.6.182} and other space-time geometries~\cite{PhysRevLett.119.141602,bistro2026graphrestrictedtensorsbuilding,fbgh-rq3l,pickering2026asymptoticallysolvablequantumcircuits,Kasim_2023}.

We would like to point out that one can also view influence-solvability from the view-point of Ruelle-Pollicott (RP) resonances~\cite{PhysRevA.46.7401,82f6-qdyd,jacoby2025spectral}: note that for dynamics like \eqref{eq:mot_classic_circuit_system} for even $t$ the time evolution operator is $\bm{U}_t = \bm{U}_2^{t/2}$ where $\bm{U}_2$ represents two rows of time-evolution. $\bm{U}_2$ is a unitary operator (on the space of quasi-local observables) and hence can be diagonalized with spectrum on the unit circle. So, naively it seems impossible to obtain an exact exponential decay $\lambda^{t/2}$, which would correspond to an eigenvalue $\lambda$ inside the unit circle $\abs{\lambda}<1$. However, if the spectrum of the unitary operator is continuous, its resolvent $\tfrac{1}{z-\bm{U}_2}$ has a branch cut on the unit-circle. The exponential decay $\lambda^{t/2}$ then corresponds to a pole $\lambda$ in the unphysical sheet obtained by analytically extending the resolvent beyond the branch cut. This is called a RP resonance. Note that with the exception of very simple dynamical systems such a~\cite{PhysRevA.46.7401} it is not possible to carry out this procedure in practice and determine the location of the RP resonance. However, in practice, there is a trick by adding a small amount of dissipation to make the system non-unitary, i.e. eigenvalues can be inside the unit-circle. In the small dissipation limit it is believed these eigenvalues do not re-approach the unit-circle, but instead converge to the RP resonances. In principle, one should be able to use any kind of dissipation (for instance Lindblad dynamics~\cite{PhysRevB.109.064311,zhang2025thermalizationratesquantumruellepollicott} or suppressing long operators~\cite{PhysRevB.105.075131} or truncating the system~\cite{TomazProsen_2002,PhysRevE.110.054204}) as long as it is somewhat physical. Note that this is also an important efficient numerical algorithm for the hard problem of long time dynamics. For instance, it can be used to efficiently compute diffusion constants~\cite{PhysRevB.105.075131,10.21468/SciPostPhys.20.2.061}. Let us now connect with influence-solvable circuits: first note that it gives exact effective subsystem dynamics generated by \eqref{eq:mot_infsolv_Geff}. Hence, any eigenvalue of \eqref{eq:mot_infsolv_Geff} inside the unit-circle is the exact location of a RP resonance. Second, we can interpret the tensor $\bm{A}$ of the influence-matrix as the optimal dissipator: truncating the system to a finite size $\ell$ using $\bm{A}$ gives exact RP resonances. Exploring this connection could lead to a much better understanding of how to choose a good dissipator, which in turn could significantly improve numerical algorithms studying long time dynamics.

\section*{Acknowledgements}
FH has received support under the Major Research Program of PSL Research University "Statistical Physics and Mathematics" launched by PSL Research University and implemented by ANR with the references ANR-10-IDEX-0001. FH acknowledges funding from the faculty of Natural, Mathematical \& Engineering Sciences at King’s College London. SWPK is supported by UKRI Engineering and Physical Sciences Research Council (EPSRC) DTP International Studentship Grant Ref. No. EP/W524475/1. Numerical computations were done in Julia~\cite{Julia-2017} and some ran on the CREATE cluster~\cite{CREATE}.

FH would like to thank the Statistical Physics Group at the University of Ljubljana for their hospitality, the many interesting discussions and specifically the useful comments of Pavel Kos and Urban Duh about this work. FH would like to thank Luca Capizzi for useful discussions and in particular for the idea presented in \cref{sec:prop_CQ}. SWPK would like to thank Katja Klobas for useful discussions, especially regarding the argument ruling out finite $\chi$ influence-solvability for diffusive systems.

\section*{Author contribution statement}
The project was initiated by FH by proving the fundamental theorem of MPS for open boundary conditions using the abstract method of equivalence classes (Appendix \ref{sec:abstract-proof}). The constructive proof (Appendix \ref{sec:explicit-proof}) was provided later by SWPK. The analysis done in Sections 4 and 5 were carried out together by FH and SWPK. Both authors contributed equally to the writing of the manuscript. The majority of the $\approx 450$ diagrams were hand-drawn by SWPK. No TikZ or AI was used in this work.

\bibliographystyle{quantum}
\bibliography{bibliography}

\appendix

\section{Proof of fundamental theorem of MPS with open boundary conditions} \label{sec:fundamental-theorem}

In this section we will proof the fundamental theorem of MPS with open boundary conditions. The theorem provides sufficient and necessary conditions for two MPS of the form $\Gamma^{\alpha_0\alpha_1 \cdots \alpha_L\alpha_{L+1}}:=\brA{\phi\ind{A}^{\alpha_0}}\bm{A}^{\alpha_1}\ldots \bm{A}^{\alpha_L}\keT{\psi\ind{A}^{\alpha_{L+1}}}$ to represent the same quantity, i.e. whenever
	\begin{align}
		\brA{\phi\ind{A}^{\alpha_0}}\bm{A}^{\alpha_1}\ldots \bm{A}^{\alpha_L}\keT{\psi\ind{A}^{\alpha_{L+1}}} &=   \brA{\phi\ind{B}^{\alpha_0}}\bm{B}^{\alpha_1}\ldots \bm{B}^{\alpha_L}\keT{\psi\ind{B}^{\alpha_{L+1}}}.
	\end{align}
	If this is true for all $L\geq 0$ we say that the tuples $(\bm{A}^\alpha,\keT{\psi\ind{A}^\alpha},\brA{\phi\ind{A}^\alpha})$ and $(\bm{B}^\alpha,\keT{\psi\ind{B}^\alpha},\brA{\phi\ind{B}^\alpha})$ are equivalent. We assume that the bond dimensions are finite (but not necessarily equal). We will denote the spaces on which the $\bm{A}^\alpha$ and $\bm{B}^\alpha$ act as $\mathscr{V}\ind{A}$ and $\mathscr{V}\ind{B}$ respectively. 

    The statement is as follows.
    \begin{theorem}[Fundamental theorem of MPS for open boundary conditions]\label{thm:app_MPS}
    Two MPS
        \begin{align}
		\brA{\phi\ind{A}^{\alpha_0}}\bm{A}^{\alpha_1}\ldots \bm{A}^{\alpha_L}\keT{\psi\ind{A}^{\alpha_{L+1}}} &=   \brA{\phi\ind{B}^{\alpha_0}}\bm{B}^{\alpha_1}\ldots \bm{B}^{\alpha_L}\keT{\psi\ind{B}^{\alpha_{L+1}}}
	\end{align}
    are equal for all $L$ if and only if there exists a map $\bm{W}:\mathscr{V}\ind{A}\to\mathscr{V}\ind{B}$ and two projectors $\bm{L}$ and $\bm{R}$ satisfying
    \begin{align}
        \bm{L}\bm{B}^\sigma\bm{W} &= \bm{W}\bm{A}^\sigma\bm{R}\label{eq:app_proof_MPS_bulk}\\
        \bm{A}^\sigma\bm{R} &= \bm{R}\bm{A}^\sigma\bm{R}\label{eq:app_proof_MPS_bulk_PR}\\
        \bm{L}\bm{B}^\sigma &= \bm{L}\bm{B}^\sigma\bm{L}\label{eq:app_proof_MPS_bulk_PL}\\
        \bm{R}\keT{\psi\ind{A}^\alpha} &= \keT{\psi\ind{A}^\alpha}\label{eq:app_proof_MPS_rightP}\\ \brA{\phi\ind{B}^\alpha}\bm{L} &= \brA{\phi\ind{B}^\alpha}\label{eq:app_proof_MPS_leftP}\\
        \bm{W}\keT{\psi^\alpha\ind{A}} &= \bm{L}\keT{\psi^\alpha\ind{B}}\label{eq:app_proof_MPS_rightW}\\
        \brA{\phi\ind{B}^\alpha}\bm{W} &= \brA{\phi\ind{A}^\alpha}\bm{R}\label{eq:app_proof_MPS_leftW}.
    \end{align}
    \end{theorem}

    Note that the backward direction can be done easily: assuming \eqref{eq:app_proof_MPS_bulk}-\eqref{eq:app_proof_MPS_leftW} we have
    \begin{align}
        &\underbrace{\brA{\phi\ind{B}^{\alpha_0}}}_{\eqref{eq:app_proof_MPS_leftP}}\bm{B}^{\alpha_1}\ldots \bm{B}^{\alpha_L}\keT{\psi\ind{B}^{\alpha_{L+1}}} = \brA{\phi\ind{B}^{\alpha_0}}\underbrace{\bm{L}\bm{B}^{\alpha_1}}_{\eqref{eq:app_proof_MPS_bulk_PL}}\ldots \bm{B}^{\alpha_L}\keT{\psi\ind{B}^{\alpha_{L+1}}}\\
        &= \brA{\phi\ind{B}^{\alpha_0}}\bm{L}\bm{B}^{\alpha_1}\bm{L}\ldots \bm{L}\bm{B}^{\alpha_L}\underbrace{\bm{L}\keT{\psi\ind{B}^{\alpha_{L+1}}}}_{\eqref{eq:app_proof_MPS_rightW}}\\
        &=\brA{\phi\ind{B}^{\alpha_0}}\bm{L}\bm{B}^{\alpha_1}\bm{L}\ldots \underbrace{\bm{L}\bm{B}^{\alpha_L}\bm{W}}_{\eqref{eq:app_proof_MPS_bulk}}\keT{\psi\ind{A}^{\alpha_{L+1}}}\\
        &=\underbrace{\brA{\phi\ind{B}^{\alpha_0}}\bm{W}}_{\eqref{eq:app_proof_MPS_leftW}}\bm{A}^{\alpha_1}\bm{R}\ldots \bm{R}\bm{A}^{\alpha_L}\bm{R}\keT{\psi\ind{A}^{\alpha_{L+1}}}\\
        &=\brA{\phi\ind{A}^{\alpha_0}}\underbrace{\bm{R}\bm{A}^{\alpha_1}\bm{R}}_{\eqref{eq:app_proof_MPS_bulk_PR}}\ldots \bm{R}\bm{A}^{\alpha_L}\bm{R}\keT{\psi\ind{A}^{\alpha_{L+1}}}\\
        &=\brA{\phi\ind{A}^{\alpha_0}}\bm{A}^{\alpha_1}\ldots \bm{A}^{\alpha_L}\underbrace{\bm{R}\keT{\psi\ind{A}^{\alpha_{L+1}}}}_{\eqref{eq:app_proof_MPS_rightP}}\\
        &=\brA{\phi\ind{A}^{\alpha_0}}\bm{A}^{\alpha_1}\ldots \bm{A}^{\alpha_L}\keT{\psi\ind{A}^{\alpha_{L+1}}}.
    \end{align}

    We will provide two proofs for the forward direction. \cref{sec:abstract-proof} make use of abstract mathematical constructions, while the latter in \ref{sec:explicit-proof} will more explicitly constructed the required objects.
	
	\subsection{Abstract proof} \label{sec:abstract-proof}
    \paragraph*{Compression of a single MPS.}
	First, we will only consider one of the two MPS and ``compress'' it to its lowest possible bond dimension, i.e. we find an equivalent MPS with smallest possible bond dimension. This can be done as follows.
	Construct for any multi-indices $\alpha=(\alpha_1,\alpha_2,\ldots,\alpha_n)$ the vectors:
	\begin{align}
		\keT{r^\alpha\ind{A}} &= \bm{A}^{\alpha_1}\cdots\bm{A}^{\alpha_{n-1}}\keT{\psi\ind{A}^{\alpha_n}} & \brA{l^\beta\ind{A}} &= \brA{\phi\ind{A}^{\beta_1}}\bm{A}^{\beta_2}\cdots\bm{A}^{\beta_n}.
	\end{align}
    With this let us define the spaces
    \begin{align}
        \mathscr{R}\ind{A} &= \mathrm{span}\qty{\keT{r\ind{A}^{\alpha}}}&\mathscr{L}\ind{A} &= \mathrm{span}\qty{\keT*{r\ind{A}^{\beta}}}.
    \end{align}

    Here the multi-indices $\alpha$ and $\beta$ can have any length. Note that, crucially, we can compute the overlap of any $\keT{r\ind{A}}\in\mathscr{R}\ind{A}$ and any $\brA{l\ind{A}}\in\mathscr{L}\ind{A}$ simply based on the MPS because
    \begin{align}
        \brAkeT*{l^\beta\ind{A}}{\psi^\alpha\ind{A}} = \Gamma^{\beta\alpha}.
    \end{align}

    In particular, they have to be the same for all equivalent MPS representations. However, there might be some unnecessary vectors in these spaces, i.e. vectors $\keT{r\ind{A}} \in \mathscr{R}\ind{A}$ that satisfy $\brAkeT{l\ind{A}}{r\ind{A}} = 0$ for all $\brA{l\ind{A}} \in \mathscr{L}\ind{A}$. To remove them we define the equivalence relations:
    \begin{align}
        \keT{r\ind{A}} &\sim \keT{r\ind{A}'} &\Leftrightarrow&& \brAkeT{l\ind{A}}{r\ind{A}}=\brAkeT{l\ind{A}}{r\ind{A}'} \text{for all} \brA{l\ind{A}}\in\mathscr{L}\ind{A}\\
        \brA{l\ind{A}} &\sim \brA{l\ind{A}'} &\Leftrightarrow&& \brAkeT{l\ind{A}}{r\ind{A}}=\brAkeT{l\ind{A}'}{r\ind{A}} \text{for all} \keT{r\ind{A}}\in\mathscr{R}\ind{A}
    \end{align}
    and define the spaces of equivalence classes
    \begin{align}
        \tilde{\mathscr{R}}\ind{A} &= \mathscr{R}\ind{A}/\sim & \tilde{\mathscr{L}}\ind{A} &= \mathscr{L}\ind{A}/\sim.
    \end{align}

    We will denote the equivalence class of $\keT{r\ind{A}}$ as $\qty[\keT{r\ind{A}}]$ and the one of $\brA{l\ind{A}}$ as $\qty[\brA{l\ind{A}}]$. Note that $\qty[\brA{l\ind{A}}]\qty[\keT{r\ind{A}}] = \brAkeT{l\ind{A}}{r\ind{A}}$.

    By construction we have $\mathrm{\dim} \tilde{\mathscr{R}}\ind{A} = \mathrm{\dim} \tilde{\mathscr{L}}\ind{A}$ and in particular $\tilde{\mathscr{L}}\ind{A} = \tilde{\mathscr{R}}\ind{A}^*$ are dual spaces of each other.

    Because of this the matrices $\bm{A}^\sigma$ give rise to well-defined maps $\tilde{\bm{A}}^\sigma:\tilde{\mathscr{R}}\to \tilde{\mathscr{R}}$ defined as $\tilde{\bm{A}}^\sigma\qty[\keT{r}\ind{A}] = \qty[\bm{A}^\sigma\keT{r\ind{A}}]$. They are well-defined since if $\keT{r\ind{A}}$ satisfies $\brAkeT{l\ind{A}}{r\ind{A}}=0$ for all $\brA{l\ind{A}}\in\mathscr{L}\ind{A}$ we have:
    \begin{align}
        \qty[\brA{l\ind{A}}]\qty[\bm{A}^\sigma\keT{r\ind{A}}] &=  \underbrace{\brA{l\ind{A}}\bm{A}^\sigma}_{\in\mathscr{L}\ind{A}}\keT{r\ind{A}} = 0.
    \end{align}
    And thus $\bm{A}^\sigma$ maps equivalence classes to equivalence classes.

    Now observe that
    \begin{align}
        \qty[\brA{\phi\ind{A}^{\alpha_0}}] \tilde{\bm{A}}^{\alpha_1} \cdots \tilde{\bm{A}}^{\alpha_L} \qty[\keT{\psi\ind{A}^{\alpha_{L+1}}}] = \brA{\phi\ind{A}^{\alpha_0}}\bm{A}^{\alpha_1} \cdots \bm{A}^{\alpha_L} \keT{\psi\ind{A}^{\alpha_{L+1}}}
    \end{align}
    meaning that the MPS constructed from $(\tilde{\bm{A}}^\alpha,\qty[\keT{\psi\ind{A}^\alpha}],\qty[\brA{\phi\ind{A}^\alpha}])$ is equivalent to $(\bm{A}^\alpha,\keT{\psi\ind{A}^\alpha},\brA{\phi\ind{A}^\alpha})$. Hence, we constructed new representation of the MPS but with (potentially) lower bond dimension. It is already intuitively clear that $(\tilde{\bm{A}}^\alpha,\qty[\keT{\psi\ind{A}^\alpha}],\qty[\brA{\phi\ind{A}^\alpha}])$ is (one of) the MPS with the smallest possible bond dimension, a fact which we will establish in the next section.
	
	\paragraph*{Equivalence of the compressed MPS.}
	Now let us consider two representations $(\bm{A}^\alpha,\keT{\psi\ind{A}^\alpha},\brA{\phi\ind{A}^\alpha})$ and $(\bm{B}^\alpha,\keT{\psi\ind{B}^\alpha},\brA{\phi\ind{B}^\alpha})$ of the same MPS $\Gamma$. For both of them we can construct $(\tilde{\bm{A}}^\alpha,\qty[\keT{\psi\ind{A}^\alpha}],\qty[\brA{\phi\ind{A}^\alpha}])$ and $(\tilde{\bm{B}}^\alpha,\qty[\keT{\psi\ind{B}^\alpha}],\qty[\brA{\phi\ind{B}^\alpha}])$ like in the last section.
    
    Let us define a map $\vu{U}: \tilde{\mathscr{R}}\ind{A} \to \tilde{\mathscr{R}}\ind{B}$ by:
	\begin{align}
		\vu{U} \qty[\keT{r\ind{A}}] &= \sum_\alpha c_\alpha \vu{U}\qty[\keT{r\ind{A}^\alpha}] = \sum_\alpha c_\alpha \qty[\keT{r\ind{B}^\alpha}]
	\end{align}
	This is well-defined. Indeed let $\keT{r\ind{A}} = \sum_\alpha c_\alpha \keT{r\ind{A}^\alpha}$ s.t. $\qty[\keT{r}\ind{A}] = 0$. Then we have for any $\brA{l\ind{B}} = \sum_\beta d_\beta \brA*{l\ind{B}^\beta}$:
	\begin{align}
		\qty[\brA{l\ind{B}}]\vu{U} \qty[\keT{r\ind{A}}] &= \sum_{\alpha,\beta} c_\alpha d_\beta \qty[\brA*{l\ind{B}^\beta}]\qty[\keT{r\ind{B}^\alpha}] =  \sum_{\alpha,\beta} c_\alpha d_\beta \brAkeT*{l\ind{B}^\beta}{r\ind{B}^\alpha}\\
        &= \sum_{\alpha,\beta} c_\alpha d_\beta \brAkeT*{l\ind{A}^\beta}{r\ind{A}^\alpha} = \brAkeT{l\ind{A}}{r\ind{A}} = 0.
	\end{align} 
	Since $\qty[\brA{l\ind{B}}]$ spans $\tilde{\mathscr{L}}\ind{B} =\tilde{\mathscr{R}}^*$, we conclude that $\vu{U} \qty[\keT{r\ind{A}}] = 0$ and thus the map $\vu{U}$ is well-defined. 

    Furthermore, it follows from this that $\vu{U} \qty[\keT{r\ind{A}}] = 0$ only if $\qty[\keT{r\ind{A}}] = 0$, implying that $\mathrm{Ker}\vu{U} = \qty{0}$ vanishes. Similarly, we also find that $\mathrm{Ker}\vu{U}^* = \qty{0}$,
    where $\vu{U}^*:\tilde{\mathscr{L}}\ind{B}\to\tilde{\mathscr{L}}\ind{A}$ is the dual map
    \begin{align}
        \vu{U}^*\qty[\brA{l\ind{B}}] &= \sum_\beta d_\beta  \vu{U}^*\qty[\brA*{l\ind{B}^\beta}] = \sum_\beta d_\beta  \qty[\brA*{l\ind{A}^\beta}].
    \end{align}
    Therefore $\vu{U}$ is an invertible map. This in particular implies that $\mathrm{dim} \tilde{\mathscr{L}}\ind{A} =\mathrm{dim} \tilde{\mathscr{L}}\ind{B}$. As a consequence the MPS generated from $\tilde{\mathscr{L}}\ind{A}$ has the smallest possible bond dimension\footnote{Assume that there would be another representation $(\bm{B}^\alpha,\keT{\psi\ind{B}^\alpha},\brA{\phi\ind{B}^\alpha})$ of the same MPS with lower bond dimension than $(\tilde{\bm{A}}^\alpha,\qty[\keT{\psi\ind{A}^\alpha}],\qty[\brA{\phi\ind{A}^\alpha}])$. Then $\mathrm{dim}\tilde{\mathscr{L}}\ind{A}=\mathrm{dim}\tilde{\mathscr{L}}\ind{B}\leq \mathrm{dim}\mathscr{V}\ind{B} < \mathrm{dim}\tilde{\mathscr{L}}\ind{A}$, which is not possible.}  

    Now observe the important relations
    \begin{align}
        \vu{U}\qty[\keT{\psi\ind{A}^\alpha}] &= \qty[\keT{\psi\ind{B}^\alpha}]\label{eq:app_FTMPS_abstract_Ucond1}\\
        \qty[\brA{\phi\ind{B}^\alpha}]\vu{U} &= \qty[\brA{\phi\ind{A}^\alpha}]\label{eq:app_FTMPS_abstract_Ucond2}\\
        \tilde{\bm{B}}^\sigma\vu{U} & = \vu{U}\tilde{\bm{A}}^\sigma.\label{eq:app_FTMPS_abstract_Ucond3}
    \end{align}

    The last identity follows from
    \begin{align}
        \vu{U}^{-1}\tilde{\bm{B}}^\sigma\vu{U}\qty[\keT{r\ind{A}}] & = \sum_\alpha c_\alpha \vu{U}^{-1}\tilde{\bm{B}}^\sigma\vu{U}\qty[\keT{r\ind{A}^\alpha}] \\
        & = \sum_\alpha c_\alpha \vu{U}^{-1}\tilde{\bm{B}}^\sigma\qty[\keT{r\ind{B}^\alpha}] \\ 
        & = \sum_\alpha c_\alpha \vu{U}^{-1}\qty[\keT{r\ind{B}^{\sigma\alpha}}] \\
        & = \sum_\alpha c_\alpha \qty[\keT{r\ind{A}^{\sigma\alpha}}] = \tilde{\bm{A}}^\sigma \qty[\keT{r\ind{A}}].
    \end{align}
    
    \paragraph*{Equivalence of the original MPS.}
    As a last step we need to rewrite \eqref{eq:app_FTMPS_abstract_Ucond1}-\eqref{eq:app_FTMPS_abstract_Ucond3} to be applicable to the original MPS $(\bm{A}^\alpha,\keT{\psi\ind{A}^\alpha},\brA{\phi\ind{A}^\alpha})$ and $(\bm{B}^\alpha,\keT{\psi\ind{B}^\alpha},\brA{\phi\ind{B}^\alpha})$ defined on the spaces $\mathscr{V}\ind{A}$ and $\mathscr{V}\ind{B}$. 

    First, pick a basis $\qty[\keT{r\ind{A}^i}]$ of $\tilde{\mathscr{R}}\ind{A}$. Second, make a specific choice of a representative for each of these vector in $\mathcal{C}\ind{A}$. Third, add further linearly independent vectors from $\mathscr{V}\ind{A}$ to obtain a basis on $\mathscr{V}\ind{A}$. Then we can easily lift $\vu{U}:\tilde{\mathscr{R}}\ind{A}\to\tilde{\mathscr{R}}\ind{B}$ to a map $\bm{V}:\mathscr{V}\ind{A}\to\mathscr{V}\ind{B}$ defined by mapping the representatives of $\qty[\keT{r\ind{A}^i}]$ onto a corresponding representative of $\qty[\keT{r\ind{B}^i}]$ and sending all other linear independent directions to $0$.

    Then, pick a projector $\bm{R}$ onto $\mathscr{R}\ind{A}$ and a projector $\bm{L}$ on $\mathscr{L}\ind{B}$ from the left, i.e. \begin{align}
        \brA{l\ind{B}}\bm{L} = \brA{l\ind{B}}
    \end{align}
    for all $\brA{l\ind{B}}\in\mathscr{L}\ind{B}$. Then, we can write \eqref{eq:app_FTMPS_abstract_Ucond1}-\eqref{eq:app_FTMPS_abstract_Ucond3} as
    \begin{align}
        \bm{W}\keT{\psi^\alpha\ind{A}} &= \bm{L}\keT{\psi^\alpha\ind{B}}\\
        \brA{\phi\ind{B}^\alpha}\bm{W} &= \brA{\phi\ind{A}^\alpha}\bm{R}\\
        \bm{L}\bm{B}^\sigma\bm{W} &= \bm{W}\bm{A}^\sigma\bm{R}, 
    \end{align}
    where $\bm{W} = \bm{L}\bm{V}\bm{R}$ is also an implementation of $\vu{U}$ on the whole space.

    The appearance of the projectors might be surprising at first. However, it is necessary to include them because the derivations so far only required equality of \eqref{eq:app_FTMPS_abstract_Ucond1}-\eqref{eq:app_FTMPS_abstract_Ucond3} if an arbitrary $\keT{r\ind{A}}\in\mathscr{R}\ind{A}$ and $\brA{l\ind{B}}\in\mathscr{L}\ind{B}$ are applied to them.
    
    Note that in addition to these conditions we also have by construction
    \begin{align}
        \bm{A}^\sigma\bm{R} &= \bm{R}\bm{A}^\sigma\bm{R} &
        \bm{L}\bm{B}^\sigma &= \bm{L}\bm{B}^\sigma\bm{L}\\
        \bm{R}\keT{\psi\ind{A}^\alpha} &= \keT{\psi\ind{A}^\alpha} & \brA{\phi\ind{A}^\alpha}\bm{L} &= \brA{\phi\ind{A}^\alpha}.
    \end{align}
    This completes the proof of \cref{thm:app_MPS}

    \subsection{Constructive proof} \label{sec:explicit-proof}
In this section we give another proof of \cref{thm:app_MPS} which is more constructive and in particular does not require us to define equivalence classes.

Consider an MPS state $\Gamma^{\alpha} := \meL{\phi^{\alpha_0}\ind{A}}{\bm{A}^{\alpha_1} \cdots \bm{A}^{\alpha_L}}{\psi\ind{A}^{\alpha_{L+1}}}$, where $\alpha := (\alpha_0, \alpha_1, \dots, \alpha_{L+1})$ of arbitrary length $L \in \mathbb{N}$. We will also define $\keT{r^{\alpha}\ind{A}} := \bm{A}^{\alpha_1} \cdots \bm{A}^{\alpha_L} \keT{\psi^{\alpha_{L+1}}\ind{A}}$ and similarly for $\brA{l^\alpha\ind{A}}$. 

By definition we have that two MPS representations A and B are equivalent if
    \begin{align}
        \meL*{\phi_A}{\bm{A}^{\alpha}}{\psi_A} = \meL*{\phi_B}{\bm{B}^{\alpha}}{\psi_B}.
    \end{align}

For the MPS A let us define the right span as $\mathscr{R}\ind{A} := \mathrm{span}\left\{\keT{r\ind{A}^\alpha} \, \forall \, \alpha \right\}$, and $\mathscr{L}^\dag\ind{A} := \mathrm{span}\left\{\brA{l\ind{A}^\alpha} \, \forall \, \alpha \right\}$.

Note that the $\mathscr{L}\ind{A}$ here is defined as the hermitian conjugate of the $\mathcal{L}\ind{A}$ in the text and in the previous proof. 

Now let us consider the structure of $\bm{A}^\alpha$. For any $\keT{r\ind{A}} \in \mathscr{R}\ind{A}$, we also have ${\bm{A}}^\alpha \keT{r\ind{A}} \in \mathscr{R}\ind{A}$. Let us define $\bm{R}$ to be the orthogonal projector onto $\mathscr{R}\ind{A}$, and similarly for other spaces. Then we have the identity $\bm{A}^\alpha \bm{R}\ind{A} = \bm{R}\ind{A} {\bm{A}}^\alpha \bm{R}\ind{A}$. Similarly, since for any $\brA{l\ind{A}} \in \mathscr{L}^\dag\ind{A}$, $\brA{l\ind{A}} {\bm{A}}^\alpha \in \mathscr{L}^\dag\ind{A}$, we must have that $\bm{L}\ind{A} \bm{A}^\alpha = \bm{L}\ind{A} \bm{A}^\alpha \bm{L}\ind{A}$.

Now consider $\keT{r\ind{A}} \in \mathscr{R}\ind{A}$. We seek a class of vectors $\keT{v\ind{A}} \in \mathscr{R}\ind{A}$ such that for any $\brA{l\ind{A}} \in \mathscr{L}^\dag\ind{A}$, $\brA{k\ind{A}}(\keT{r\ind{A}} + \keT{v\ind{A}}) = \brAkeT{l\ind{A}}{r\ind{A}}$. Such vectors must be orthogonal to $\mathscr{L}\ind{A}$, and therefore we have $\keT{v\ind{A}} \in \mathscr{L\ind{A}}^\perp$. Therefore $\keT{v\ind{A}} \in \mathscr{R\ind{A}} \cap \mathscr{L\ind{A}}^\perp$.

To project out this space from $\mathscr{R}\ind{A}$, we take the space $\tilde{\mathscr{R}\ind{A}} := \mathscr{R}\ind{A} \cap (\mathscr{R}\ind{A} \cap \mathscr{L}\ind{A}^\perp)^\perp = \mathscr{R}\ind{A} \cap (\mathscr{R}\ind{A}^\perp + \mathscr{L}\ind{A})$, where we used the De Morgan identity in the last equality. Similarly, we can define $\tilde{\mathscr{L}}\ind{A} = \mathscr{L}\ind{A} \cap (\mathscr{L}\ind{A}^\perp + \mathscr{R}\ind{A})$.
Then we have that

\begin{equation}
\begin{aligned}
  \brAkeT{l\ind{A}}{r\ind{A}} = \brA{l} \tilde{\bm{R}}\ind{A} \keT{r\ind{A}}, \quad \tilde{\bm{R}}\ind{A} \keT{r\ind{A}} \in \mathscr{R}\ind{A}, \\ \brAkeT{l\ind{A}}{r\ind{A}} = \brA{l\ind{A}} \tilde{\bm{L}}\ind{A} \keT{r\ind{A}}, \\ 
  \quad \brA{l\ind{A}} \tilde{\bm{L}}\ind{A} \in \mathscr{L}\ind{A}^\dag \quad \forall \brA{l\ind{A}} \in \mathscr{L}\ind{A}^\dag, \keT{r\ind{A}} \in \mathscr{R}\ind{A}.
\end{aligned}    
\end{equation}

This is equivalent to

\begin{align}
  \bm{L}\bm{R} = \bm{L} \tilde{\bm{R}} \bm{R} = \bm{L} \bm{R} \tilde{\bm{R}}, \quad \bm{L}\bm{R} = \bm{L} \tilde{\bm{L}} \bm{R} = \tilde{\bm{L}} \bm{L} \bm{R}.
\end{align}

Then we have
\begin{align}
  \meL{\phi\ind{A}^{\alpha_0}}{\bm{A}^{\alpha_1}& \cdots \bm{A}^{\alpha_L}}{\psi\ind{A}^{\alpha_{L+1}}}\\
  & = \meL{\phi\ind{A}^{\alpha_0}}{\tilde {\bm{R}}\ind{A} \bm{A}^{\alpha_1} \tilde {\bm{R}}\ind{A} \cdots \tilde {\bm{R}}\ind{A} \bm{A}^{\alpha_L} \tilde {\bm{R}}\ind{A}}{\psi\ind{A}^{\alpha_{L+1}}} \\
  & = \meL{\phi\ind{A}^{\alpha_0}}{\tilde {\bm{L}} \bm{A}^{\alpha_1} \tilde {\bm{L}}\ind{A} \cdots \tilde {\bm{L}}\ind{A} \bm{A}^{\alpha_L} \tilde {\bm{L}}\ind{A}}{\psi\ind{A}^{\alpha_{L+1}}}.
\end{align}

We now prove that $\tilde{\mathscr{L}}\ind{A} = \bm{L}\ind{A} (\mathscr{R}\ind{A})$, $\tilde{\mathscr{R}}\ind{A} = \bm{R}\ind{A} (\mathscr{L}\ind{A})$ and therefore $\dim \tilde{\mathscr{L}}\ind{A} = \dim \tilde{\mathscr{R}}\ind{A}$. 

Consider an element $\keT{x\ind{A}} \in \tilde{\mathscr{R}}\ind{A} = \mathscr{R}\ind{A} \cap (\mathscr{R}\ind{A}^\perp + \mathscr{L}\ind{A})$. Then $\keT{x\ind{A}} = \keT*{r^\perp\ind{A}} + \keT{l\ind{A}}$. Then $\bm{R}\ind{A} \keT{x\ind{A}} = \bm{A}\ind{A}(\keT{l\ind{A}} + \keT*{r^\perp\ind{A}})$. Consider the LHS. Since $\keT{x\ind{A}} \in \mathscr{R}\ind{A}$, $\bm{R} \keT{x\ind{A}} = \keT{x\ind{A}}$. Consider the RHS. $\bm{R}\ind{A}$ is distributive. $\keT*{r^\perp\ind{A}} \in \mathscr{R}\ind{A}^\perp$ so $\bm{R}\ind{A} \keT*{r^\perp\ind{A}} = 0$. Therefore we have $\keT{x\ind{A}} = R\ind{A} \keT{l\ind{A}}$. Therefore $\keT{x\ind{A}} \in \bm{R}\ind{A} (\mathscr{L}\ind{A})$ and $\tilde{\mathscr{R}\ind{A}} \subseteq \bm{R}\ind{A} (\mathscr{L}\ind{A})$.

Now for the converse. Consider $\keT{x\ind{A}} \in \bm{R}\ind{A}(\mathscr{L}\ind{A})$. It can be written as $\keT{x\ind{A}} = \bm{R}\ind{A} \keT{l\ind{A}}$. $\keT{l\ind{A}} = \bm{R}\ind{A} \keT{l\ind{A}} + \bm{P}_{\mathscr{R}\ind{A}^\perp} \keT{l\ind{A}}$. Therefore $\keT{l\ind{A}} = \keT{x\ind{A}} + \bm{P}_{\mathscr{R}\ind{A}^\perp} \keT{l\ind{A}}$ and $\keT{x\ind{A}} = -\keT{l\ind{A}} + \bm{P}_{\mathscr{R}\ind{A}^\perp} \keT{l\ind{A}}$. Therefore $\keT{x\ind{A}} \in \mathscr{L}\ind{A} + \mathscr{R}\ind{A}^\perp$. $\keT{x\ind{A}} \in \mathscr{R}\ind{A}$ by assumption. Therefore $\bm{R}\ind{A}(\mathscr{L}\ind{A}) \subseteq \mathscr{R}\ind{A} \cap (\mathscr{L}\ind{A} + \mathscr{R}\ind{A}^\perp)$.

Therefore $\tilde{\mathscr{R}}\ind{A} = \mathscr{R}\ind{A} \cap (\mathscr{R}\ind{A}^\perp + \mathscr{L}\ind{A}) = R(\mathscr{L}\ind{A})$. By symmetry, we have $\tilde{\mathscr{L}}\ind{A} = \bm{L}\ind{A}(\mathscr{R}\ind{A})$. 

Now their dimensions are equal as $\dim \tilde{\mathscr{R}\ind{A}} = \dim \mathrm{image} \bm{R}\ind{A}\bm{L}\ind{A} = \rank \bm{R}\ind{A}\bm{L}\ind{A} = \rank \bm{L}\ind{A}\bm{R}\ind{A} = \dim \mathrm{image} \bm{L}\ind{A}\bm{R}\ind{A} = \dim \tilde{\mathscr{L}}\ind{A}$.

Consider two equivalent MPS representations A and B. Define an additive map $\tilde{\bm{W}}: \tilde{\mathcal{L}}\ind{A} \rightarrow \tilde{\mathcal{R}}\ind{B}$ that acts as
\begin{align}
  \tilde{W} : \tilde{\bm{L}}^<\ind{A} \keT{r^{\alpha}\ind{A}} \rightarrow \tilde{\bm{R}}^<\ind{B} \keT{r^{\alpha}\ind{B}}.
\end{align}

Here, the projector $\tilde{\bm{L}}\ind{A}^< : \mathscr{V}\ind{A} \rightarrow \tilde{\mathscr{L}}\ind{A}$ is a rectangular orthogonal projector. Projector for the other spaces are similarly defined. The inequality reminds us the dimension of the spaces on the left and right. On the other hand, $\tilde{\bm{R}}\ind{A}^> = (\tilde{\bm{R}}\ind{A}^<)^\top : \tilde{\mathscr{R}}\ind{A} \rightarrow \mathscr{V}\ind{A}$.

Consider a vector $\keT{c\ind{A}}$ in $\tilde{\mathscr L}\ind{A}$. Recall that $\bm{L}\ind{A} (\mathscr{R}\ind{A}) = \tilde{\mathscr{L}}\ind{A}$ and $\tilde{\bm{L}}\ind{A} \bm{L}\ind{A} = \tilde{\bm{L}}\ind{A}$. Therefore we have $\tilde{\bm{L}}\ind{A} (\mathscr{R}\ind{A}) = \tilde{\mathscr{L}}\ind{A}$. Since $\{\keT{r^{\alpha}\ind{A}}\}_{\alpha}$ is an over-complete basis on $\mathscr{R}\ind{A}$, there exists a convention of coefficients such that $\keT{c\ind{A}} = \sum_{\alpha} c^\alpha \tilde{\bm{L}}^<\ind{A} \keT{r^\alpha\ind{A}}$. Choose a convention. We claim that $\tilde{\bm{W}}$ is linear. Consider any two $\alpha \keT{c\ind{A}}$ and $\beta \keT{c'\ind{A}}$. Applying the additive property of $\tilde{\bm{W}}$, we see that $\tilde{\bm{W}}$ is linear.

We can see that $\Im \tilde{\bm{W}} = \tilde{\mathscr{R}}\ind{B}$ in the following. Consider a possible target vector $\keT{c\ind{B}} \in \tilde{\mathscr{R}}\ind{B}$. Again, since $\{\keT{r^{\alpha}\ind{B}}\}_{\alpha}$ is an overcomplete basis, there exists a representation $\keT{c\ind{B}} = \sum_\alpha c^\alpha \tilde{\bm{R}}^<\ind{B} \keT{r^\alpha\ind{B}}$. Then, the vector $\sum_\alpha c^\alpha \tilde{\bm{L}}^<\ind{A} \keT{r^\alpha\ind{A}}$ maps onto $\keT{c\ind{B}}$. By rank-nullity theorem, $\ker \tilde{\bm{W}} = \{0\}$, and therefore $\tilde{\bm{W}}$ is an invertible linear map. Hence, we also must have that $\dim \tilde{\mathscr{L}}\ind{A} = \dim \tilde{\mathscr{R}}\ind{B}$ and $\tilde{\mathscr{L}}\ind{A} = \tilde{\mathscr{R}}\ind{B}$.

We may now define a lifted version of $\tilde{\bm{W}}$ that acts as $\mathscr{V}\ind{A} \rightarrow \mathscr{V}\ind{B}$, defined as
\begin{align}
  \bm{W} = \bm{L}\ind{B} \tilde{\bm{R}}^>\ind{B} \tilde{\bm{W}} \tilde{\bm{L}}^<\ind{A} \bm{R}\ind{A}.
\end{align}

We will define $\tilde{\bm{R}}\ind{B} \bm{B}^\alpha \tilde{\bm{R}}\ind{B} =: \bar{\bm{B}}^\alpha$ and $\tilde{\bm{R}}\ind{B}^< \bm{B}^\alpha \tilde{\bm{R}}\ind{B}^> =: \tilde{\bm{B}}^\alpha$, and $\tilde{\bm{L}}\ind{A} \bm{A}^\alpha \tilde{\bm{L}}\ind{A} =: \bar{\bm{A}}^\alpha$ and $\tilde{\bm{L}}\ind{A}^< \bm{A}^\alpha \tilde{\bm{L}}\ind{A}^> =: \tilde{\bm{A}}^\alpha$.

Then we have (brakets denote which matrices will alter in the following step)

\begin{align*}
  \bm{L}\ind{B} \bm{B}^\alpha \bm{W} 
  & = (\bm{L}\ind{B} \bm{B}^\alpha \bm{L}\ind{B}) \tilde{\bm{R}}^>\ind{B} \tilde{\bm{W}} \tilde{\bm{L}}^<\ind{A} \bm{R}\ind{A} \\
  & = \bm{L}\ind{B} \bm{B}^\alpha (\tilde{\bm{R}}^>\ind{B}) \tilde{\bm{W}} \tilde{\bm{L}}^<\ind{A} \bm{R}\ind{A} \\
  & = \bm{L}\ind{B} (\bm{B}^\alpha \bm{R}\ind{B}) \tilde{\bm{R}}^>\ind{B} \tilde{\bm{W}} \tilde{\bm{L}}^<\ind{A} \bm{R}\ind{A} \\
  & = (\bm{L}\ind{B} \bm{R}\ind{B}) \bm{B}^\alpha \bm{R}\ind{B} \tilde{\bm{R}}^>\ind{B} \tilde{\bm{W}} \tilde{\bm{L}}^<\ind{A} \bm{R}\ind{A} \\
  & = \bm{L}\ind{B} \tilde{\bm{R}}\ind{B} (\bm{R}\ind{B} \bm{B}^\alpha\bm{R}\ind{B}) \tilde{\bm{R}}^>\ind{B} \tilde{\bm{W}} \tilde{\bm{L}}^<\ind{A} \bm{R}\ind{A} \\
  & = \bm{L}\ind{B} \tilde{\bm{R}}\ind{B} \bm{B}^\alpha (\bm{R}\ind{B} \tilde{\bm{R}}^>\ind{B}) \tilde{\bm{W}} \tilde{\bm{L}}^<\ind{A} \bm{R}\ind{A} \\
  & = \bm{L}\ind{B} \tilde{\bm{R}}\ind{B} \bm{B}^\alpha (\tilde{\bm{R}}^>\ind{B}) \tilde{\bm{W}} \tilde{\bm{L}}^<\ind{A} \bm{R}\ind{A} \\
  & = \bm{L}\ind{B} (\tilde{\bm{R}}\ind{B} \bm{B}^\alpha \tilde{\bm{R}}\ind{B}) \tilde{\bm{R}}^>\ind{B} \tilde{\bm{W}} \tilde{\bm{L}}^<\ind{A} \bm{R}\ind{A} \\
  & = \bm{L}\ind{B} (\bar{\bm{B}}^\alpha \tilde{\bm{R}}\ind{B}^>) \tilde{\bm{W}} \tilde{\bm{L}}^<\ind{A} \bm{R}\ind{A} \\
  & = \bm{L}\ind{B} \tilde{\bm{R}}\ind{B}^> (\tilde{\bm{B}}^\alpha \tilde{\bm{W}}) \tilde{\bm{L}}^<\ind{A} \bm{R}\ind{A} \\
  & = \bm{L}\ind{B} \tilde{\bm{R}}\ind{B}^> \tilde{\bm{W}} (\tilde{\bm{A}}^\alpha \tilde{\bm{L}}^<\ind{A}) \bm{R}\ind{A} \\
  & = \bm{L}\ind{B} \tilde{\bm{R}}\ind{B}^> \tilde{\bm{W}} \tilde{\bm{L}}^<\ind{A} (\bar{\bm{A}}^\alpha) \bm{R}\ind{A} \\
  & = \bm{L}\ind{B} \tilde{\bm{R}}\ind{B}^> \tilde{\bm{W}} (\tilde{\bm{L}}^<\ind{A} \tilde{\bm{L}}\ind{A}) \bm{A}^\alpha (\tilde{\bm{L}}\ind{A}) \bm{R}\ind{A} \\
  & = \bm{L}\ind{B} \tilde{\bm{R}}\ind{B}^> \tilde{\bm{W}} (\tilde{\bm{L}}^<\ind{A}) \bm{A}^\alpha (\bm{L}\ind{A} \tilde{\bm{L}}\ind{A} \bm{R}\ind{A}) \\
  & = \bm{L}\ind{B} \tilde{\bm{R}}\ind{B}^> \tilde{\bm{W}} \tilde{\bm{L}}^<\ind{A} (\bm{L}\ind{A} \bm{A}^\alpha \bm{L}\ind{A}) \bm{R}\ind{A} \\
  & = \bm{L}\ind{B} \tilde{\bm{R}}\ind{B}^> \tilde{\bm{W}} (\tilde{\bm{L}}^<\ind{A} \bm{L}\ind{A}) (\bm{A}^\alpha \bm{R}\ind{A}) \\
  & = (\bm{L}\ind{B} \tilde{\bm{R}}\ind{B}^> \tilde{\bm{W}} \tilde{\bm{L}}^<\ind{A} \bm{R}\ind{A}) \bm{A}^\alpha \bm{R}\ind{A} \\
  & = \bm{W} \bm{A}^\alpha \bm{R}\ind{A}. 
\end{align*}

We can also see that by definition of $\tilde{\bm{W}}$ and that $\keT{\psi\ind{A}^\alpha} = \keT{r\ind{A}^\alpha}$, we have 

\begin{equation}
\begin{aligned}
  \bm{W} \keT{\psi^\alpha\ind{A}} & = \bm{L}\ind{B} \tilde{\bm{R}}\ind{B} \keT{\psi^\alpha\ind{B}} = \bm{L}\ind{B} \tilde{\bm{R}}\ind{B} \bm{R}\ind{B} \keT{\psi^\alpha\ind{B}} \\
  & = \bm{L}\ind{B} \bm{R}\ind{B} \keT{\psi^\alpha\ind{B}} = \bm{L}\ind{B} \keT{\psi^\alpha\ind{B}}.
\end{aligned}
\end{equation}

Similarly, we can show that $\brA{\phi^\alpha\ind{B}} \bm{W} = \brA{\phi^\alpha\ind{A}} \bm{R}\ind{A}$.

We therefore have these necessary conditions for MPS equivalence as claimed in \cref{thm:app_MPS}.

\section{Details on spatially-invertible $\chi=1$ solvable gates} \label{app:spat-inv}
\subsection{Derivation of $\chi=1$ influence-solvability for spatially invertible gates}
Consider the class of gates invertible in space. For such gates, in equations such as, \cref{diag:L2}, we may apply the spatially inverted gate $\pic{space-inverse}$ from the right. For example, for $\mathscr{L}_1$, we have
% \begin{align}
%     \includegraphics[width=0.75\linewidth]{figs/l1-space-invertible.png}
% \end{align}
\begin{align}
    \mathscr{L}_1 & = \mathrm{span}\qty{\pic{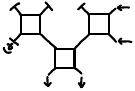}} \\
    & = \mathrm{span}\qty{\pic{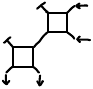}} \; \pic{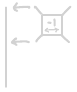} \\
    & = \mathrm{span}\qty{\pic{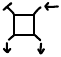}}.
\end{align}
Then for $\mathscr{L}_2$, we have
% \begin{align}
%     \includegraphics[width=0.75\linewidth]{figs/l2-spatial-invertible.png}
% \end{align}
\begin{align}
    \mathscr{L}_2 = \mathrm{span}\qty{\pic{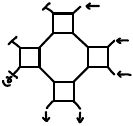}} = \mathrm{span} \qty{\pic{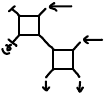}}.
\end{align}
Finally, for $\mathscr{L}_3$, we can use similar techniques to find
% \begin{align}
%     \includegraphics[width=0.6\linewidth]{figs/l3-spatial-invertible.png}
% \end{align}
\begin{equation}
\begin{aligned}
    \mathscr{L}_3 & = \mathrm{span}\qty{\pic{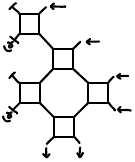}} = \mathrm{span} \qty{\pic{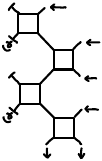}} \\
    & = \mathrm{span} \qty{\pic{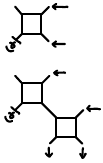}} = \mathrm{span} \qty{\pic{appB/6}} \\
    & = \mathscr{L}_2.
\end{aligned}
\end{equation}
Therefore, we have $\mathscr{L} = \mathscr{L}_2$.

Now that $\mathscr{L}$ is known, we can substitute $\mathscr{L}$ into the solvability condition \eqref{eq:zerobond_cond_b} and apply the spatial inverse from the right to find
% \begin{align}
%     \includegraphics[width=0.75\linewidth]{figs/spatial-invertible-solvability.png}
% \end{align}
\begin{align}
    & \pic{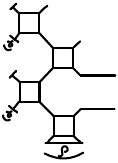} = \pic{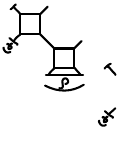}.
\end{align}
Now, defining 
\begin{align}
    \pic{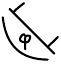} := \pic{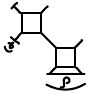},
\end{align}
% $\includegraphics[width=0.4\linewidth]{figs/phi.png}$, 
we get
% \begin{align} \label{eq:solvability-phi}
%     \includegraphics[width=0.65\linewidth]{figs/solvability-phi.png}
% \end{align}
\begin{align} \label{eq:solvability-phi}
    \pic{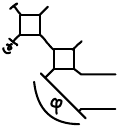} = \pic{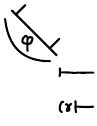}.
\end{align}
For all vectors $\bm{v}$ orthogonal to $\keT{\gamma}$, we can apply it to the lowest leg to find that
% \begin{align}
%     \includegraphics[width=0.75\linewidth]{figs/apply-v.png}
% \end{align}
\begin{align}
    \pic{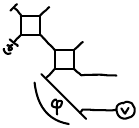} & = \pic{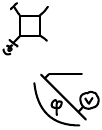} = 0 \\
    \nonumber \\
    \implies \pic{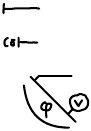} & = 0.
\end{align}
This means that $\keT{\phi}$ is a product state, i.e. $\keT{\phi} = \keT{\beta} \otimes \keT{\gamma}$.

Finally, replacing $\keT{\phi}$ in \eqref{eq:solvability-phi}, we find the two identities
\begin{align} \label{eq:gamma-and-beta-from-left}
    \pic{gamma-lower-left-gate} = \pic{beta-upper-left-identity} \quad \text{ and } \quad \pic{beta-lower-left-gate} = \pic{gamma-upper-left-identity}\,.
\end{align}
% \begin{align} \label{eq:gamma-and-beta-from-left}
%     \includegraphics[width=0.75\linewidth]{figs/beta-and-gamma.png}
% \end{align}

\subsection{A family of $\chi=1$ influence-solvable and spatially invertible gates that is not dual-unitary}
In \cref{sec:d=2-space-invertible}, we show that all generalised solvable gates for qubits ($d=2$) are dual unitary. To prove that this is not the case for $d \geq 3$, below we provide an example of a gate that is solvable and space invertible but not dual-unitary. Here, we consider the case where $\keT{\beta} = \keT{\gamma}$. In this case we have the single requirement
\begin{align}
    \pic{page31/page3112} = \pic{page31/page3115}.
\end{align}
% \begin{align}
%     \includegraphics[width=0.5\linewidth]{figs/gamma=beta-case.png}
% \end{align}
Now consider the pure state $\hat \gamma = \ketbra{\gamma}{\gamma}$, and a gate that, if $\ket{\gamma}$ is inserted on the lower left, acts as a SWAP in space from left to right, i.e.
% \begin{align}
%     \includegraphics[width=0.4\linewidth]{figs/gamma-SWAP-gate.png} \; \forall \; i = \{0, 1, \dots, d-1\}.
% \end{align}
\begin{align}
    \pic{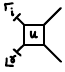} = \pic{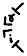} \quad \forall \; i \in \{0, 1, \dots, d-1\}.
\end{align}
Here we can identify $\ket{\gamma} = \ket{0}$. Then, applying the folding formalism, and summing over $i$, we have
\begin{align}
    \sum_i \pic{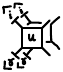} = \sum_i \pic{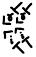} \\
    \implies \pic{page31/page3112} & = \pic{page31/page3115}.
\end{align}
% \begin{align}
%     \includegraphics[width=0.6\linewidth]{figs/fold-and-sum-i.png}
% \end{align}
Due to the unitary condition, the gate must as a SWAP for the set $\{\ket{0, i}, \ket{i, 0} : i = 0, 1, \dots, d-1\}$. For example, for $d=3$, we have
\begin{align}
    \setlength{\tabcolsep}{2.5pt}
    u = 
    \footnotesize \left(
    \begin{tabular}{c|ccccccccc}
       &00&01&10&02&20&11&12&21&22 \\
    \hline
    00 & 1&  &  &  &  &  &  &  &   \\
    01 &  &  & 1&  &  &  &  &  &   \\
    10 &  & 1&  &  &  &  &  &  &   \\
    02 &  &  &  &  & 1&  &  &  &   \\
    20 &  &  &  & 1&  &  &  &  &   \\
    11 &  &  &  &  &  & $*$& $*$& $*$& $*$ \\
    12 &  &  &  &  &  & $*$& $*$& $*$& $*$ \\
    21 &  &  &  &  &  & $*$& $*$& $*$& $*$ \\
    22 &  &  &  &  &  & $*$& $*$& $*$& $*$ \\
    \end{tabular}
    \right)
\end{align}
In general the matrix is a block diagonal matrix of a SWAP-like part of size $(d+1) \times (d + 1)$ and an arbitrary part of size $(d^2 - (d+1)) \times (d^2 - (d+1))$.  

The starred entries can be arbitrary, as long as it is not singular in space (left to right spatially), and not dual-unitary itself. This can be seen by considering the condition for dual unitarity,
\begin{align}
    \centering
    \includegraphics[width=0.6\linewidth]{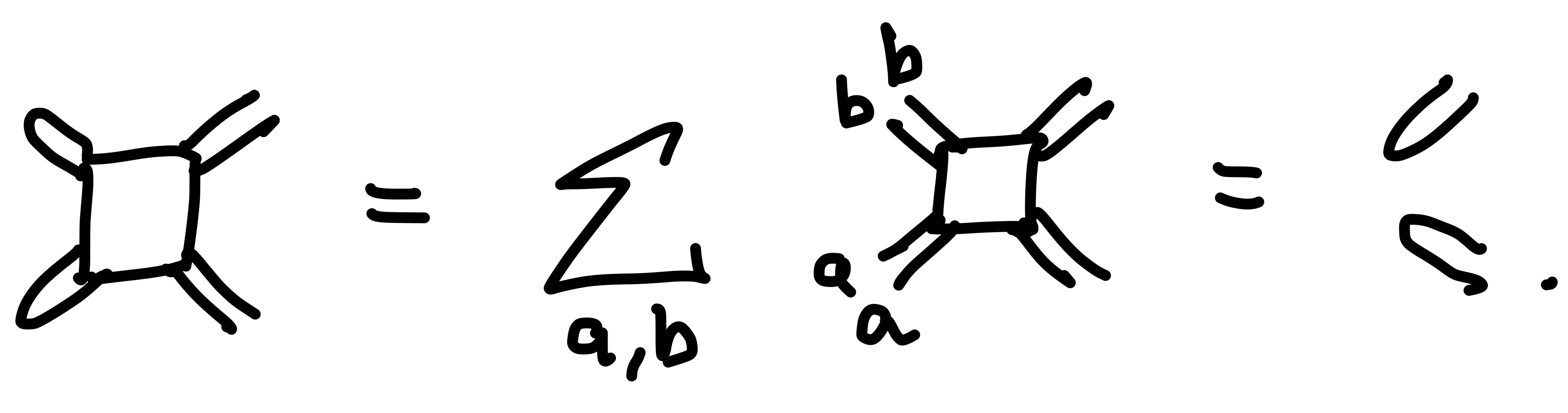}
\end{align}
Now define the identity on the folded subspace $\mathrm{span}\{\ket{i} : i \neq 0\}^{\otimes 2}$ as 
\begin{align}
    \pic{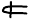} := \sum_{a \neq 0} \pic{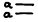} = \pic{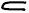} - \pic{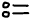}.
\end{align}
Then we have
\begin{align}
    \includegraphics[width=0.75\linewidth]{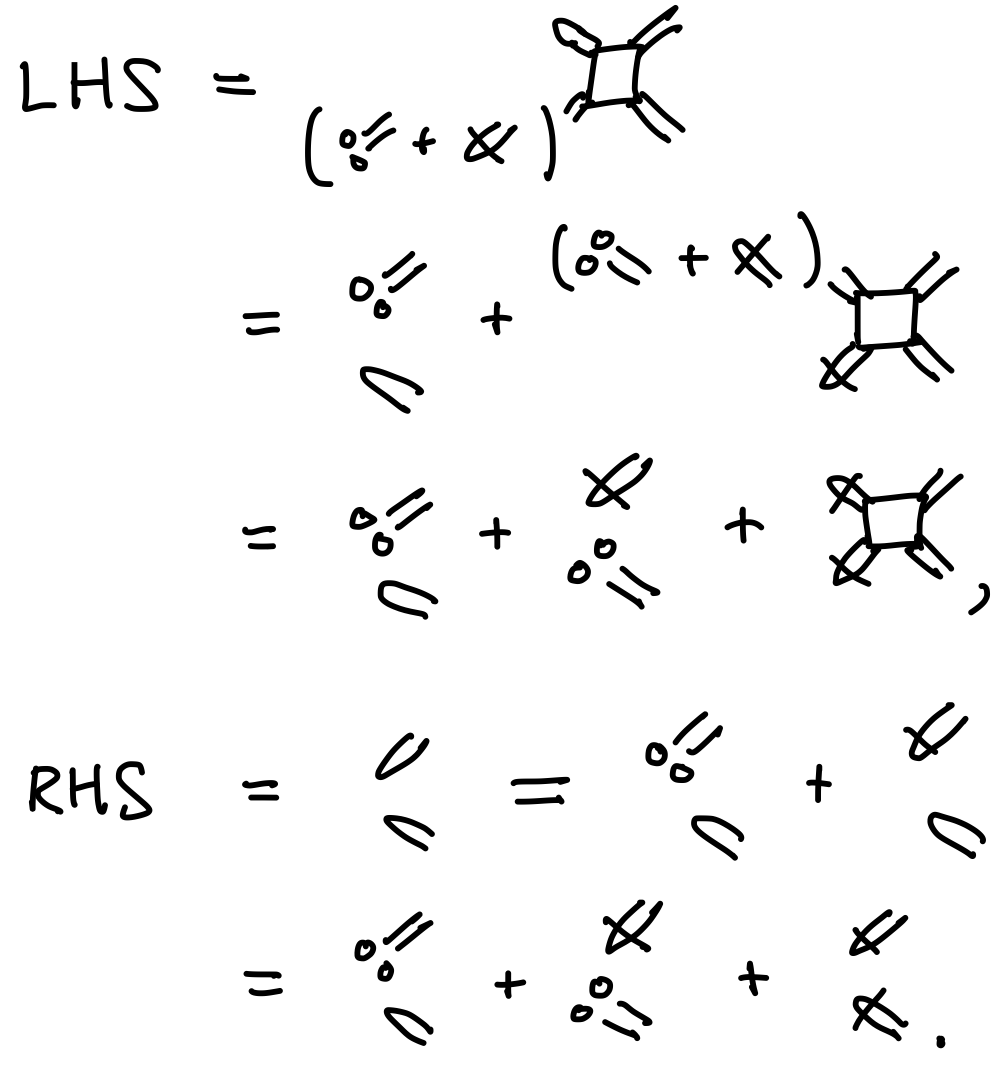}
\end{align}
Therefore for the gate to be not dual-unitary, we require
\begin{align}
    \includegraphics[width=0.35\linewidth]{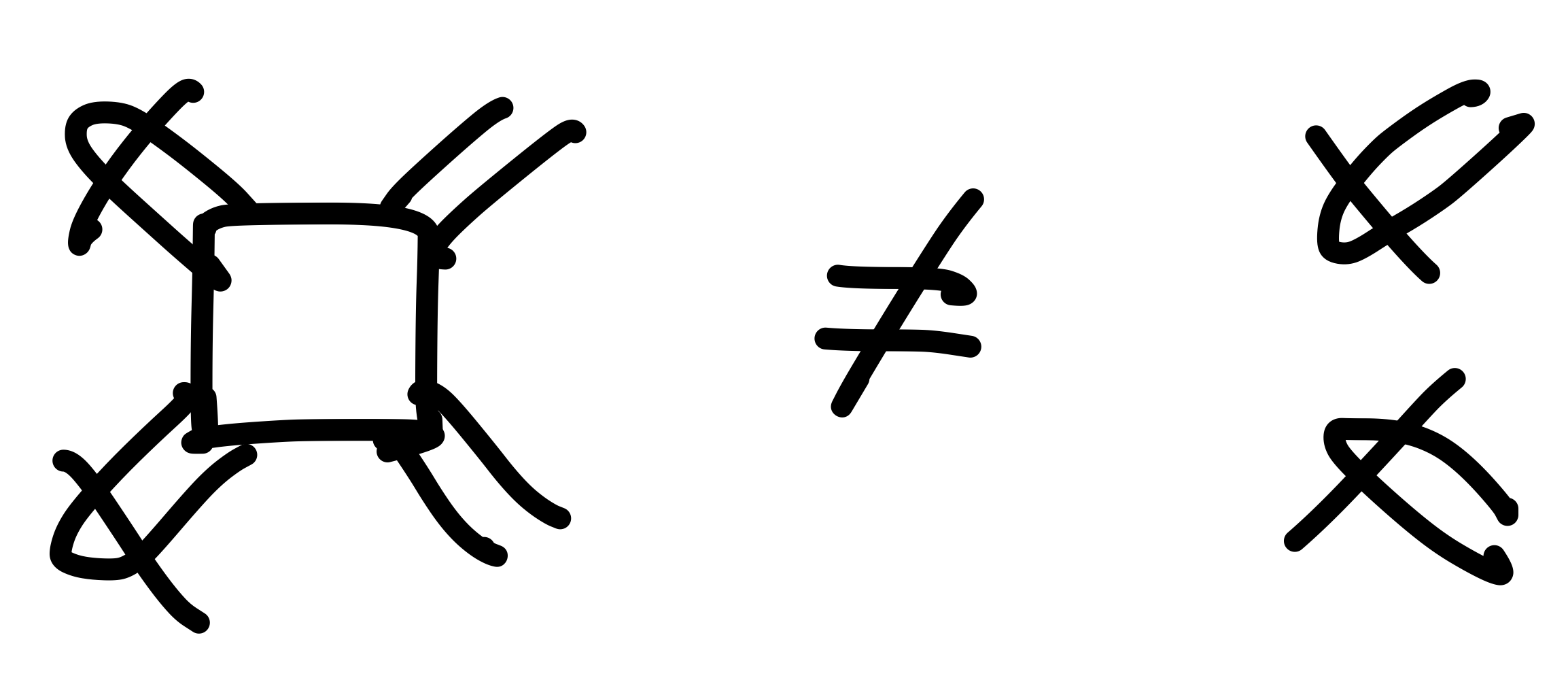}.
\end{align}

The condition for this gate to be dual-unitary is given by the following. 

With $d=2$ we see that the arbitrary part is only a $1 \times 1$ matrix (i.e. a phase), preventing it from not dual-unitary.

\section{Details on classification of $d=2$ quantum space-invertible unitary gates} \label{sec:d=2-space-invertible}
We first recall the exhaustive parameterisations of $d=2$ two-site unitaries given by \eqref{eq:zerobond_quantum_decomp_XYZ}.

Ref.~\onlinecite{bertini2019exact} parametrised all possible dual-unitaries. It turns out that they are given by the case when \eqref{eq:xyz-in-x-y-z} holds.

Now recall that for any $\chi=1$ influence-solvable and spatially invertible gates, that \cref{eq:gamma-beta-equation} holds. Let us define the superprojector
\begin{align}
    \pic{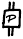} := \pic{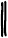} - \frac{1}{d} \pic{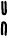}.
\end{align}
Applying this on the lower-left leg of \cref{eq:gamma-beta-equation}, we get
\begin{align}
    \pic{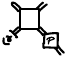} = 0.
\end{align}
Now using the exhaustive parametersation for two-site qubit gates, by applying unitaries on the open legs and recalling that the trace operation can absorb unitaries, we have
\begin{align}
    \pic{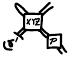} = 0,
\end{align}
where $\keT{\gamma'}$ is some other state. Therefore to have be influence-solvable, it is necessary that the null-space of the following matrix
\begin{align}
    \pic{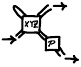}
\end{align}
exists. Looking for such solutions, we find that we require the same restrictions as dual-unitaries, i.e. \eqref{eq:xyz-in-x-y-z}.

Now we show that a subset of these can have more solvable states than dual-unitaries. Without a loss of generality, let us restrict to the case where $\pic{XYZ} = \pic{Z}$.

With the definitions given by the main text, \cref{eq:gamma-beta-equation} can be written as 
\begin{align}
    \frac{I}{2} + \gamma'_Z Z = \frac{I}{2} + \beta'_Z \sqdiamond \triangle Z \triangle^\dagger \sqdiamond^\dagger.
\end{align}
This requires that $\beta'_Z \sqdiamond \triangle Z \triangle^\dagger \sqdiamond^\dagger = \alpha Z$. Therefore we have $\beta'_Z = \gamma'_Z \alpha$, and similarly $\gamma'_Z = \beta'_Z \alpha$. Therefore we have $\gamma'_Z = \gamma'_Z \alpha^2$, which gives $\alpha = \pm 1$ if $\gamma'_Z \neq 0$. $\gamma'_Z = 0$ would give the infinite temperature state, i.e. the solvable state for dual-unitaries.

First consider the case $\alpha = 1$. Then we require
\begin{align}
    \triangle \sqdiamond Z \sqdiamond^\dagger \triangle^\dagger = 0 \Longleftrightarrow \qty[\triangle \sqdiamond, Z] = 0.
\end{align}
The only unitaries that commute with $Z$ is $\triangle \sqdiamond = e^{- i \phi Z}$, and therefore we have the condition
\begin{align}
    e^{i \phi Z} \triangle = \sqdiamond^\dagger. 
\end{align}

Now consider $\alpha = -1$. Then this requires that $\gamma'_Z = - \beta'_Z$. Therefore we have
\begin{align}
    \triangle \sqdiamond Z \sqdiamond^\dagger \triangle^\dagger = -Z \Longleftrightarrow \qty{\triangle \sqdiamond, Z} = 0.
\end{align}
Writing $\triangle \sqdiamond = e^{-i \frac{\phi}{2} \vec{n} \cdot \vec{\sigma}} = \cos \frac{\phi}{2} I - i (\vec{n} \cdot \vec{\sigma}) \sin \frac{\phi}{2}$, we require $\phi = \pi$ and the normalised vector $\vec{n} = (n_X, n_Y, 0) = (\cos \psi, \sin \psi, 0)$ since the identity commutes with $Z$. Therefore we have the requirement (renaming $\psi$ to $\phi$)
\begin{align}
    e^{i \frac{\pi}{2} (X \cos \phi + Y \sin \phi)} \triangle = \sqdiamond^\dagger.
\end{align}

\section{Details on classification of $d=2$ quantum non-space-invertible unitary gates}\label{app:quantum_d2_space_notinv}

\subsection{Identification of non-invertible gates}
First, we need to identify all unitary gates that are not invertible in space. We can use the gauge transformation \eqref{eq:zerobond_quantum_gauge} to write any two qubit gate as
\begin{align}
    \pic{unitary} = \pic{XYZdecomp} \mathop{\simeq}_\text{gauge-fix} \pic{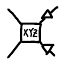}\label{eq:app_quantum_noninv_gate_after_gauge}
\end{align}
Here $\hat{u}\upd{XYZ}(x,y,z)$ is given by \eqref{eq:zerobond_quantum_XYZ}. The determinant of the gate $\hat{u}\upd{XYZ}$ viewed as a matrix in space is given by

\begin{widetext}
    \begin{equation}
    \begin{aligned}
        \det\hat{u}\upd{XYZ}\ind{space}(x,y,z) & = \frac{1}{16} e^{-4 i (x+y+z)} \left(e^{4 i (x+y+z)}-2 e^{2 i (x+y+2 z)}+2 e^{2 i (x+y)}-e^{4 i x}-e^{4 i y}+e^{4 i z}\right)\\
        & \times \left(e^{4 i (x+y+z)}+2 e^{2 i (x+y+2 z)}-2 e^{2 i (x+y)}-e^{4 i x}-e^{4 i y}+e^{4 i z}\right).
    \end{aligned}
    \end{equation}
\end{widetext}

\twocolumngrid

We would like to find the conditions for the above to be equal to zero, which would imply that the gate is non-invertible in space. Let us introduce $A=e^{2ix}$, $B=e^{2iy}$ and $C=e^{2iz}$. We can write
\begin{align}
	\det\hat{u}\upd{XYZ}\ind{space}(x,y,z) &= \frac{1}{16 A^2B^2C^2} f_1(A,B,C)f_2(A,B,C)
\end{align}
with
\begin{align}
	f_1(A,B,C)&= A^2B^2C^2-2ABC^2+2AB-A^2-B^2+C^2 \nonumber \\
	&= (1-AB)^2C^2 - (A-B)^2\\ 
	f_2(A,B,C)&=A^2B^2C^2+2ABC^2-2AB-A^2-B^2+C^2 \nonumber \\
	&= (1+AB)^2C^2 - (A+B)^2 = f_1(-A,B,C).
\end{align}
Therefore we require either $f_1(A,B,C) = 0$ or $f_2(A,B,C) = 0$. Let us assume that $f_1(A,B,C) = 0$. Since $\abs{C} = 1$, this is only possible if
\begin{align}
	\abs{1-AB} &= \abs{A-B} = \abs{1-B/A} 
\end{align}
Using $\abs{1-e^{i\phi}}^2 = 2-2\cos{\phi}$ we conclude
\begin{align}
	\cos(2(x+y)) &= \cos(2(y-x))
\end{align}
implying
\begin{align}
	\sin(2x)\sin(2y) &= 0
\end{align}
and therefore either $x\in \pi/2\mathbb{Z}$ or $y\in \pi/2\mathbb{Z}$. Note that we obtain exactly the same conclusion if $f_2(A,B,C)=0$.

In particular, this implies that either $A=\pm 1$ or $B=\pm 1$. Since either
\begin{align}
	C^2 &= \frac{(A-B)^2}{(1-AB)^2} &\mathrm{or}&& C^2 &= \frac{(A+B)^2}{(1+AB)^2}
\end{align}
this implies $C=\pm 1$, unless both either $A=\pm B$ (in which case $C$ is arbitrary). Hence, we conclude that the gate is only not invertible in space if out of $A,B,C$ two are $\pm 1$ and the remaining one is arbitrary. 

By symmetry we will restrict ourselves to $A=\pm 1$ and $B=\pm 1$ and $C$ arbitrary. Next we show that we can restrict to $A=B=1$, i.e. $x=y=0$. For that observe
\begin{align}
	e^{-i\tfrac{\pi}{2}\hat{\sigma}\ind{x}\otimes \hat{\sigma}\ind{x}} = - i \hat{\sigma}\ind{x}\otimes \hat{\sigma}\ind{x}
\end{align}
and similar for $\hat{\sigma}\ind{y}$. Therefore
\begin{align}
	\hat{u}\upd{XYZ}(x+\pi/2,y,z) &= \hat{u}\upd{XYZ}(x,y,z) (- i \hat{\sigma}\ind{x}\otimes \hat{\sigma}\ind{x})
\end{align}
which is $\hat{u}\upd{XYZ}$ multiplied by single site unitaries which can be absorbed into the single site unitaries in \eqref{eq:zerobond_quantum_decomp_XYZ}.

Since $A=\pm 1$ implies $x \in \tfrac{\pi}{2} \mathbb{Z}$, the gate $\hat{u}\upd{XYZ}(x,y,z)$ is equivalent to $\hat{u}\upd{XYZ}(0,y,z)$. Using a similar argument for $y$ we conclude that (up to gauge freedom) the only XYZ gates not invertible in space are Ising gates:
\begin{align}
	\hat{u}\upd{XYZ}(0,0,z) &= \exp(-iz\hat{\sigma}\ind{z}\otimes \hat{\sigma}\ind{z}).
\end{align}
These can be viewed as controlled phase gates. After including the single site unitaries in \eqref{eq:app_quantum_noninv_gate_after_gauge} we thus find that the only two-qubit gates not invertible in space are (up to gauge freedom)
\begin{align}
	\hat{u}' &= \ketbra{0}{0} \otimes \hat{\Gamma}_0 + \ketbra{1}{1} \otimes \hat{\Gamma}_1.
\end{align}
or in circuit notation:
\begin{align}
    \pic{u}  = \pic{u0} + \pic{u1}.\label{eq:app_d2quantum_gate}
\end{align}
These gates are controlled gates. If the left bit is $0$, then one applies the single site unitary $\hat{\Gamma}_0$ on the right, otherwise if the left bit is $1$, then one applies $\hat{\Gamma}_1$.

Using the representation \eqref{eq:app_d2quantum_gate} we can explicitly compute the quantum channel $\bm{A}$ representing the bath:
\begin{align}
    \fpic{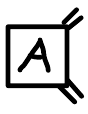} = \gamma_{00}\fpic{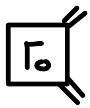} + \gamma_{11}\fpic{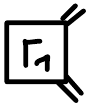}.
\end{align}
Equivalently, as formula, $\bm{A}$ applied to a single qubit density matrix $\hat{\omega}$ is given by
\begin{align}
	\bm{A}[\hat{\omega}] &= \gamma_{00} \hat{\Gamma}_0\hat{\omega}\hat{\Gamma}_0^\dagger + \gamma_{11} \hat{\Gamma}_1\hat{\omega}\hat{\Gamma}_1^\dagger. \label{eq:app_d2quantum_A}
\end{align}

We will see that it will be important for solvability to identify whether the quantum channel $\bm{A}$ is invertible (i.e. whether $\bm{A}$ written as a matrix is invertible).

As in the main text we will denote by $\hat{\gamma}$ and $\hat{\rho}$ the density matrices corresponding to $\keT{\gamma}$ and $\keT{\rho}$ in the folded picture. Also, we will denote the entries of $\hat{\Gamma}_i$ as $\qty[\hat{\Gamma}_i]_{ab} = \Gamma_{iab}$.

Note that we know that the special case $\hat{\Gamma}_1=\hat{\Gamma}_0$ is always solvable for any $\hat{\gamma}$. In the following analysis we will exclude this case.

\subsection{Invertible quantum channel}
Assuming that the quantum channel is invertible we proceed as follows: first observe that the span of the $\bm{B}$ tensors can be simplified to
\begin{align}
    \mathop{\mathrm{span}}_{abcd}\qty{\fpic{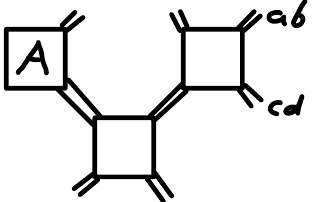}}=\mathop{\mathrm{span}}_{ab}\qty{\fpic{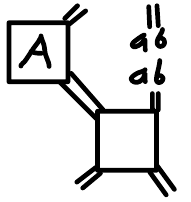}}.
\end{align}

With this \eqref{eq:zerobond_classic_L2_a} and \eqref{eq:zerobond_classic_L2_b} become
\begin{align}
    \fpic{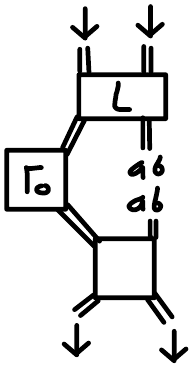} &\in \mathscr{L}\\
    %&\includegraphics[width=\linewidth]{figs/fred/Screenshot 2026-03-01 150717.png}\\
    \fpic{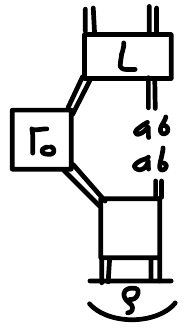} &= \fpic{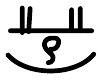}\gamma_{ab}\delta_{ab}\label{eq:app_d2quantum_condb1}
    %&\includegraphics[width=\linewidth]{figs/fred/Screenshot 2026-03-01 150734.png}
\end{align}

We can define the spaces $\mathscr{L}_{ab}$ and $\mathscr{L}_{n,ab}$ as
\begin{align}
    \mathscr{L}_{ab} &= \mathrm{span}\qty{\fpic{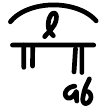}; \brA{l} \in \mathscr{L}}\\
    \mathscr{L}_{n,ab} &= \mathrm{span}\qty{\fpic{Ainv/app_quantumnoninv_Ainv1}; \brA{l} \in \mathscr{L}_n}
    %&\includegraphics[width=\linewidth]{figs/fred/Screenshot 2026-03-01 141212.png}\\
    %&\includegraphics[width=\linewidth]{figs/fred/Screenshot 2026-03-01 141224.png}.
\end{align}

Note that we can always decompose any $\brA{l} \in \mathscr{L}$ as $\fpic{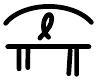} =\sum_{ab} \fpic{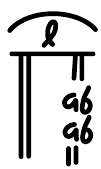}$ and hence $\mathscr{L} \subset \sum_{ab} \mathscr{L}_{ab} \brA{a,b}$.

However, due to the invertibility of the quantum channel $\bm{A}$, we also have that
\begin{align}
	\mathrm{dim} \qty[\sum_{ab} \mathscr{L}_{ab} \brA{ab}] = \mathrm{dim} \mathscr{L} 
\end{align}
and hence $\mathscr{L} = \sum_{ab} \mathscr{L}_{ab} \brA{a,b}$.

This simple observation has an important consequence. Since we know that the infinite temperature state $\mathbb{1}\otimes \mathbb{1} \in \mathscr{L}$, we have $\mathbb{1} \in \mathscr{L}_{aa}$. In graphical notation this is
\begin{align}
    \fpic{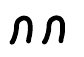} \in \mathscr{L} \Rightarrow \fpic{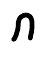} \in \mathscr{L}_{aa}.
\end{align}
Hence, from \eqref{eq:app_d2quantum_condb1} we find
\begin{align}
    \fpic{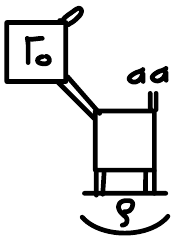} = \fpic{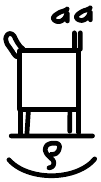} = \gamma_{aa} \overset{!}{=} \fpic{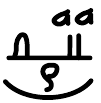} \gamma_{aa}. 
\end{align}
Here we used the definition \eqref{eq:zerobond_gamma_def} of $\gamma$. From this we follows that for each $a=0,1$ either $\gamma_{aa} = 0$ or $\beta_{aa}=1$, where $\beta_{aa} := \fpic{Ainv/app_quantumnoninv_Ainv11}$.

However, since $\beta_{aa}\geq 0$ and $\beta_{00}+\beta_{11} = 1$, we require that only one of the $\beta_{aa} = 1$ and the other one vanishes. Because of that also one of the $\gamma_{aa}$ has to vanish, implying that the other $\gamma_{aa}=1$. This implies that either
\begin{align}
	\hat{\gamma} &= \ketbra{0}{0} &&\mathrm{or}& \hat{\gamma} &= \ketbra{1}{1}. 
\end{align}
In particular $\hat{\gamma}$ is a pure state. Note that we can always flip $0 \leftrightarrow 1$, by gauge transforming the gate using single site $\hat{\sigma}\ind{x}$ matrices. Hence, we can restrict our analysis to the case $\hat \gamma = \ketbra{0}{0}$.

Note that in this case the quantum channel $\bm{A}$ applied to a single site density matrix $\hat{\omega}$ is given by
\begin{align}
	\bm{A}[\hat{\omega}] &= \hat{\Gamma}_0\hat{\omega}\hat{\Gamma}_0^\dagger.
\end{align}

This implies we can write \eqref{eq:app_d2quantum_condb1} equivalently as
\begin{align}
    \fpic{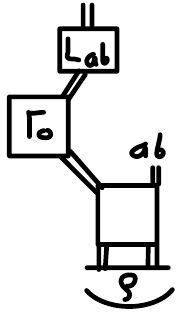} = \fpic{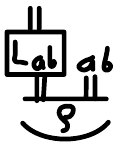}\delta_{a0}\delta_{b0}\label{eq:app_d2quantum_condb2}
\end{align}

\paragraph*{Removing the projector $\bm{L}$.}
The inconvenience of \eqref{eq:app_d2quantum_condb1} is the appearance of the projector $\bm{L}$, which projects on a space $\mathscr{L}$ that will depend in a complicated way on the specific gate. Fortunately, in this case \eqref{eq:app_d2quantum_condb1} is equivalent to the same condition but with $\bm{L}$ dropped. 

To see this let us choose an orthogonal basis $\brA{l_i}$ of $\mathscr{L}$. First observe that we can any $\brA{l} \in \mathscr{L}$ as
\begin{align}
    \fpic{Ainv/app_quantumnoninv_Ainv30} = \sum_{iab}\alpha_{iab}\fpic{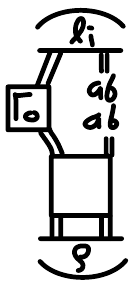}    
\end{align}
Denoting $\bm{B}_{ab} := \fpic{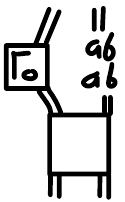}$ we observe that
\begin{align}
    \fpic{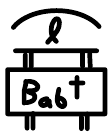} = \sum_{icd} \alpha_{icd}\fpic{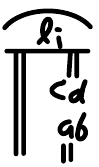}\delta_{ac}\delta_{bd} =  \sum_{i} \alpha_{iab}\fpic{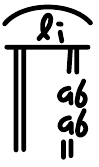} \in \mathscr{L}%\\
\end{align}
In formulas this means that
\begin{align}
	\bm{L}\bm{B}_{ab}^\dagger &= \bm{L}\bm{B}_{ab}^\dagger\bm{L}.
\end{align}
If we choose the projector $\bm{L}$ to be orthogonal, then we conclude
\begin{align}
	\bm{L}\bm{B}_{ab}\bm{L} &= \bm{B}_{ab}\bm{L}.
\end{align}
Note that in this notation \eqref{eq:app_d2quantum_condb1} is given by 
\begin{align}
	\bm{L}\bm{B}_{ab}\keT{\rho} &= \keT{\rho} \delta_{a0}\delta_{b0}.
\end{align}

Now observe that
\begin{align}
	\bm{L}\bm{B}_{ab}\keT{\rho} &= \bm{L}\bm{B}_{ab}\bm{L}\keT{\rho} = \bm{B}_{ab}\bm{L}\keT{\rho} = \bm{B}_{ab}\keT{\rho},
\end{align}
where we used $\bm{L}\bm{B}_{ab} = \bm{L}\bm{B}_{ab}\bm{L}$ and $\keT{\rho} = \bm{P}\keT{\rho}$.

Thus, we conclude $\bm{B}_{ab}w = w \delta_{a0}\delta_{b0}$, which, in circuit notation is nothing but \eqref{eq:app_d2quantum_condb1} without $\bm{L}$:
\begin{align}
    \fpic{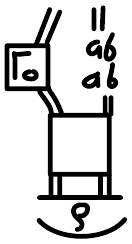} = \fpic{Ainv/app_quantumnoninv_Ainv4}\delta_{a0}\delta_{b0}\label{eq:app_d2quantum_condb3}
\end{align}

\paragraph*{Analysis of \eqref{eq:app_d2quantum_condb3}.}
Therefore, we need to find $\keT{\rho}$, s.t.
\begin{align}
    \fpic{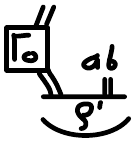} = \fpic{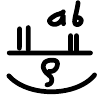}\delta_{a0}\delta_{b0}
\end{align}
where
\begin{align}
    \fpic{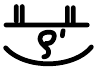} = \fpic{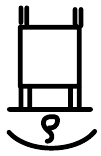}
\end{align}
This in particular implies that
\begin{align}
    \fpic{Ainv/app_quantumnoninv_Ainv21} = \fpic{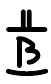}\;\fpic{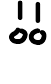}
\end{align}
and thus
\begin{align}
    \fpic{Ainv/app_quantumnoninv_Ainv4} = \fpic{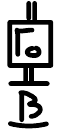}\;\fpic{Ainv/app_quantumnoninv_Ainv26} =: \fpic{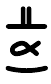}\;\fpic{Ainv/app_quantumnoninv_Ainv26}\label{eq:app_d2quantum_inv_inv_eq1}
\end{align}
From this it follows that
\begin{align}
    \fpic{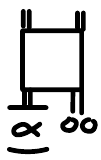} = \fpic{Ainv/app_quantumnoninv_Ainv25}\;\fpic{Ainv/app_quantumnoninv_Ainv26}
\end{align}
which written in components gives
\begin{align}
	\alpha_{ab} \hat{\Gamma}_a\ketbra{0}{0}\hat{\Gamma}_b^\dagger &= \beta_{ab}\ketbra{0}.\label{eq:app_d2quantum_inv_inv_eq2}
\end{align}
Hence, either $\hat{\Gamma}_0\ket{0} \sim \ket{0}$ or $\hat{\Gamma}_1\ket{0} \sim \ket{0}$. Note that any 2x2 unitary $\hat{\Gamma}$ that satisfies $\hat{\Gamma}\ket{0} \sim \ket{0}$ is diagonal in the $\ket{0},\ket{1}$ basis.

First, let us assume that $\hat{\Gamma}_0$ is not diagonal. Then $\hat{\Gamma}_1$ is necessarily diagonal and we have
\begin{align}
	\hat{\alpha} &= \ketbra{1}{1}\\
	\hat{\beta} &= \ketbra{1}{1}.
\end{align}
However, from \eqref{eq:app_d2quantum_inv_inv_eq1} we also require that 
\begin{align}
    \hat{\alpha} = \hat{\Gamma}_0\hat{\beta}\hat{\Gamma}_0^\dagger,\label{eq:app_d2quantum_inv_inv_eq3}
\end{align}
which contradicts that $\hat{\Gamma}_0$ is not diagonal.

Hence, we conclude that $\hat{\Gamma}_0$ is a diagonal matrix. If $\hat{\Gamma}_1$ is not diagonal the only possible solution to \eqref{eq:app_d2quantum_inv_inv_eq2} is
\begin{align}
	\hat{\alpha} &= \ketbra{0}{0}\\
	\hat{\beta} &= \ketbra{0}{0}.
\end{align}
This is compatible with $\hat{\alpha} = \hat{\Gamma}_0\hat{\beta}\hat{\Gamma}_0^\dagger$. If additionally $\hat{\Gamma}_1$ is also diagonal then
\begin{align}
	\beta_{ab} = \Gamma_{a00}\bar{\Gamma}_{b00}\alpha_{ab}.
\end{align}
From \eqref{eq:app_d2quantum_inv_inv_eq3} we then find that
\begin{align}
	\alpha_{ab} &= \Gamma_{0aa}\beta_{ab}\bar{\Gamma}_{0bb} = \Gamma_{0aa}\bar{\Gamma}_{0bb}\Gamma_{a00}\bar{\Gamma}_{b00}\alpha_{ab}.
\end{align}
Note that for $a=b=0$ we always have $\Gamma_{000}\bar{\Gamma}_{000}\Gamma_{000}\bar{\Gamma}_{000} = \abs{\Gamma_{000}}^4= 1$, allowing us to choose $\alpha_{00}=1$. 

If for more $a,b$ we have $\Gamma_{0aa}\bar{\Gamma}_{0bb}\Gamma_{a00}\bar{\Gamma}_{b00} = 1$, then we can also arbitrarily choose $\alpha_{ab}$, otherwise necessarily $\alpha_{ab}=0$.

To conclude, we only find solvability if and only if $\hat{\Gamma}_0$ is a diagonal matrix. In this case we can always choose $\hat{\rho} = \ketbra{0}{0}\otimes \ketbra{0}{0}$.

This establishes one possible option for $\hat{\rho}$ independently of $\bm{L}$. However, since $\keT{W}=\bm{L}\keT{\rho}$, the space of possible $\hat{\rho}$ depends on $\bm{L}$, which in turn depends on the specific gate.

\subsection{Non-invertible quantum channel}

The arguments above only apply to the case where the quantum channel $\bm{A}$ given by \eqref{eq:app_d2quantum_A} is invertible. Let us now identify the cases where $\bm{A}$ is not invertible. This happens when there exists a $\hat{\omega}\neq 0$ such that
\begin{align}
	\bm{A}[\hat{\omega}] &= \gamma_{00} \hat{\Gamma}_0\hat{\omega}\hat{\Gamma}_0^\dagger + \gamma_{11} \hat{\Gamma}_1\hat{\omega}\hat{\Gamma}_1^\dagger = 0
\end{align}
implying that
\begin{align}
	\gamma_{00} (\hat{\Gamma}_1^\dagger\hat{\Gamma}_0)\hat{\omega}(\hat{\Gamma}_1^\dagger\hat{\Gamma}_0)^\dagger &= -\gamma_{11} \hat{\omega}\label{eq:A_inv_A_inv_basic}
\end{align}
In particular, it follows
\begin{align}
	\gamma_{00}^2 \tr \hat{\omega}^\dagger\hat{\omega} &= \gamma_{11}^2 \tr \hat{\omega}^\dagger\hat{\omega},
\end{align}
implying that $\gamma_{00}=\gamma_{11}=1/2$ (recall that $0\leq \gamma_{aa} \leq 1$ and $\gamma_{00}+\gamma_{11}=1$). Thus we conclude
\begin{align}
	\qty{\hat{\Gamma}_1^\dagger\hat{\Gamma}_0,\hat{\omega}} &= 0.\label{eq:A_inv_A_inv_AC}
\end{align}
By taking trace of \eqref{eq:A_inv_A_inv_basic} we also find
\begin{align}
	\tr \hat{\omega} &= 0
\end{align}
Therefore, we can write
\begin{align}
	\hat{\omega} &= \sum_{i=1}^3 \omega_i \sigma^i,
\end{align}
where $\omega_i\in\mathbb{C}$ are arbitrary coefficients. Similarly, we know that any single site unitary can be written as
\begin{align}
	\hat{\Gamma}_1^\dagger\hat{\Gamma}_0 &= e^{i\phi'}e^{i(\vec{a}\cdot \vec{\sigma})} = e^{i\phi'}\cos(\abs{a}) + i e^{i\phi'}\sin(\abs{a}) \hat{\sigma}_{\vec{m}},
\end{align}
where $\vec{a} = \abs{a}\vec{m}$ and $\hat{\sigma}_{\vec{m}} = \sum_i m_i \sigma^i$. Plugging this into \eqref{eq:A_inv_A_inv_AC} we obtain
\begin{align}
	\qty(\sum_i \omega_i\sigma^i) \cos(\abs{a}) + i\sin(\abs{a}) \sum_i m_i \omega_i &= 0
\end{align}
This implies that $\cos(\abs{a}) = 0$ and hence
\begin{align}
	\hat{\Gamma}_1^\dagger\hat{\Gamma}_0 &=  e^{-i\phi} \hat{\sigma}_{\vec{m}},
\end{align}
from which we conclude
\begin{align}
		\hat{\Gamma}_1 &= e^{i\phi}\hat{\sigma}_{\vec{n}}\hat{\Gamma}_0,
\end{align}
where\footnote{Recall that $\hat{\Gamma}_0\hat{\sigma}_{\vec{m}}\hat{\Gamma}_0^\dagger$ can be seen as a rotation of the 3d vector $\vec{m}$. In particular, $\vec{n}$ is also a normalized vector.} $\hat{\sigma}_{\vec{n}} = \hat{\Gamma}_0\hat{\sigma}_{\vec{m}}\hat{\Gamma}_0^\dagger$.

Therefore, we conclude that the quantum channel is only not invertible if and only if 
\begin{align}
	\hat{\Gamma}_1 &= e^{i\phi}\hat{\sigma}_{\vec{n}}\hat{\Gamma}_0,
\end{align}
in which case $\hat{\gamma} = \tfrac{1}{2}\mathbb{1}$ is necessarily the infinite temperature state. We can explicitly check:
\begin{align}
	\bm{A}[\mathbb{1}] &= \mathbb{1}\\
	\bm{A}[\hat{\Gamma}_0^\dagger\hat{\sigma}_{\vec{n}}\hat{\Gamma}_0] &= \hat{\sigma}_{\vec{n}}\\
	\bm{A}[\hat{\Gamma}_0^\dagger\hat{\sigma}_{\vec{v}}\hat{\Gamma}_0] &= 0,
\end{align}
where in the last line $\vec{v}\cdot\vec{n}$ = 0. 

In fact, we can write
\begin{align}
    \fpic{Anoninv/app_quantumnoninv_Anoninv32} &= \fpic{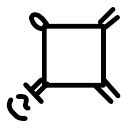} = \fpic{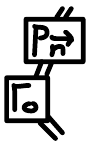} = \fpic{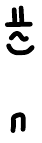} + \frac{1}{2} \fpic{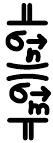}\\
    \fpic{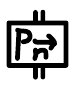} &= \fpic{Anoninv/app_quantumnoninv_Anoninv36} + \frac{1}{2} \fpic{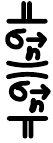}
    %&\includegraphics[width=\linewidth]{figs/fred/Screenshot 2026-03-17 104417},
\end{align}
where $\bm{P}_{\vec{n}}$ is the quantum channel projecting on $\mathbb{1}$ and $\hat{\sigma}_{\vec{n}}$.

We now do a case by case analysis.
\paragraph*{Case 1: $\bra{0}\hat{\sigma}_{\vec{n}}\ket{0} = 0$.}
In this case we also have $\bra{1}\hat{\sigma}_{\vec{n}}\ket{1} = 0$ since $\tr \hat{\sigma}_{\vec{n}} = 0$.

Given this identity it is a simple matter to show that
\begin{align}
    \fpic{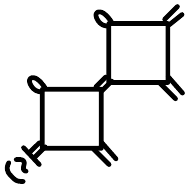} = \fpic{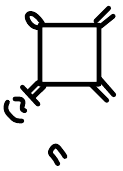} + \frac{1}{2} \overbrace{\fpic{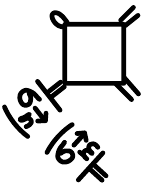}}^{=0} = \fpic{Anoninv/app_quantumnoninv_Anoninv4} 
\end{align}
meaning that the gate satisfies one of the $\mathrm{DU}(2)$ conditions and hence it is influence-solvable.

\paragraph*{Case 2: $\bra{0}\hat{\sigma}_{\vec{m}}\ket{0} = 0$.}

Again, this also implies $\bra{1}\hat{\sigma}_{\vec{m}}\ket{1} = 0$, from which we find
\begin{align}
    \fpic{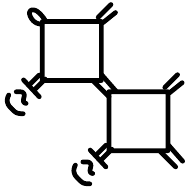} = \fpic{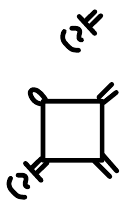} + \frac{1}{2} \underbrace{\fpic{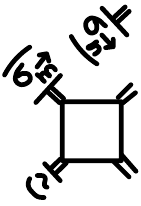}}_{=0} = \fpic{Anoninv/app_quantumnoninv_Anoninv7}
\end{align}
that the gate satisfies (one of the) $\mathrm{DU}(2)$ conditions and hence it is influence-solvable.

\paragraph*{Case 3: $\bra{0}\hat{\sigma}_{\vec{n}}\ket{0} \neq 0$ and $\bra{0}\hat{\sigma}_{\vec{m}}\ket{0} \neq 0$.}

In this case there is no influence-solvability. To show this we need to perform a general analysis. First, note that because of the projector $\bm{P}_{\vec{n}}$ the subspace $\bm{L}$ becomes effectively lower dimensional.

To make use of this, we define the spaces
\begin{align}
    \mathscr{Q} &= \qty{\fpic{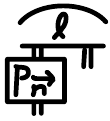}; \brA{l} \in \mathscr{L}} = \mathop{\mathrm{span}}_{ab}\qty{\fpic{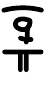}\fpic{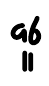}; \brA{q} \in \mathscr{Q}_{ab}}\\
    \mathscr{Q}_{ab} &= \qty{\fpic{Anoninv/app_quantumnoninv_Anoninv9}; \brA{l} \in \mathscr{L}_{ab}} = \qty{\fpic{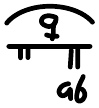}\fpic{Anoninv/app_quantumnoninv_Anoninv13}; \brA{q} \in \mathscr{Q}}
    %&\includegraphics[width=\linewidth]{figs/fred/Screenshot 2026-03-17 105044}
\end{align}
and note that $\mathscr{Q}_{ab}\subset\mathrm{span}\qty{\mathbb{1},\hat{\sigma}_{\vec{n}}}$ is at most two-dimensional. In particular, we already know that $\mathbb{1} \in \mathscr{Q}_{aa}$.

Since $\bm{A}$ contains a projector we can write:
\begin{align}
    \fpic{Ainv/app_quantumnoninv_Ainv21} := \fpic{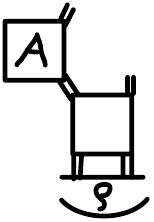} = \fpic{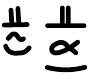} + \frac{1}{2}\fpic{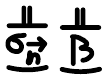}
\end{align}
using which we can rewrite \eqref{eq:zerobond_cond_b} into
\begin{align}
    \fpic{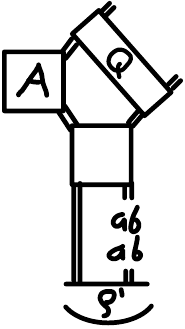} = \frac{1}{2}\fpic{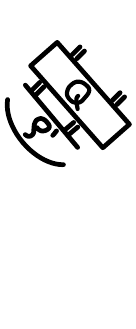}\delta_{ab}
\end{align}
or equivalently:
\begin{align}
    \fpic{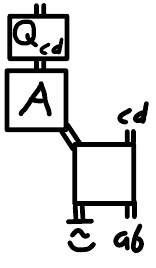} \alpha_{ab} + \fpic{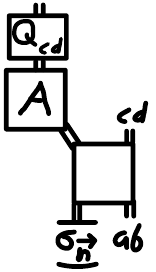} \beta_{ab} = \frac{1}{2} \delta_{ab} \qty(\fpic{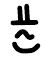} \alpha_{cd} + \frac{1}{2} \fpic{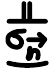} \beta_{cd})\label{eq:app_d2quantum_notinv_notinv_condb}.
\end{align}

Note that since $\hat{\gamma} = \tfrac{1}{2}\mathbb{1}$ we have that $\alpha_{aa} = \tfrac{1}{2}$. In the following we are going to analyse \eqref{eq:app_d2quantum_notinv_notinv_condb} for $a=b$ and $c=d$:
\begin{align}
    \fpic{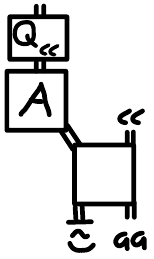} + \fpic{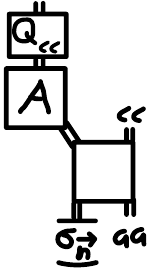}\beta_{aa} = \frac{1}{2} \fpic{Anoninv/app_quantumnoninv_Anoninv37} + \frac{1}{2} \fpic{Anoninv/app_quantumnoninv_Anoninv39}\beta_{cc}\label{eq:app_d2quantum_notinv_notinv_condb_ac}
\end{align}
which will already be sufficient to determine all solvable gates. 

Let us first briefly discuss the case where $\fpic{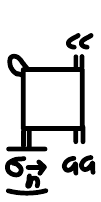} = 0$. Note that if this holds for any $a$ and $c$ it holds for all $a$ and $c$. In this case we can use that $\mathbb{1}\in\mathscr{Q}_{cc}$ to obtain $\fpic{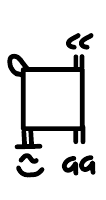} = \frac{1}{2}$. This implies
\begin{align}
     0 = \fpic{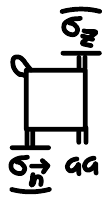} = \frac{1}{2}\underbrace{\Tr(\hat{\sigma}_{\vec{n}}\hat{\sigma}\ind{z})}_{=2n\ind{z} \neq 0} \underbrace{\bra{a}\hat{\sigma}_{\vec{m}}\ket{a}}_{\neq 0}
\end{align}
which is not possible since the RHS does not vanish due to the assumption that $\hat{\sigma}_{\vec{n}}$ and $\hat{\sigma}_{\vec{m}}$ have non-zero diagonal terms.

Therefore, in order to find influence-solvability we need to have that
\begin{align}
    \fpic{Anoninv/app_quantumnoninv_Anoninv31} \neq 0.\label{eq:app_d2quantum_notinv_notinv_cond1}
\end{align}
This immediately implies that $\mathscr{Q}_{aa} = \mathrm{span}=\qty{\mathbb{1},\hat{\sigma}_{\vec{n}}}$ is the maximum possible space. Thus \eqref{eq:app_d2quantum_notinv_notinv_condb_ac} splits into the two equations
\begin{align}
    \fpic{Anoninv/app_quantumnoninv_Anoninv24} + \fpic{Anoninv/app_quantumnoninv_Anoninv31} \beta_{aa} &= \frac{1}{2}\label{eq:app_d2quantum_notinv_notinv_eq1}\\
    \fpic{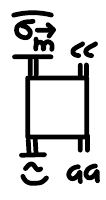} + \fpic{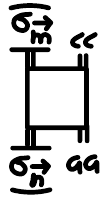}\beta_{aa} = \beta_{cc}\label{eq:app_d2quantum_notinv_notinv_eq2}.
\end{align}
Taking the difference of the first equation for $c=0$ and $c=1$ we obtain
\begin{align}
    \beta_{aa} = - \fpic{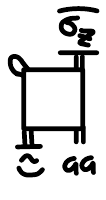}\Bigg/\fpic{Anoninv/app_quantumnoninv_Anoninv29} =: \beta
\end{align}
Note that the denominator is non-zero due to \eqref{eq:app_d2quantum_notinv_notinv_cond1} and that the dependence of $a$ disappears. Note that $\hat{\beta}$ cannot be zero since otherwise either $\bra{0}\hat{\sigma}_{\vec{n}}\ket{0} = 0$ or $\bra{1}\hat{\sigma}_{\vec{n}}\ket{1} = 0$. Inserting the fact that $\beta_{00}=\beta_{11}=\beta$ into \eqref{eq:app_d2quantum_notinv_notinv_eq2} and summing over all $a$ and $c$ we obtain
\begin{align}
    \fpic{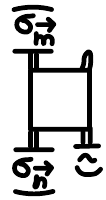} = 2 \bra{0}\hat{\sigma}_{\vec{n}} \ket{0}\bra{0}\hat{\sigma}_{\vec{m}} \ket{0} \overset{!}{=} 2
\end{align}
Since $\abs{\bra{0}\hat{\sigma}_{\vec{n}}\ket{0}} = \abs{n\ind{z}} \leq 1$ the only two options are 
\begin{align}
    \hat{\sigma}_{\vec{n}}=\hat{\sigma}_{\vec{m}} = \pm \hat{\sigma}\ind{z}.
\end{align}
However, this can only be true if $\hat{\Gamma}_0$ is a diagonal matrix. Therefore, the LHS of \eqref{eq:app_d2quantum_notinv_notinv_eq1} vanishes for $c=0$ and $a=1$, which contradicts its RHS.

\section{Proof of \eqref{eq:gamma-and-beta-from-below}} \label{apdx:proof-of-gamma-and-beta-from-below}

Let $\hat{\rho}_\mathrm{lr} = \hat \gamma \otimes \hat \sigma$. After the application of the gate, we have:
\begin{equation*}
    \hat \rho'_\mathrm{lr} = \hat u(\hat \gamma \otimes \hat \sigma) \hat u^\dagger
\end{equation*}

Since the gate is unitary, entropy is conserved:
\begin{equation*}
    S^\mathrm{in} = S_\gamma + S_\sigma = S^\mathrm{out} = S_{\rho'_{LR}}.
\end{equation*}

But by \eqref{eq:gamma-and-beta-from-left}, $S_{\rho'_R} = S_{\beta}$. Using the subadditivity of entropy, $S_{AB} \le S_A + S_B$, we have
\begin{equation*}
    S_\gamma + S_\sigma \le S_{(\gamma \otimes \sigma)'_L} + S_\beta.
\end{equation*}

Also considering the right hand side of \eqref{eq:gamma-and-beta-from-left}, we have
\begin{equation*}
    S_\beta + S_\rho \le S_{(\beta \otimes \rho)'_L} + S_\gamma.
\end{equation*}

Combining these inequalities, we have
\begin{align*}
    \Rightarrow \quad & \cancel{S_\gamma} + S_\sigma \le S_{(\gamma \otimes \sigma)'_L} + S_{(\beta \otimes \rho)'_L} - S_\rho + \cancel{S_\gamma} \\
    \Rightarrow \quad & S_\sigma + S_\rho \le S_{(\gamma \otimes \sigma)'_L} + S_{(\beta \otimes \rho)'_L} \le 2\ln d.
\end{align*}

Let $\sigma, \rho = \mathbb{1}/d$. Then we have
\begin{equation*}
    2\ln d \le S_{(\gamma \otimes \mathbb{1}/d)'} + S_{(\beta \otimes \mathbb{1}/d)'} \le 2\ln d.
\end{equation*}

This implies that
\begin{align*}
    & S_{(\gamma \otimes \mathbb{1}/d)'} = S_{(\beta \otimes \mathbb{1}/d)'} = \ln d \\
    \Rightarrow \quad & (\gamma \otimes \mathbb{1}/d)'_L = (\beta \otimes \mathbb{1}/d)'_L = \mathbb{1}/d.
\end{align*}

Now, substituting back, we have
\begin{equation*}
    S_\gamma + \cancel{\ln d} = S_{(\gamma \otimes \mathbb{1}/d)'} \le \cancel{\ln d} + S_\beta.
\end{equation*}
By symmetry, we also have $S_\beta \le S_\gamma$, implying that $S_\beta = S_\gamma$.

Finally, we have
\begin{align*}
    & \cancel{S_\gamma} + {\ln d} = S_{(\gamma \otimes \mathbb{1}/d)'} \leq {\ln d} + \overset{S_\gamma}{\cancel{S_\beta}} = S_{(\gamma \otimes \mathbb{1}/d)'_L} + S_{(\gamma \otimes \mathbb{1}/d)'_R} \\
    & \implies S_{(\gamma \otimes \mathbb{1}/d)'} = S_{(\gamma \otimes \mathbb{1}/d)'_L} + S_{(\gamma \otimes \mathbb{1}/d)'_R}.
\end{align*}

Equality holds only if $(\gamma \otimes \mathbb{1}/d)'$ is a product state. Therefore we have
\begin{equation*}
    \Rightarrow (\gamma \otimes \mathbb{1}/d)' = \mathbb{1}/d \otimes \beta,
\end{equation*}
i.e. LHS of \eqref{eq:gamma-and-beta-from-below}. RHS follows by symmetry.

\section{Details on influence-solvable $d=2$ classical gates}\label{app:classic_d2}
In \cref{tab:app_d2} we give a full list of all $d=2$ classical gates and the $\keT{\gamma}$ for which they are (left) influence-solvable.
\begin{table}[!h]
    \centering
    \begin{tabular}{ccc}
        \toprule
        $\sigma$ & Comment & $\keT{\gamma}$\\
        \midrule
        0 & identity & all \\
        1 & west & $(1,0)^\top$ $(1/2,1/2)^\top$ \\
        2 & SWAP & all \\
        3 & & all \\
        4 & & $(1/2,1/2)^\top$ \\
        5 & east & all \\
        6 & & $(0,1)^\top$ $(1/2,1/2)^\top$ \\
        7 & & all \\
        8 & & $(1/2,1/2)^\top$ \\
        9 & & all \\
        10 & & all \\
        11 & & all \\
        12 & & all \\
        13 & & all \\
        14 & & all \\
        15 & & $(1/2,1/2)^\top$ \\
        16 & & all \\
        17 & & $(0,1)^\top$ $(1/2,1/2)^\top$ \\
        18 & & all \\
        19 & & $(1/2,1/2)^\top$ \\
        20 & & all \\
        21 & & all \\
        22 & & $(1,0)^\top$ $(1/2,1/2)^\top$ \\
        23 & & all\\
        \bottomrule
    \end{tabular}
    \caption{A full list of all left solvable $\gamma$ for each $d=2$ classical gate labeled by their permutation index $\sigma$.}
    \label{tab:app_d2}
\end{table}

\section{Proof of existence of solvable initial states for generalized $\mathrm{DU}(2)$}\label{app:DU2_bottom_initstate}
The two generalizations of $\mathrm{DU}(2)$ of \cref{res:zerobond_DU1_DU2} can be used to simplify the circuit and establish influence-solvability similar to \eqref{eq:FTMPS_rel_DU2_bottom_simplify}. However, this only works if there is a compatible initial state. For the left condition of \eqref{eq:zerobond_DU2_gen} any initial state is solvable. However, for the right condition of \eqref{eq:zerobond_DU2_gen}
\begin{align}
    \fpic{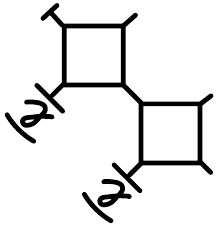} = \fpic{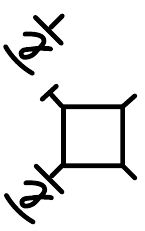}
\end{align}
one additionally needs a compatible initial state. Such a condition is for instance $\fpic{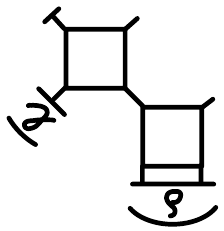} = \fpic{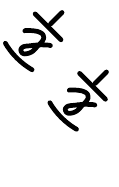}$.

Here we want to show that such a compatible initial state always exists. We start by defining $\keT{\alpha}$ as
\begin{align}
    \fpic{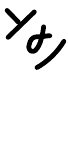} = \fpic{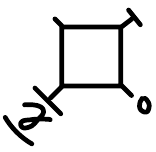}.
\end{align}
This satisfies
\begin{align}
\fpic{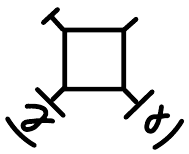} = \fpic{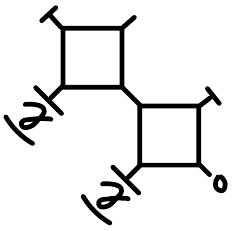} = \fpic{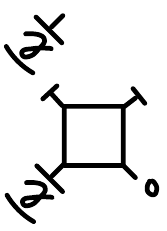} = \fpic{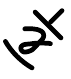}.
\end{align}
Therefore if we define the initial state $\keT{\rho}$ via 
\begin{align}
    \fpic{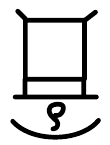} = \fpic{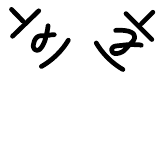}
\end{align}%\includegraphics[width=2cm]{figs/fred/Screenshot 2026-03-28 190151}
we conclude
\begin{align}
    \fpic{compinit/app_compinit4} = \fpic{compinit/app_compinit8}\;\fpic{compinit/app_compinit11} = \fpic{compinit/app_compinit5}.
\end{align}
Thus, we conclude that it is always possible to find a compatible initial state $\keT{\rho}$.

\section{Review of translation invariant MPS states}\label{app:mps_space}

Translation invariant MPS states~\cite{Fannes1992} are a simple way to obtain a translation invariant state in the thermodynamic limit. They are defined on a finite system of size $L$ as follows
\begin{align}
    \fpic{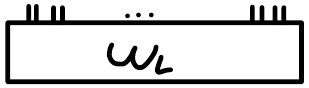} = \fpic{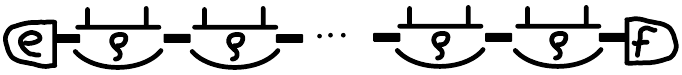}
\end{align}

In order to define them on the infinite system, one restricts to local observables and defines
\begin{align}
    \omega(A) = \fpic{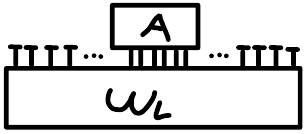} = \fpic{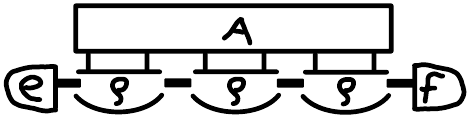}.
\end{align}
Here, $L$ can be any system size larger than the support of $A$. In order to ensure that we obtain the same result for any $L$ we need to assume that the left and right boundary states are eigenvectors of
\begin{align}
    \fpic{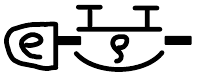} &= \fpic{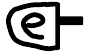} & \fpic{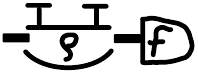} &= \fpic{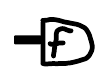}.\label{eq:app_mps_space_boundary_vec_cond}
\end{align}
This way, one can compute the expectation value $\omega(A)$ of any observable $A$ which is sufficient to formally define a state. In order to ensure that the state is normalized we need that
\begin{align}
    \omega(1) = \fpic{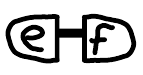} = 1.
\end{align}

These states have many useful properties allowing for efficient computation. For instance, consider a correlation function
\begin{align}
    \expval{A_0B_x} &=\omega(A_0 B_x)\nonumber\\
    &= \fpic{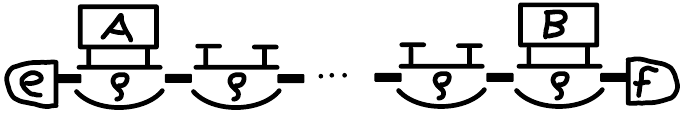}.%\\
\end{align}
By defining the matrices
\begin{align}
    \bm{\rho}^A &= \fpic{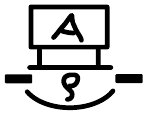} & \bm{\rho}^B &= \fpic{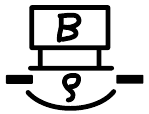} &
    \bar{\bm{\rho}} &= \fpic{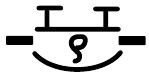}
\end{align}
we can write this as
\begin{align}
    \expval{A_0B_x} &= \vec{e}^\top\bm{\rho}^A\bar{\bm{\rho}}^{x-1}\bm{\rho}^B\vec{f}\\
    &= \sum_i \lambda_i^{x-1}\qty(\vec{e}^\top\bm{\rho}^A\vec{v}_i)\qty(\vec{v}_i^\top\bm{\rho}^B\vec{f}).
\end{align}
Here $\vec{v}_i$ and $\lambda_i$ are the eigenvalues of $\bar{\bm \rho}$ (this assumes that $\bar{\bm \rho}$ is diagonalizable, otherwise one has to use its Jordan normal form). This way we can efficiently compute correlation functions in translation invariant MPS states.

To be physically meaningful all eigenvalues must be $\abs{\lambda_i} \leq 1$ (otherwise correlation functions would explode). Note that we know from \eqref{eq:app_mps_space_boundary_vec_cond} that at least one $\lambda_i=1$. In the case that all other $\abs{\lambda_i} < 1$ the state is clustering, i.e. as $x\to \infty$ the state decorrelates:
\begin{align}
    \expval{A_0B_x}\upd{c} :=\expval{A_0B_x}-\expval{A_0}\expval{B_x} \to 0.
\end{align}
In fact the decay is exponential and given by the gap $\bar{\lambda}^x$, where $\bar{\lambda}$ is the largest eigenvalue smaller than $1$.

Note that translation invariant MPS states are defined with open boundary conditions. Therefore, crucially, the corresponding MPS with periodic boundary conditions
\begin{align}
    \fpic{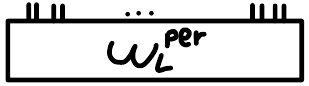} = \fpic{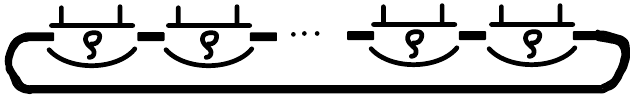}
\end{align}
are not necessarily states for any finite $L$. For instance, there is no reason for them to be normalized. In fact we have
\begin{align}
    \omega_L\upd{per}(1) &= \tr \bar{\bm\rho}^L = \sum_i \lambda_i^L,
\end{align}
which in general only approaches $1$ as $L\to \infty$. The only option to have a proper normalized state for periodic systems of any size $L$ is to have $\tr \bar{\bm\rho}^L=1$. Note that this implies that $\bar{\bm\rho}$ can only have a single non-zero eigenvalue. In fact, unless the state is factorized, this means that $\bar{\bm\rho}$ is not diagonalizable, but consists of Jordan blocks. In particular, connected correlation functions do not decay exponentially, but instead identically vanish after a finite number of lattice sites.

Due to the clustering property translation invariant MPS states can be used to approximate any clustering translation invariant state~\cite{Fannes1992}. An important example of clustering states are Gibbs states $e^{-\beta \hat{H}}/Z$. In fact, they can efficiently be approximated by MPS states~\cite{PhysRevB.73.085115,Verstraete01032008,PhysRevB.91.045138}.

\end{document}